\documentclass[12pt]{article}
\pdfoutput=1

\usepackage{graphics,amssymb,float}
\usepackage{graphicx}
\usepackage{rotating}
\usepackage{dcolumn}
\usepackage{bm}
\usepackage{cite}
\usepackage{amsmath}
\usepackage{indentfirst} 
\usepackage{epstopdf}

\def\CU{C_U}
\def\CD{C_D}
\def\CL{C_L}
\def\CV{C_V}
\def\CG{C_g}
\def\CP{C_\gamma}
\def\cu{\CU}
\def\cd{\CD}
\def\cv{\CV}
\def\cg{\CG}
\def\cp{\CP}
\def\sina{\sin\alpha}
\def\cosa{\cos\alpha}
\def\tanb{\tan\beta}
\def\sinb{\sin\beta}
\def\cosb{\cos\beta}
\def\cotb{\cot\beta}

\def\chimin{\chi^2_{\rm min}}
\def\dof{{\rm d.o.f.}}
\def\gam{\gamma}
\def\ie{{\it i.e.}}
\def\eg{{\it e.g.}}
\def\what{\widehat}
\def\dcg{\Delta \CG}
\def\dcp{\Delta \CP}
\def\cgb{\anti \CG}
\def\cpb{\anti \CP}
\def\cl{\CL}
\def\gev{~{\rm GeV}}
\def\chisq{\chi^2}
\def\anti{\overline}
\def\fbi{~{\rm fb}^{-1}}

\def\bit{\begin{itemize}}
\def\eit{\end{itemize}}
\def\ben{\begin{enumerate}}
\def\een{\end{enumerate}}
\def\beq{\begin{equation}}
\def\eeq{\end{equation}}
\def\mw{m_W}
\def\mz{m_Z}
\def\ggf{{\rm ggF}}
\def\vbf{{\rm VBF}}
\def\vh{{\rm VH}}
\def\tth{{\rm ttH}}
\def\Eq#1{Eq.~(\ref{#1})}

\def\lsim{\mathrel{\raise.3ex\hbox{$<$\kern-.75em\lower1ex\hbox{$\sim$}}}}
\def\gsim{\mathrel{\raise.3ex\hbox{$>$\kern-.75em\lower1ex\hbox{$\sim$}}}}
\def\ifmath#1{\relax\ifmmode #1\else $#1$\fi}

\begin{document}
\begin{titlepage}
\begin{center}

\vspace*{-1cm}
\begin{flushright}
LAPTH-061/12\\
LPSC12350\\
LPT Orsay 12-119\\
NSF-KITP-12-236
\end{flushright}

\vspace*{1.6cm}
{\Large\bf Higgs Couplings at the End of 2012} 

\vspace*{1cm}\renewcommand{\thefootnote}{\fnsymbol{footnote}}

{\large 
G.~B\'elanger$^{1}$\footnote[1]{Email: belanger@lapp.in2p3.fr},
B.~Dumont$^{2}$\footnote[2]{Email: dumont@lpsc.in2p3.fr},
U.~Ellwanger$^{3}$\footnote[3]{Email: Ulrich.Ellwanger@th.u-psud.fr},
J.~F.~Gunion$^{4,5}$\footnote[4]{Email: jfgunion@ucdavis.edu}, 
S.~Kraml$^{2}$\footnote[5]{Email: sabine.kraml@lpsc.in2p3.fr}
} 

\renewcommand{\thefootnote}{\arabic{footnote}}

\vspace*{1cm} 
{\normalsize \it 
$^1\,$LAPTH, Universit\'e de Savoie, CNRS, B.P.110, F-74941 Annecy-le-Vieux Cedex, France\\[1mm]
$^2\,$Laboratoire de Physique Subatomique et de Cosmologie, UJF Grenoble 1,
CNRS/IN2P3, INPG, 53 Avenue des Martyrs, F-38026 Grenoble, France\\[1mm]
$^3\,$Laboratoire de Physique Th\'eorique, UMR 8627, CNRS and
Universit\'e de Paris--Sud, F-91405 Orsay, France\\[1mm]
$^4\,$Department of Physics, University of California, Davis, CA 95616, USA\\[1mm]
$^5\,$Kavli Institute for Theoretical Physics, University of California, Santa Barbara,\\ CA 93106-4030, USA}

\vspace{1cm}

\begin{abstract}
Performing a fit to all publicly available data, we analyze the extent
to which the latest results from the LHC and Tevatron constrain the
couplings of the Higgs boson-like state at $\sim 125\gev$. To this end
we assume that only Standard Model (SM) particles appear in the Higgs
decays, but tree-level Higgs couplings to the up-quarks, down-quarks and
vector bosons, relative to the SM are free parameters. We also assume
that the leptonic couplings relative to the SM are the same as for the
down-quark, and a custodial symmetry for the $V=W,Z$ couplings. In the
simplest approach, the effective Higgs couplings to gluons and photons
are computed in terms of the previous parameters. This approach is also
applied to Two-Higgs-Doublet Models of Type~I and Type~II. However, we
also explore the possibility that the net Higgs to $gg$ and $\gam\gam$
couplings have extra loop contributions coming from Beyond-the-Standard
Model physics. We find that the SM $p$-value $\sim 0.5$ is more than
$2\sigma$ away from fits in which: a) there is some non-SM contribution
to the $\gam\gam$  coupling of the Higgs; or b) the sign of the top
quark coupling to the Higgs is opposite that of the W coupling. In both
these cases $p$-values $\sim 0.9$ can be achieved. Since option b) is
difficult to realize in realistic models, it would seem that new physics
contributions to the effective couplings of the Higgs are preferred.
\end{abstract}

\end{center}

\end{titlepage}

\section{Introduction}

The recent discovery~\cite{atlas:2012gk, cms:2012gu} of a new particle with properties consistent with a Standard Model (SM) Higgs boson is clearly the most significant news from the Large Hadron Collider (LHC). 
This discovery was supported by evidence for a Higgs boson found by the CDF and D0 collaborations at the
Tevatron \cite{tevatron:2012zzl} and completes our picture of the SM. 
However, the SM leaves many fundamental questions open---perhaps the most pressing issue being that the SM does not explain the value of the electroweak scale, \ie\  the Higgs mass, itself.  Clearly, a prime goal after the discovery is to thoroughly test the SM nature for this Higgs-like signal. 

The SM makes precise predictions for the production cross sections
of the Higgs boson $H$ (via gluon-gluon fusion (ggF), vector boson
fusion (VBF), associated production with an electroweak gauge
boson $V=W,Z$ (VH) and associated production with a $t\anti t$ pair (\tth)), 
and its decay branching fractions into various
final states ($\gamma \gamma$, $ZZ^{(*)}$, $WW^{(*)}$, $b\anti b$, and
$\tau \tau$) as a function of its unpredicted mass $M_H$. The observation
of the Higgs boson at the LHC is based primarily on the 
$\gamma\gamma$ \cite{ATLAS-CONF-2012-168,CMS-PAS-HIG-12-015}, 
$ZZ^{(*)}$ \cite{ATLAS-CONF-2012-169,CMS-PAS-HIG-12-041}
and $WW^{(*)}$ \cite{ATLAS-CONF-2012-162,CMS-PAS-HIG-12-042} decay modes. 
The Higgs boson mass, $M_H$, is quite precisely measured to be in the 125--126~GeV 
range using the high resolution  $\gamma \gamma$ and $ZZ^{(*)}$ final states~\cite{ATLAS-CONF-2012-170,CMS-PAS-HIG-12-045}.\footnote{Although ATLAS finds a lower value of $M_H\simeq123.5$~GeV in the $ZZ$ channel, the combined value from ATLAS is $M_H=125.2\pm 0.3\ {\rm (stat)} \pm 0.6\ {\rm (sys)}$~GeV, in very good agreement with $M_H=125.8\pm 0.4\ {\rm (stat)} \pm 0.4\ {\rm (sys)}$~GeV measured by CMS. We find it reasonable to assume that the lower $M_H$ value from the ATLAS $H\to ZZ$ measurement is due to a statistical fluctuation or unknown systematics.}
The evidence for the Higgs boson at the Tevatron is based principally on the
$b\anti b$ decay mode \cite{Aaltonen:2012qt,tevatron:2012zzl,HCPtevBBtalk}, the observed enhancements
being consistent with a large range of possible Higgs masses.

With the measurements in various channels, a comprehensive study of the properties of the Higgs-like state becomes possible and has the potential for revealing whether or not the Higgs sector is as simple as envisioned in the SM.  In particular it is crucial to determine the Higgs couplings to gauge bosons and to fermions as defined by the Lagrangian
\begin{equation}
  {\cal L} =  g\left[  C_W\,  \mw W_\mu W^\mu + C_Z{\mz\over \cos\theta_W} Z_\mu Z^\mu - \sum_F C_F {m_F\over 2\mw} \bar F F\, \right] H\,, 
\label{ldef}
\end{equation}
where the $C_I$ are scaling factors for the couplings relative to their SM values, introduced to test possible deviations in the data from SM expectations. In principle all the $C_I$ are independent, in particular the $C_F$ can be different for up- and down-type quarks and/or leptons. A significant deviation of any $C_I$ from unity would imply new physics beyond the SM. 

While fits to various combinations of $C_I$'s are performed by the experimental collaborations 
themselves~\cite{ATLAS-CONF-2012-127, CMS-PAS-HIG-12-045}, 
we find it important to develop our own scheme in order to bring all results from ATLAS, CMS 
and the Tevatron experiments together and test not only the SM but also specific models beyond.  
Such fits by theorists, using various parametrizations, were performed previously 
in \cite{Carmi:2012yp,Azatov:2012bz,Espinosa:2012ir,Klute:2012pu,Azatov:2012wq,
Carmi:2012zd,Low:2012rj,Corbett:2012dm,Giardino:2012dp,Ellis:2012hz,Montull:2012ik,
Espinosa:2012im,Carmi:2012in,Banerjee:2012xc,Bonnet:2012nm,Plehn:2012iz,
Espinosa:2012in,Elander:2012fk,Djouadi:2012rh,Altmannshofer:2012ar,Dobrescu:2012td,Chang:2012ve,
Moreau:2012da,Cacciapaglia:2012wb,Bechtle:2012jw,Corbett:2012ja,Masso:2012eq,Azatov:2012qz}. 
Here, we go beyond these works by including all publicly available data as of the end of 2012. 
In particular we take into account the updates presented at the Hadron Collider Physics 
Symposium in Nov.~2012 (HCP2012) \cite{HCP} and at the 
Open Session of the CERN Council in Dec.~2012 \cite{cerncouncil}.

Our parametrization is as follows. We treat the couplings to up-type and down-type fermions, $\cu$ and $\cd$, as independent parameters (but we  only consider the case $\cl=\cd$, and we assume that the $C_F$ are family universal). 
Moreover, we assume a custodial symmetry in employing a single $C_W=C_Z\equiv\CV$ in Eq.~(\ref{ldef}). 
The structure we are testing thus becomes
\begin{equation}
   {\cal L} =  g\left[  \CV \left(\mw W_\mu W^\mu+{\mz\over \cos\theta_W} Z_\mu Z^\mu \right)  
   - \CU {m_t\over 2\mw} \bar tt  - \CD    {m_b\over 2\mw} \bar bb - \CD {m_\tau\over 2 \mw}\bar\tau\tau \right ]H\,. 
   \label{ourldef}
\end{equation}
In general, the $C_I$ can take on negative as well as positive values; there is one overall sign ambiguity which we fix by taking $\CV>0$.
Even in this restricted context, various types of deviations of these three  $C_I$ from unity are possible in extended theories such as Two-Higgs-Doublet Models (2HDMs), models with singlet-doublet mixing, and supersymmetric models such as the Minimal Supersymmetric Standard Model (MSSM) and the Next-to-Minimal Supersymmetric Standard Model (NMSSM). 

In addition to the tree-level couplings given above, the $H$ has couplings to $gg$ and $\gam\gam$ that are first induced at one loop and are completely computable in terms of $\cu$, $\cd$ and $\cv$ if only loops containing SM particles are present. We define $\anti \cg$ and $\anti \cp$ to be the ratio of these couplings so computed to the  SM (\ie\ $\cu=\cd=\cv=1$) values.
However, in some of our fits we will also allow for additional loop contributions $\dcg$ and $\dcp$ from new particles; in this case $\cg=\anti \cg+\dcg$ and $\cp=\anti \cp +\dcp$. The largest set of independent parameters in our fits is thus
\begin{equation}
   \cu,~\cd,~\cv,~\dcg,~\dcp\,.
\end{equation}

In this study, we focus on models in which the Higgs decays only to SM particles, in particular not allowing for invisible (e.g. $H\to \tilde \chi^0_1 \tilde \chi^0_1$, where $\tilde \chi^0_1$ is the lightest SUSY particle) or undetected decays (such as $H\to aa$, where $a$ is a light CP-odd, perhaps singlet scalar). This approach, when we allow in the most general case for the $\cu$, $\cd$, $\cv$, $\cp$ and $\cg$ couplings to be fully independent, encompasses a very broad range of models, including in particular those  in which the Higgs sector consists of any number of doublets  + singlets, the only proviso being the absence of decays of the observed $\sim 125\gev$ state to non-SM final states. 
(A fit for invisible Higgs decays was performed early on in \cite{Espinosa:2012vu}.) 
This approach however does not cover models  such as composite models and Higgs-radion mixing models
for which the $VVH$ coupling has a more complicated tensor structure than that given in \Eq{ourldef}.
Our procedure will also be inadequate should the observed signal at $\sim 125\gev$ actually arise from two 
or more degenerate Higgs bosons (see {\it e.g.}~\cite{Gunion:2012gc,Ferreira:2012nv}).   
Although the success of our fits implies that there is no need for such extra states, the explicit tests for degenerate states 
developed in \cite{Gunion:2012he} should be kept in mind as a  means to test directly for two or more Higgs bosons
contributing to the signal at 125--126 GeV.

This paper is organized as follows. The experimental inputs and our fitting procedure are described in Section~2. The results of  three generic fits are presented in Section~3 together with the results of a fit in Two-Higgs-Doublet models. Section~4 contains our conclusions.

\clearpage
\section{Experimental inputs and fitting procedure}

We perform fits employing all production/decay channels for which results are available from the ATLAS and CMS collaborations at the LHC, as well as the Tevatron CDF+D0 Higgs results. 
The experimental results are given in terms of signal strengths $\mu(X,Y)$, the ratio of the observed rate for some process $X\to H \to Y$ relative to the prediction for the SM Higgs.  
Often it is the case that several production processes contribute to a given experimental channel.  For example, both vector boson fusion and gluon fusion can contribute to the ``VBF" channels (or ``categories'') that are defined by a given set of experimental cuts. In comparing theory to experiment it is thus important to incorporate the estimates from the experiments of the relative contributions of the theoretically distinct production/decay processes.  
The values for the signal strengths in the various (sub)channels as reported by the experiments and used in this analysis, together with the estimated decompositions into production channels are given in Tables~\ref{ATLASresults}--\ref{Tevatronresults}. 

We adopt the simple technique of computing the $\chi^2$ associated with a given choice of the input parameters following the standard definition: 
\beq
  \chi^2=\sum_k {(\overline\mu_k-\mu_k)^2\over \Delta \mu_k^2}\,,
\label{chisqdef}
\eeq
where $k$ runs over all the experimentally defined production/decay channels employed, $\mu_k$ is the observed signal strength for channel $k$, $\overline\mu_k$ is the value predicted for that channel for a given choice of parameters and $\Delta \mu_k$ is the experimental error for that channel.  The $\overline\mu_k$ associated with each experimentally defined channel is further decomposed as 
\beq
  \overline\mu_k=\sum T^i_k \what \mu_i
  \label{murel}
\eeq
where the $T^i_k$ give the amount of contribution to the experimental channel $k$ coming from the theoretically defined channel $i$ and $\what \mu_i$ is the prediction for that channel for a given choice of $\CU$, $\CD$, $\CV$ and (for fits where  treated as independent) $\cp$ and $\cg$.  
For the computation of the $\what \mu_i$ including NLO corrections we follow the procedure recommended by the 
LHC Higgs Cross Section Working Group in~\cite{LHCHiggsCrossSectionWorkingGroup:2012nn}. In particular 
we include all the available QCD corrections for $C_g$ using \texttt{HIGLU}~\cite{Spira:1995mt,Spira:1996if} 
and for $C_\gamma$ using \texttt{HDECAY}~\cite{Spira:1996if,Djouadi:1997yw}, and we switch off the
electroweak corrections. 
The $T^i_k$ depend on the specific analysis and hence differ from experiment to experiment. 
Often, the $T^i_k$ are determined from simulations of a SM Higgs signal.  
In some cases, the experiments have done the unfolding of theoretical vs.\ experimental 
channels from the data and provide directly experimental results for the theoretically relevant $\what\mu_i$'s 
and the correlations between them. \\


\begin{table}\centering
\begin{tabular}{|c|c|c|cccc|}
\hline
Channel & Signal strength $\mu$ & $M_H$ (GeV) & \multicolumn{4}{c|}{Production mode} \\
& & & ggF & VBF & VH & ttH \\
\hline
\multicolumn{7}{|c|}{$H \rightarrow \gamma\gamma$ ($4.8\fbi$ at 7 TeV + $13.0\fbi$ at 8 TeV)~\cite{ATLAS-CONF-2012-168}} \\
\hline
$\mu({\rm ggF+ttH},\gamma\gamma)$ & $1.85 \pm 0.52$ & 126.6 & 100\% & -- & -- & -- \\
$\mu({\rm VBF+VH} ,\gamma\gamma)$ & $2.01 \pm 1.23$ & 126.6 & -- & 60\% & 40\% & -- \\
\hline
\multicolumn{7}{|c|}{$H \rightarrow ZZ$ ($4.6\fbi$ at 7 TeV + $13.0\fbi$ at 8 TeV)~\cite{ATLAS-CONF-2012-169,ATLAS-CONF-2012-170}} \\
\hline
Inclusive & $1.01^{+0.45}_{-0.40}$  & 125 & 87\% & 7\% & 5\% & 1\% \\
\hline
\multicolumn{7}{|c|}{$H \rightarrow WW$ $(13.0\fbi$ at 8 TeV)~\cite{ATLAS-CONF-2012-158,ATLAS-CONF-2012-170}} \\
\hline
$e\nu\mu\nu$ & $1.42^{+0.58}_{-0.54}$ & 125.5 & 95\% & 3\% & 2\% & -- \\
\hline
\multicolumn{7}{|c|}{$H \rightarrow b\bar{b}$ ($4.7\fbi$ at 7 TeV + $13.0\fbi$ at 8 TeV)~\cite{ATLAS-CONF-2012-161,ATLAS-CONF-2012-170}} \\
\hline
VH tag & $-0.39 \pm 1.02$ & 125.5 & -- & -- & 100\% & -- \\
\hline
\multicolumn{7}{|c|}{$H \rightarrow \tau\tau$ ($4.6\fbi$ at 7 TeV + $13.0\fbi$ at 8 TeV)~\cite{ATLAS-CONF-2012-160}} \\
\hline
$\mu({\rm ggF},\tau\tau)$ & $\phantom{-}2.41 \pm 1.57$ & 125 & 100\% & -- & -- &-- \\
$\mu(\mathrm{VBF}+\mathrm{VH},\tau\tau)$ & $-0.26 \pm 1.02$ & 125 & -- & 60\% & 40\% & -- \\
\hline
\end{tabular}
\caption{ATLAS results as employed in this analysis. The correlations included in the fits are $\rho = -0.37$ for the $\gamma\gamma$ and $\rho = -0.50$ for the $\tau\tau$ channels.}
\label{ATLASresults}
\end{table}

\begin{table}\centering
\begin{tabular}{|c|c|c|cccc|}
\hline
Channel & Signal strength $\mu$ & $M_H$ (GeV) & \multicolumn{4}{c|}{Production mode} \\
& & & ggF & VBF & VH & ttH \\
\hline
\multicolumn{7}{|c|}{$H \rightarrow \gamma\gamma$ ($5.1\fbi$ at 7 TeV + $5.3\fbi$ at 8 TeV)~\cite{cms:2012gu,CMS-PAS-HIG-12-015,CMS-PAS-HIG-12-045}} \\
\hline
$\mu({\rm ggF+ttH},\gamma\gamma)$ & $0.95 \pm 0.65$ & 125.8 & 100\% & -- & -- & -- \\
$\mu({\rm VBF+VH},\gamma\gamma)$ & $3.77 \pm 1.75$ & 125.8 & -- & 60\% & 40\% & --\\
\hline
\multicolumn{7}{|c|}{$H \rightarrow ZZ$ ($5.1\fbi$ at 7 TeV + $12.2\fbi$ at 8 TeV)~\cite{CMS-PAS-HIG-12-045,CMS-PAS-HIG-12-041}} \\
\hline
Inclusive & $\phantom{-}0.81^{+0.35}_{-0.28}$ & 125.8 & 87\% & 7\% & 5\% & 1\% \\
\hline
\multicolumn{7}{|c|}{$H \rightarrow WW$ (up to $4.9\fbi$ at 7 TeV + $12.1\fbi$ at 8 TeV)~\cite{CMS-PAS-HIG-12-039,CMS-PAS-HIG-12-042,CMS-PAS-HIG-12-045}} \\
\hline
0/1 jet & $\phantom{-}0.77^{+0.27}_{-0.25}$ & 125.8 & 97\% & 3\% & -- & -- \\
VBF tag & $-0.05^{+0.74}_{-0.55}$ & 125.8 & 17\% & 83\% & -- & -- \\
VH tag & $-0.31^{+2.22}_{-1.94}$ & 125.8 & -- & -- & 100\% & -- \\
\hline
\multicolumn{7}{|c|}{$H \rightarrow b\bar{b}$ (up to $5.0\fbi$ at 7 TeV + $12.1\fbi$ at 8 TeV)~\cite{CMS-PAS-HIG-12-044,CMS-PAS-HIG-12-025,CMS-PAS-HIG-12-045}} \\
\hline
VH tag & $\phantom{-}1.31^{+0.65}_{-0.60}$ & 125.8 & -- & -- & 100\% & -- \\
ttH tag & $-0.80^{+2.10}_{-1.84}$ & 125.8 & -- & -- & -- & 100\% \\
\hline
\multicolumn{7}{|c|}{$H \rightarrow \tau\tau$ (up to $5.0\fbi$ at 7 TeV + $12.1\fbi$ at 8 TeV)~\cite{CMS-PAS-HIG-12-043,CMS-PAS-HIG-12-051,CMS-PAS-HIG-12-045}} \\
\hline
0/1 jet & $\phantom{-}0.85^{+0.68}_{-0.66}$ & 125.8 & 76\% & 16\% & 7\% & 1\% \\
VBF tag & $\phantom{-}0.82^{+0.82}_{-0.75}$ & 125.8 & 19\% & 81\% & -- & -- \\
VH tag & $\phantom{-}0.86^{+1.92}_{-1.68}$ & 125.8 & -- & -- & 100\% & -- \\
\hline
\end{tabular}
\caption{CMS results as employed in this analysis. The correlation included for the $\gamma\gamma$ channel is $\rho = -0.54$.}
\label{CMSresults}
\end{table}

\begin{table}\centering
\begin{tabular}{|c|c|c|cccc|}
\hline
Channel & Signal strength $\mu$ & $M_H$ (GeV) & \multicolumn{4}{c|}{Production mode} \\
& & & ggF & VBF & VH & ttH \\
\hline
\multicolumn{7}{|c|}{$H \rightarrow \gamma\gamma$~\cite{HCPHiggsTevatron}} \\
\hline
Combined & $6.14^{+3.25}_{-3.19}$ & 125 & 78\% & 5\% & 17\% & -- \\
\hline
\multicolumn{7}{|c|}{$H \rightarrow WW$~\cite{HCPHiggsTevatron}} \\
\hline
Combined & $0.85^{+0.88}_{-0.81}$ & 125 & 78\% & 5\% & 17\% & -- \\
\hline
\multicolumn{7}{|c|}{$H \rightarrow b\bar{b}$~\cite{HCPtevBBtalk}} \\
\hline
VH tag & $1.56^{+0.72}_{-0.73}$ & 125 & -- & -- & 100\% & -- \\
\hline
\end{tabular}
\caption{Tevatron results for up to $10\fbi$ at $\sqrt{s} = 1.96$~TeV, as employed in this analysis.}
\label{Tevatronresults}
\end{table}


With this framework programmed,  our fitting procedure is as follows.  We first scan over a fine grid of the free 
parameters of the scenario considered, for example, $\CU$, $\CD$, $\CV$ with $\CG,\CP=\anti\CG,\anti\CP$ 
as computed from the SM-particle loops (this will be Fit {\bf II} below). 
We obtain the value of $\chisq$ associated with each point in the grid 
and thus determine the values of the parameters associated with the approximate minimum (or minima).  
To get the true minimum $\chisq$, $\chimin$, and the associated ``best-fit" values and the one-standard 
deviation ($1\sigma$) errors on them we employ 
MINUIT~\cite{James:1975dr}.  
(The errors on parameters which are not input, \ie\ $\CG$ and $\CP$, are determined from the grid data.)
For plotting distributions of $\chisq$ as a function of any one variable, we use the above grid data 
together with the best fit value, to profile the minimal $\chisq$ value with respect to the remaining 
unconstrained parameters. The 68\%, 95\% and 99.7\% confidence level (CL) intervals are then given 
by $\chisq=\chimin+1$, $+4$ and  $+9$, respectively. 
Two-dimensional $\chisq$ distributions are obtained analogously from a grid in the two parameters of interest, profiling over the other, unseen parameters; 
in this case, we show contours of $\chisq$ corresponding to the 68\% ($\chisq=\chimin+2.30$), 
95\% ($\chisq=\chimin+6.18$) and 99.7\% ($\chisq=\chimin+11.83$) confidence levels for 2 parameters treated jointly. \\

Before presenting our results, a couple of comments are in order. First of all, we stress that in models of new physics beyond-the-SM (BSM), both the branching ratios and the production cross sections  and distributions (and indeed the number of Higgs particles) may differ from SM expectations.
For any BSM interpretation of the Higgs search results  it is absolutely crucial to have as precise and complete channel-by-channel information as possible~\cite{Kraml:2012sg}. 
Unfortunately, not all the experimental analyses give all the necessary details. 
Below we comment on how we use the currently available information from the experiments. 
The ideal case would of course be that the experiments consistently do the unfolding of theoretical vs.\ experimental channels from the data as mentioned above and always provide directly the experimental results for the theoretically relevant $\what\mu_i$'s (see \Eq{murel}) and the correlations between them.

\subsubsection*{ATLAS}

\begin{itemize}
\item $H \to \gamma\gamma$: we take our information from Fig.~4 of~\cite{ATLAS-CONF-2012-168}. 
This figure shows the results after unfolding to obtain the experimental results for the $\mu$'s as defined theoretically.   
Fig.~4 does make the approximation that VBF and VH can be lumped together (\ie\ have the same efficiencies after cuts) and that  ggF and ttH can be similarly lumped together (note that ttH contributes less than $1\%$).  
We fit the 68\% CL contour assuming that the $\Delta\chi^2$ follows a bivariate normal distribution.\footnote{We thank Guillaume Drieu La Rochelle for providing this fit, cf.\ Table~4 in version~2 of \cite{Cacciapaglia:2012wb}.} 
With this, the correlation $\rho = -0.37$ between the ggF and VBF+VH channels is automatically taken into account.
We also note that while Fig.~4 of~\cite{ATLAS-CONF-2012-168} is for 126.6 GeV, Fig.~12 (right) in the same paper shows that there is a broad ``plateau''  as a function of the mass when the energy scale uncertainty is taken into account, implying that  the results should not depend too much on the mass. 
\item $H \to ZZ$: the signal strength in this channel reported by ATLAS \cite{ATLAS-CONF-2012-169,ATLAS-CONF-2012-170}
is $\mu=1.3^{+0.53}_{-0.48}$ with a best fit mass of $M_H = 123.5\pm 0.9\ {\rm (stat)} \pm 0.3\ {\rm (sys)}$~GeV. 
At $M_H=125$~GeV, the signal strength is $\mu=1.01^{+0.45}_{-0.40}$, see Fig.~10 in \cite{ATLAS-CONF-2012-170}.  
Assuming that the discrepancy in the Higgs mass determined from the $\gam\gam$ and the 4~lepton final states is 
due to a statistical fluctuation (rather than unknown systematics) we use $\mu(ZZ)$ at $M_H=125$~GeV, \ie\  
close to the combined best fit mass from ATLAS, in our fits. 

Alternatively, one could rescale the value of $\mu=1.3^{+0.53}_{-0.48}$ at $M_H=123.5$~GeV for a Higgs mass of 125~GeV. This would give $\mu(ZZ)=1.15^{+0.53}_{-0.48}$ at $M_H=125$~GeV (or $\mu(ZZ)=1.11^{+0.53}_{-0.48}$ at $M_H=125.5$~GeV). We checked that taking this alternative approach has only marginal influence on our results.  

Regarding the decomposition in production modes, no statement is made in the conference note or paper. 
However, as it is an inclusive analysis, 
we take the relative ratios of production cross sections for a SM Higgs as a reasonable approximation. 
To this end, we use the ratios given by the LHC Higgs Cross Section Working Group~\cite{HXSWG}.  

\item $H \to WW$: we adopt relative contributions of 95\% \ggf\ and 5\% VBF~\cite{ATLAS-CONF-2012-158}. We do not include any result for 7 TeV because the update presented at HCP is a combination of 7 and 8 TeV.
\item $H \to \tau\tau$: ATLAS provides only an overall signal strength with no information on the decomposition with respect to production modes. However, the conference note \cite{ATLAS-CONF-2012-160} contains the results of unfolding to the theory-level $\mu$'s via a plot (Fig.~19) of the experimental results for 
$ \mu_{\rm ggF} \times B/B_{\rm SM}$ versus $\mu_{\mathrm{VBF}+VH} \times B/B_{\rm SM}$ at $M_H=125$~GeV. We fit the contours with the same procedure as for $H \to \gamma\gamma$. The correlation (included in the fit) is $\rho = -0.50$.
\end{itemize}

\subsubsection*{CMS}

\begin{itemize}
\item $H \to \gamma\gamma$: we follow the same procedure as for ATLAS $H \to \gamma \gamma$, using Figure 11 from~\cite{CMS-PAS-HIG-12-045}. The correlation (included in the fit) is $\rho = -0.54$.
\item $H \to ZZ$: no decomposition with respect to production modes is given in the conference note or paper. 
As it is a fully inclusive analysis, we use the relative ratios of production cross sections given by the LHC 
Higgs Cross Section Working Group~\cite{HXSWG} as a good approximation~\cite{albert}. 
\item $H \rightarrow WW$: the information provided in the conference note and papers is incomplete; our decomposition into 
production modes is based on \cite{albert}. 
Our combination (weighted mean) agrees within 9\% with that given by CMS ($\mu_{\rm comb} = 0.64 \pm 0.24$ instead of $0.70^{+0.24}_{-0.23}$).
\item $H \rightarrow b\bar{b}$: 
as there is no information on possible contaminations by other production modes, we assume 100\% VH or 100\% ttH production for the respective categories.
\item $H \rightarrow \tau\tau$: 
for the 0/1 jet and VBF tag categories we extract the decomposition into production modes from~\cite{CMS-PAS-HIG-12-043}, assuming that there is no significant change in the efficiencies between $M_H = 125$~GeV and $M_H = 125.8$~GeV. We use the efficiencies from the first three categories ($\mu\tau_{h}+X$, $e\tau_{h}+X$ and $e\mu+X$) because they are the most sensitive ones; they lead to very similar decompositions which we use in our analysis.  
Our combination (weighted mean) agrees within 6\% with that given by CMS ($\mu_{\rm comb} = 0.83 \pm 0.49$ instead of $0.88^{+0.51}_{-0.48}$). 
\end{itemize}

\subsubsection*{Tevatron}

\begin{itemize}
\item $H \to \gamma\gamma$ and $H\to WW$: no decomposition into production modes is given by the experiments. 
We assume that the analyses are inclusive and we thus employ the ratios of the  theoretical predictions for the (SM) Higgs production cross sections.  
\item $H \to b\bar{b}$: we use the new results from HCP2012~\cite{HCPtevBBtalk} assuming 100\% VH.
\end{itemize}

\section{Results}

\subsection{General coupling fits}

\subsubsection*{\boldmath Fit I: $\CU=\CD=\CV=1$, $\dcg$ and $\dcp$ free}

For a first test of the SM nature of the observed Higgs boson, we take $\cu=\cd=\cv=1$ (\ie\ quark, lepton and $W,Z$ vector boson couplings to the Higgs 
are required to be SM-like) but we allow for additional new physics contributions  to the $\gam\gam$ and $gg$ couplings, parameterized by $\dcg$ and $\dcp$, coming from loops involving non-SM particles or from anomalies. 
This fit, which we refer to as Fit~{\bf I}, is designed to determine if the case where all tree-level Higgs couplings 
are equal to their SM values can be consistent with the data.    For example, such a fit is relevant in the context of UED models where the tree-level couplings of the Higgs are SM-like~\cite{Petriello:2002uu,Belanger:2012mc}.

\begin{figure}[h!]\centering
\includegraphics[scale=0.56]{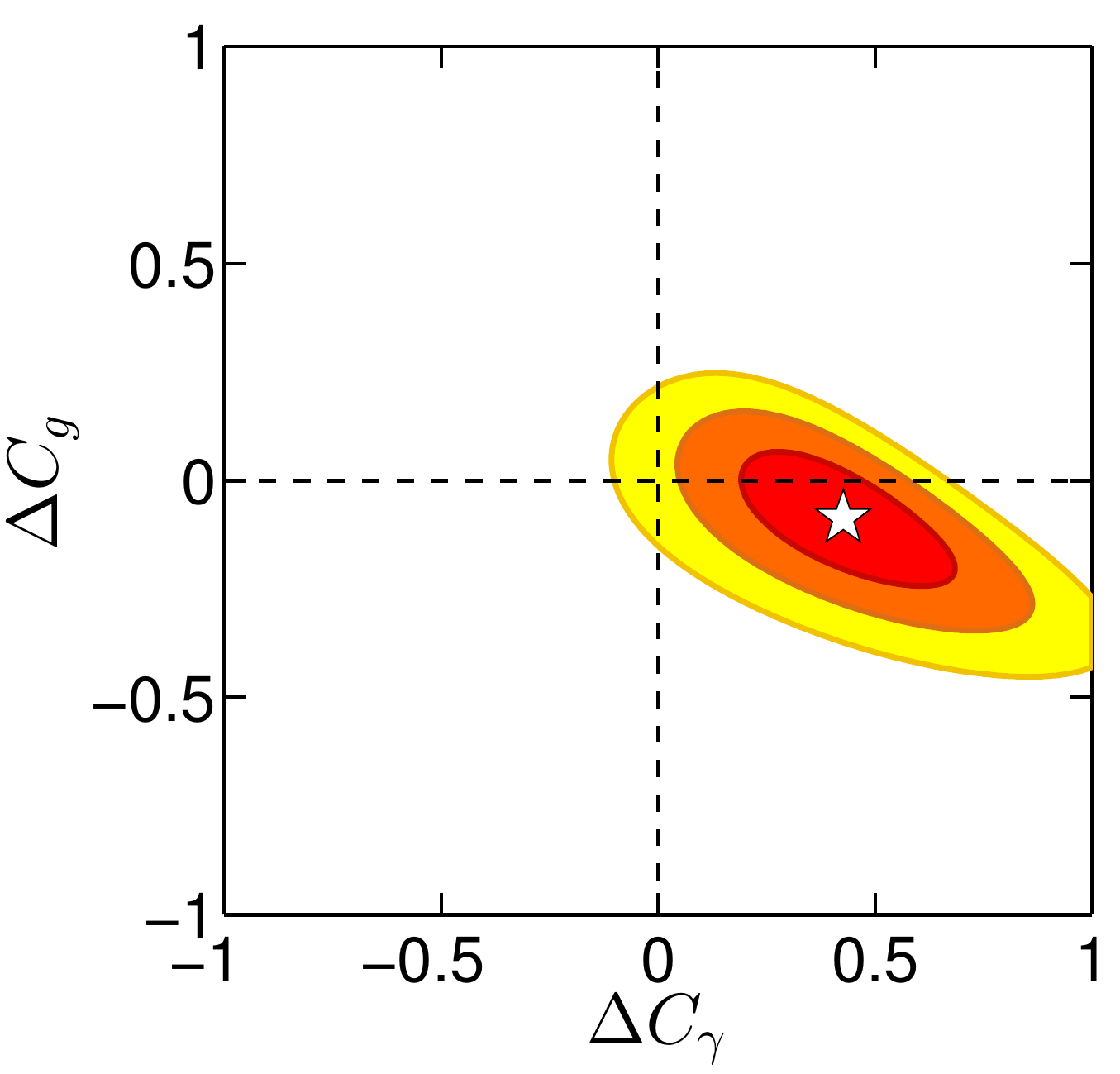}
\caption{Two parameter fit of $\Delta C_\gamma$ and $\Delta C_g$, assuming $\CU=\CD=\CV=1$ (Fit~{\bf I}). 
The red, orange and yellow ellipses show the 68\%, 95\% and 99.7\%  CL regions, respectively. 
The white star marks the best-fit point $\dcp=0.426$, $\dcg=-0.086$.
\label{fit1} }
\end{figure}

Figure~\ref{fit1} displays the results of this fit in the  $\dcg$ versus $\dcp$ plane.  
The best fit is obtained for $\dcp\simeq0.43$, $\dcg\simeq-0.09$,  and has $\chimin=12.31$ for 19 degrees of freedom (\dof), 
giving a $p$-value of $0.87$. 
The results of this fit are summarized in Table~\ref{chisqmintable}, together with the results of the other fits of this section.

We note that the SM (\ie\ $\CU=\CD=\CV=1$, $\dcg=\dcp=0$) has $\chi^2= 20.2$ and is hence more 
than $2\sigma$ away from the best fit in Fig.~\ref{fit1}.
The number of degrees of freedom for the SM fit is 21,  
implying a $p$-value of $0.51$. The largest  $\chi^2$ contributions come from the $H\to\gamma\gamma$ channels 
from ATLAS ($\Delta\chi^2=5.06$), CMS ($\Delta\chi^2=3.36$) and Tevatron ($\Delta\chi^2=2.60$), followed 
by the VBF result for $H\to WW$ from CMS with $\Delta\chi^2=2.01$. 

\subsubsection*{\boldmath Fit~II: varying $\cu$, $\cd$ and $\cv$ ($\dcp=\dcg=0$)}

\begin{figure}[p] 
\includegraphics[scale=0.32]{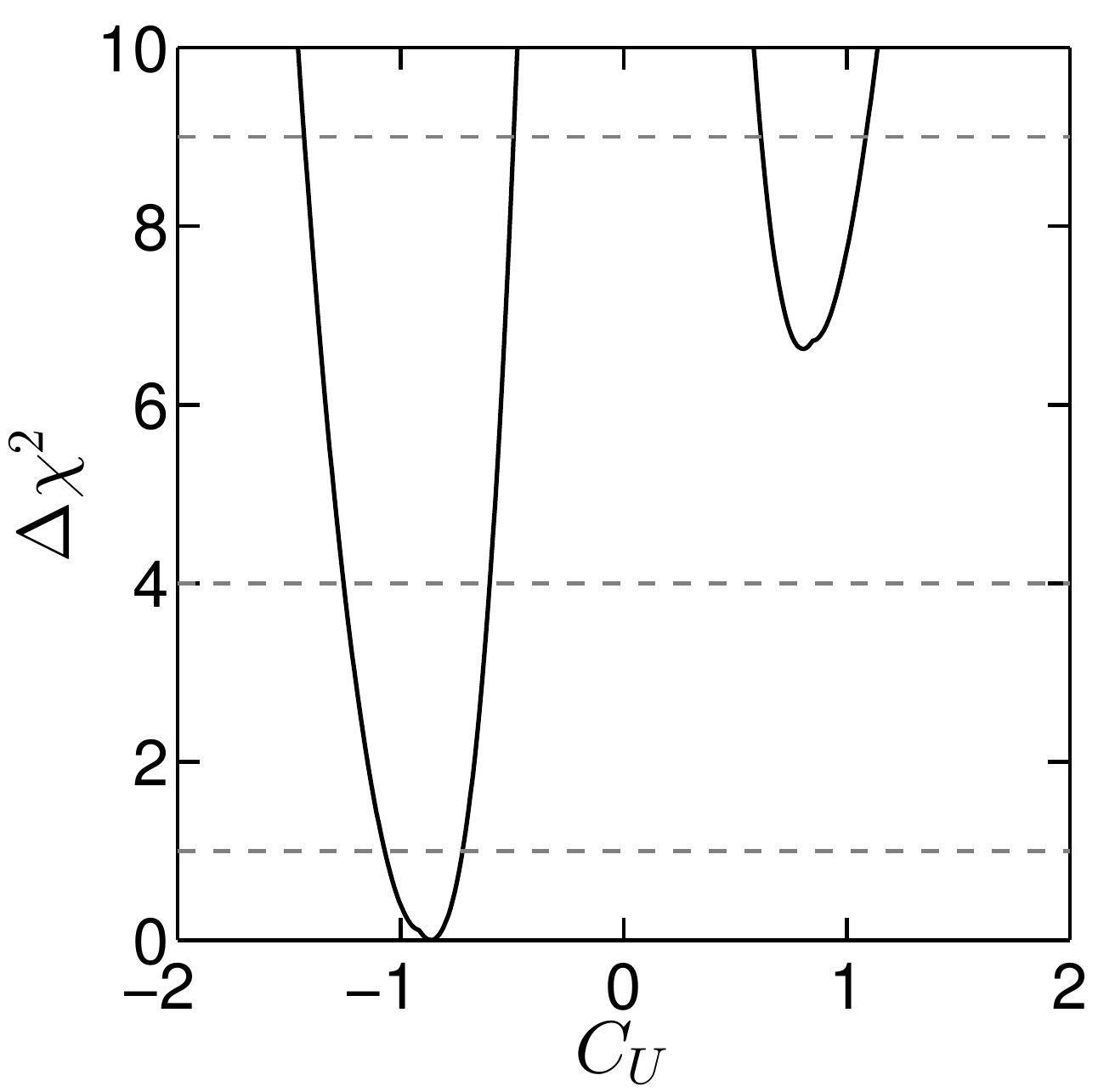}\includegraphics[scale=0.32]{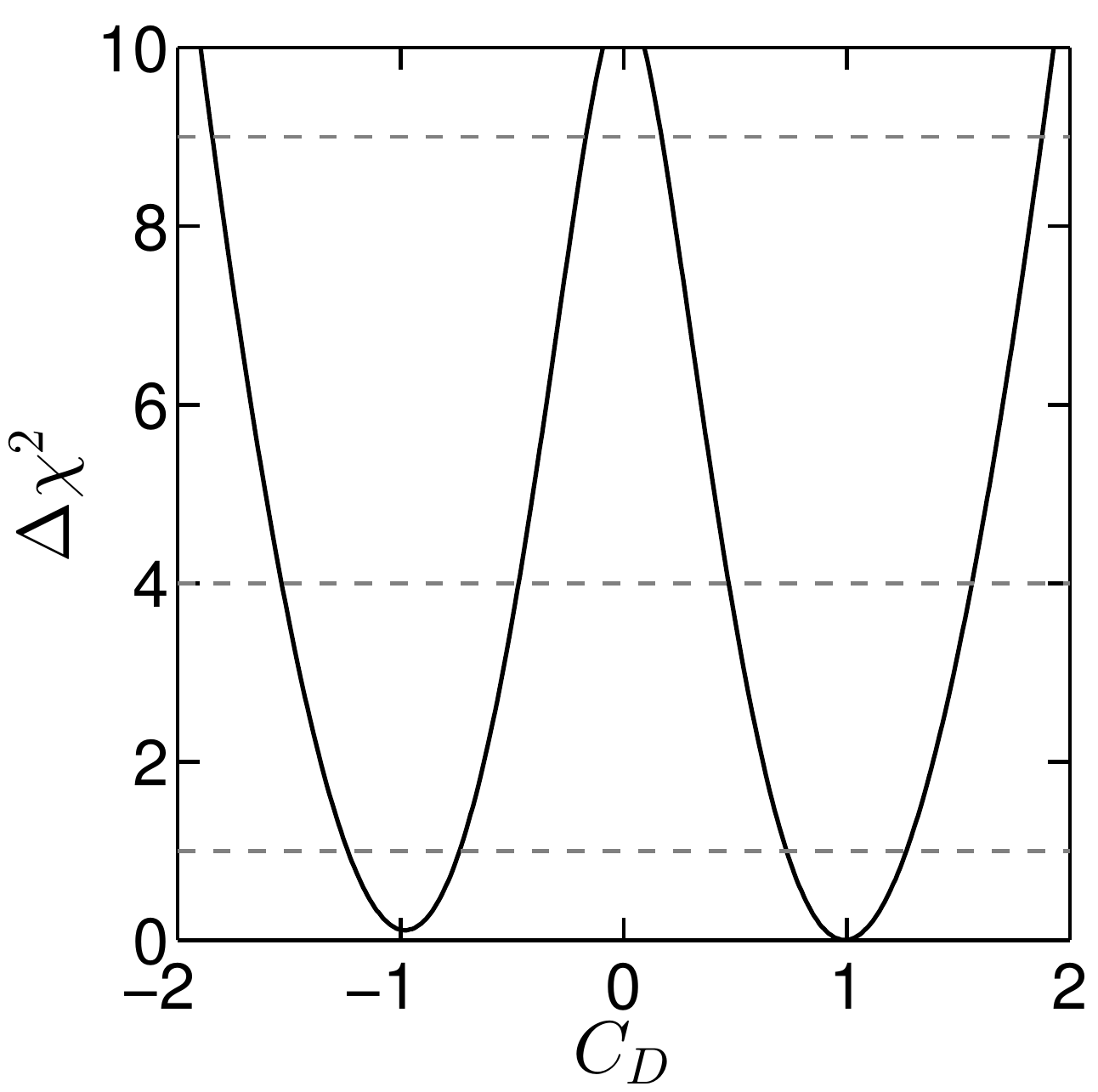}\includegraphics[scale=0.32]{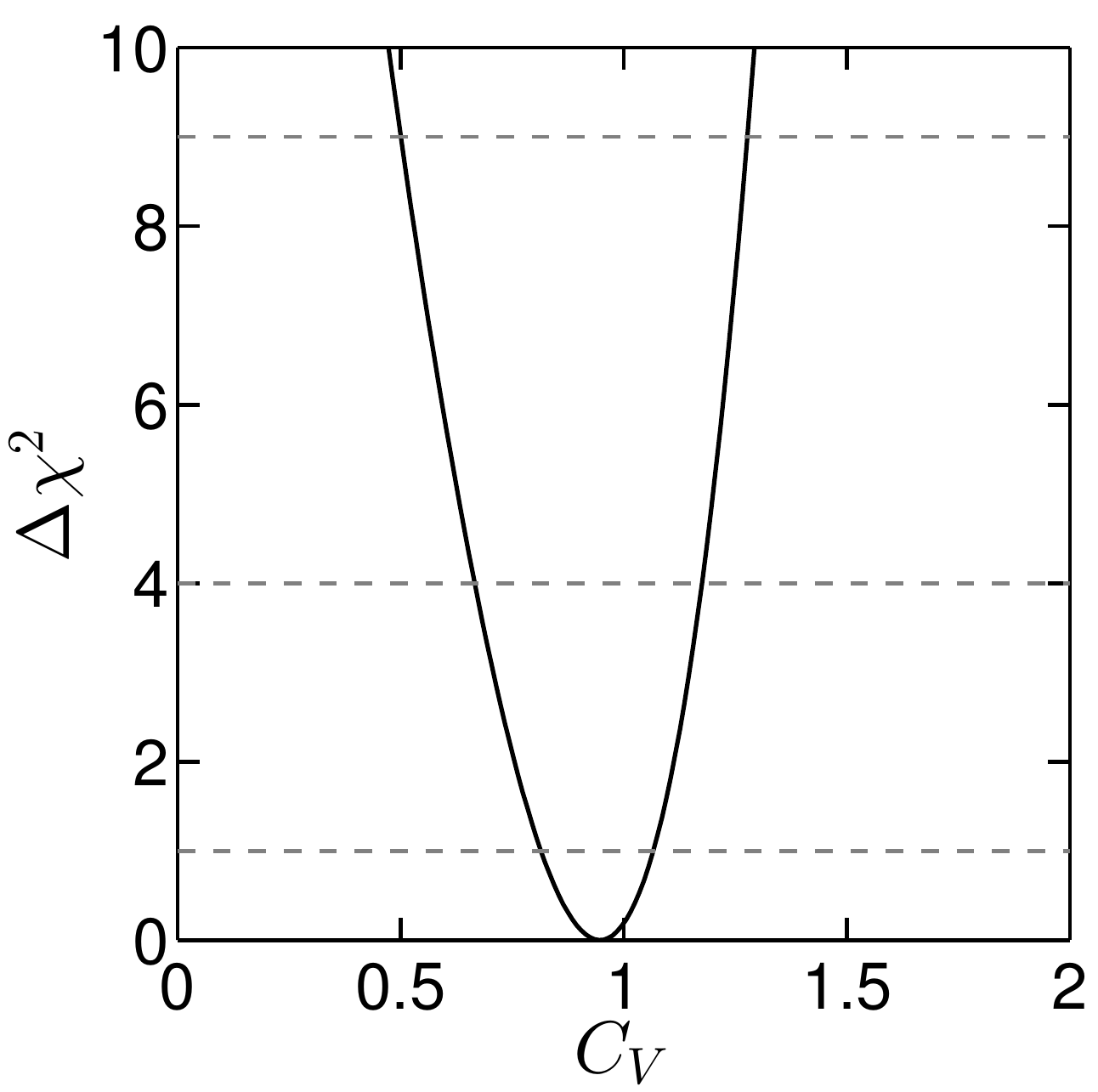}\includegraphics[scale=0.32]{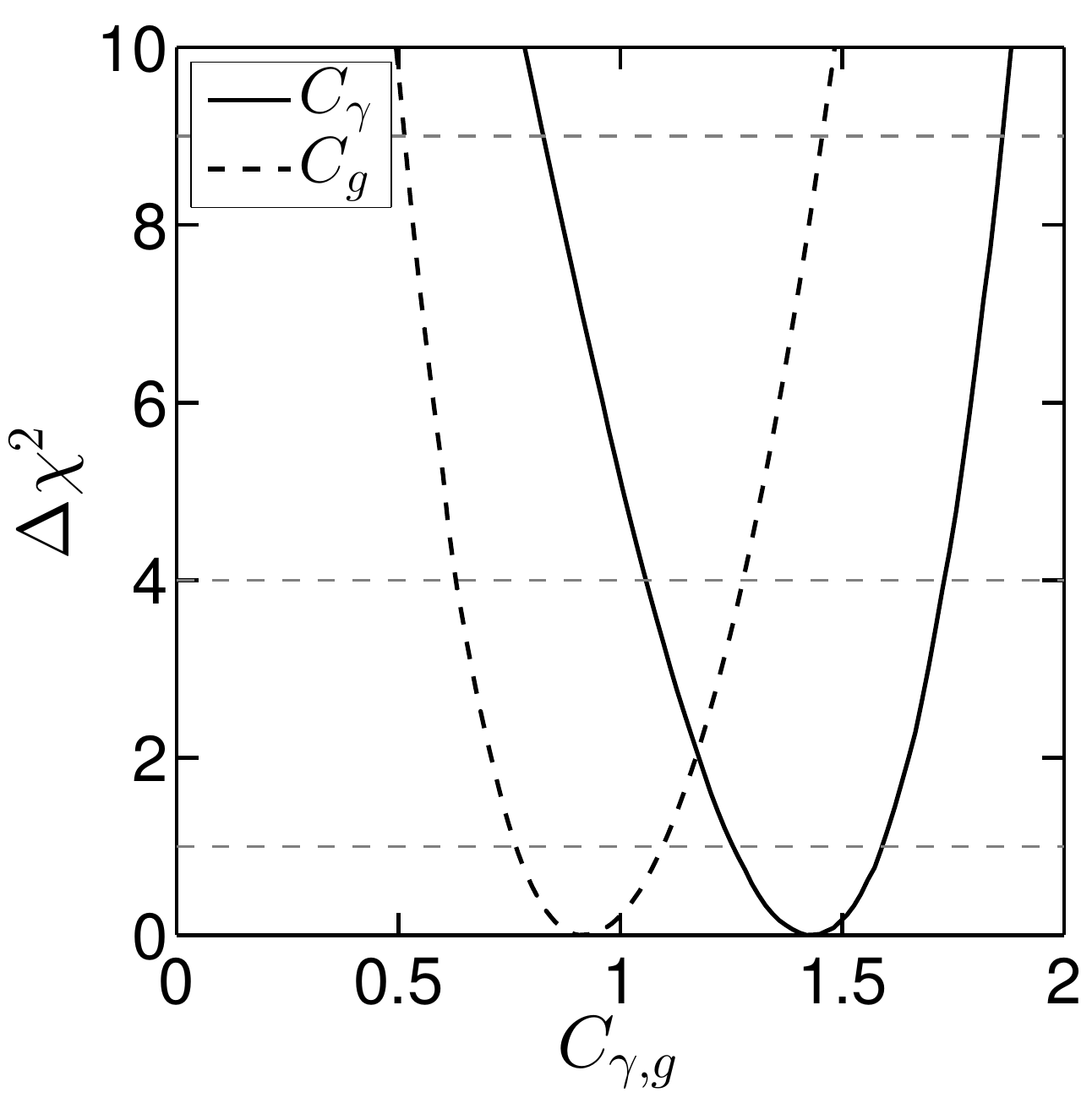}
\caption{One-dimensional $\chisq$ distributions for the three parameter fit, Fit~{\bf II},  of $\CU$, $\CD$, $\CV$ with $\cp=\cpb$ and $\cg=\cgb$ as computed in terms of $\cu,\cd,\cv$. 
\label{fit2-1d} }
\end{figure}

\begin{figure}[p]\centering
\includegraphics[scale=0.4]{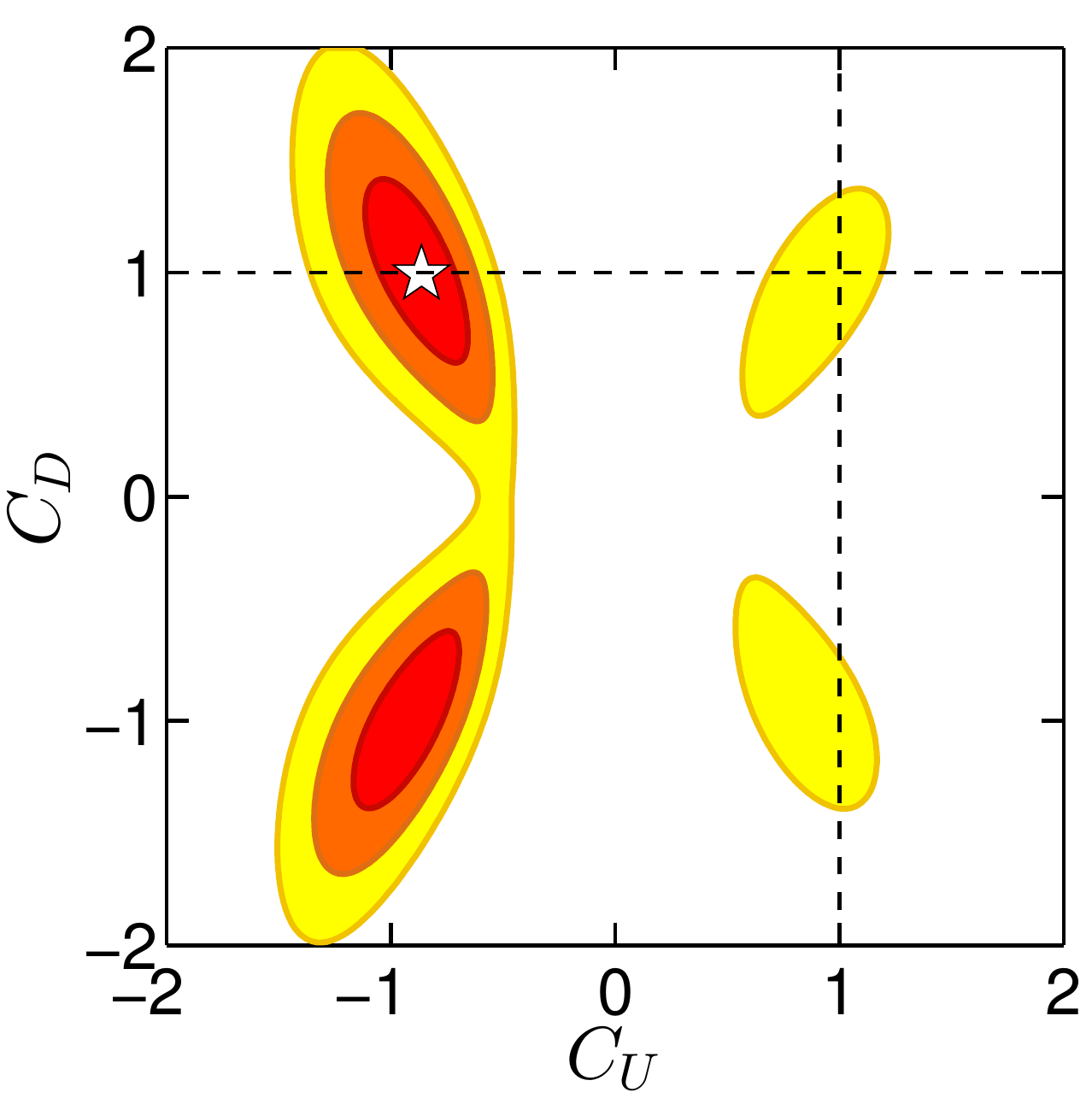}\includegraphics[scale=0.4]{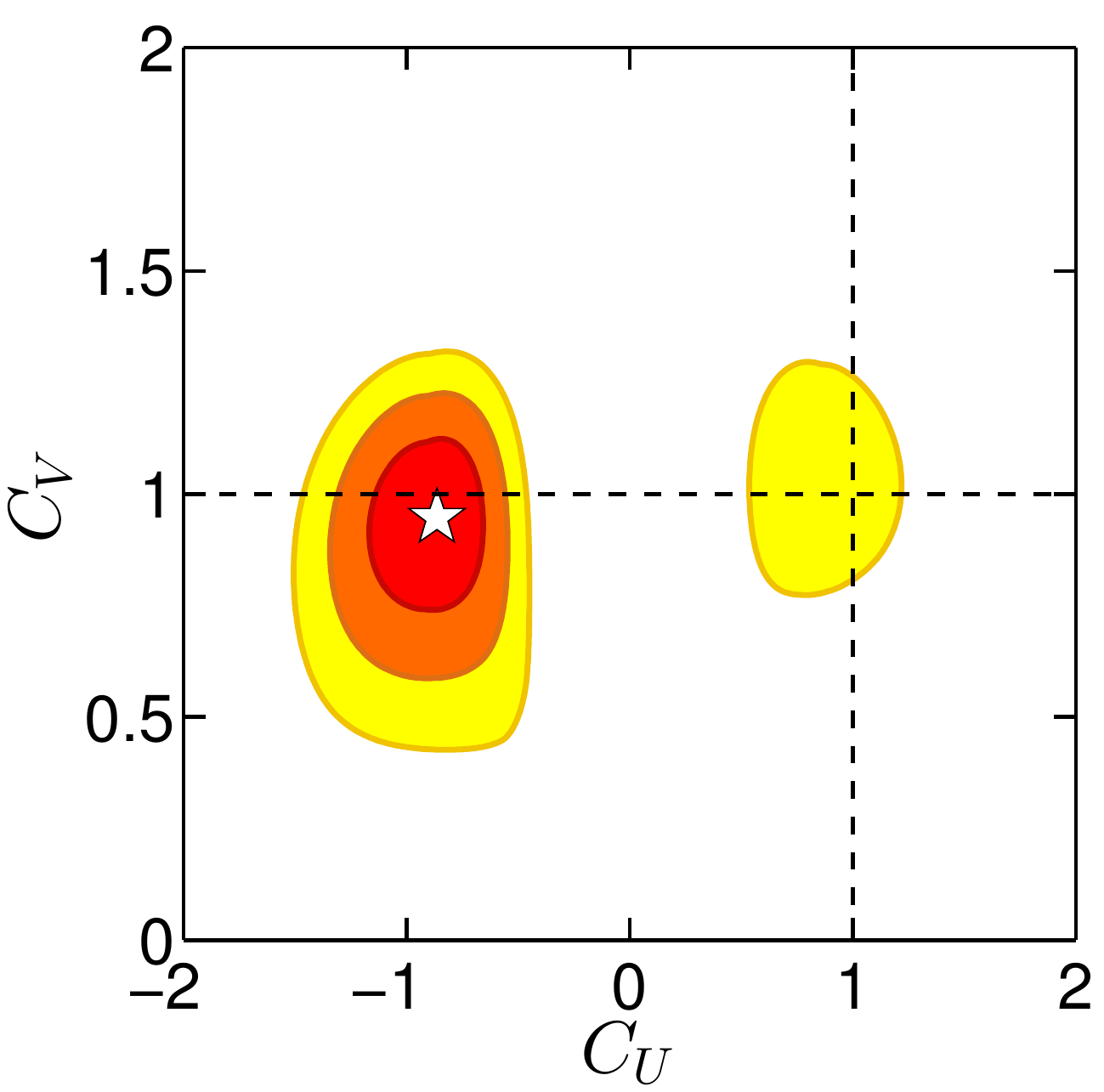}\includegraphics[scale=0.4]{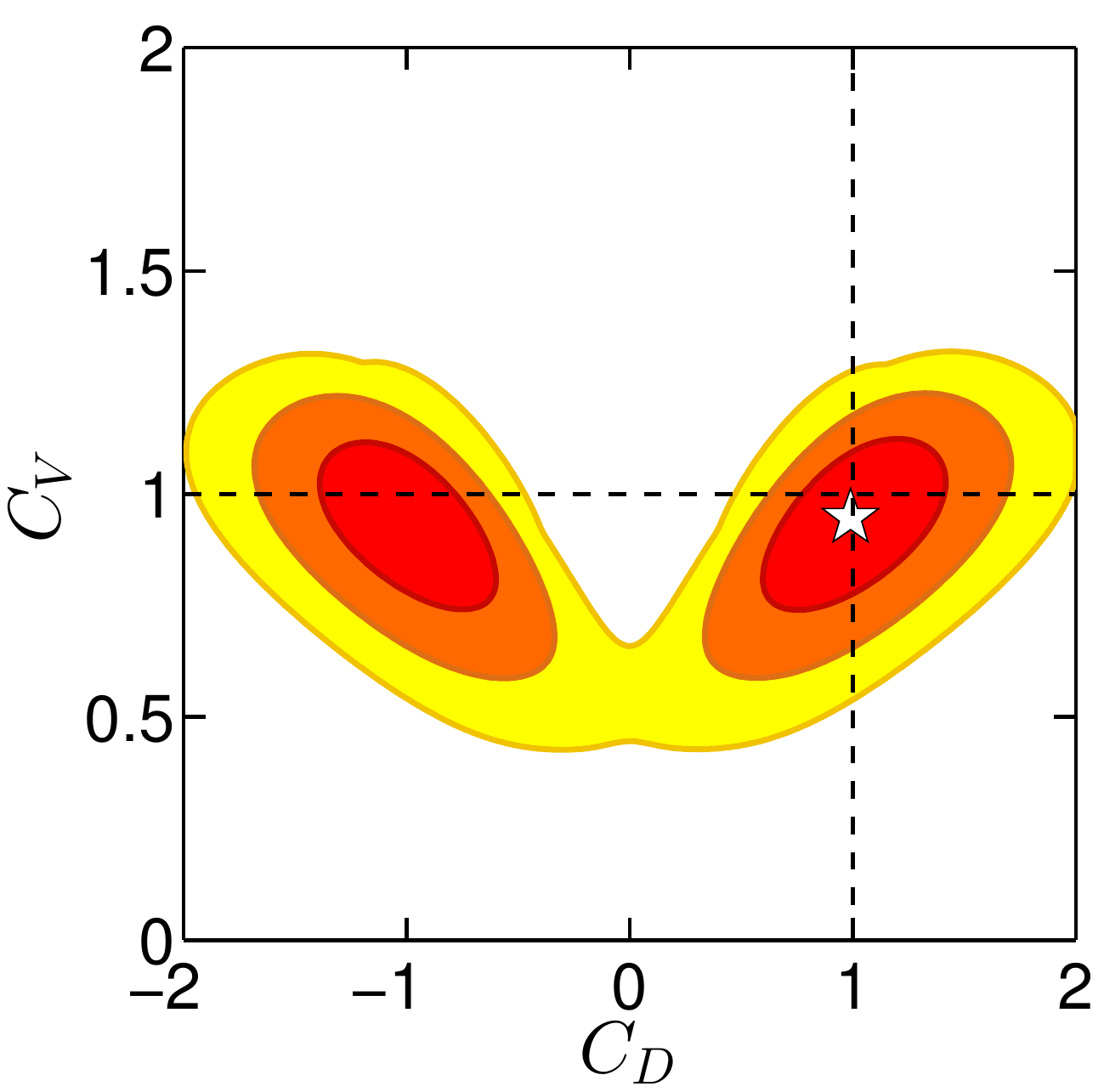}\\
\includegraphics[scale=0.4]{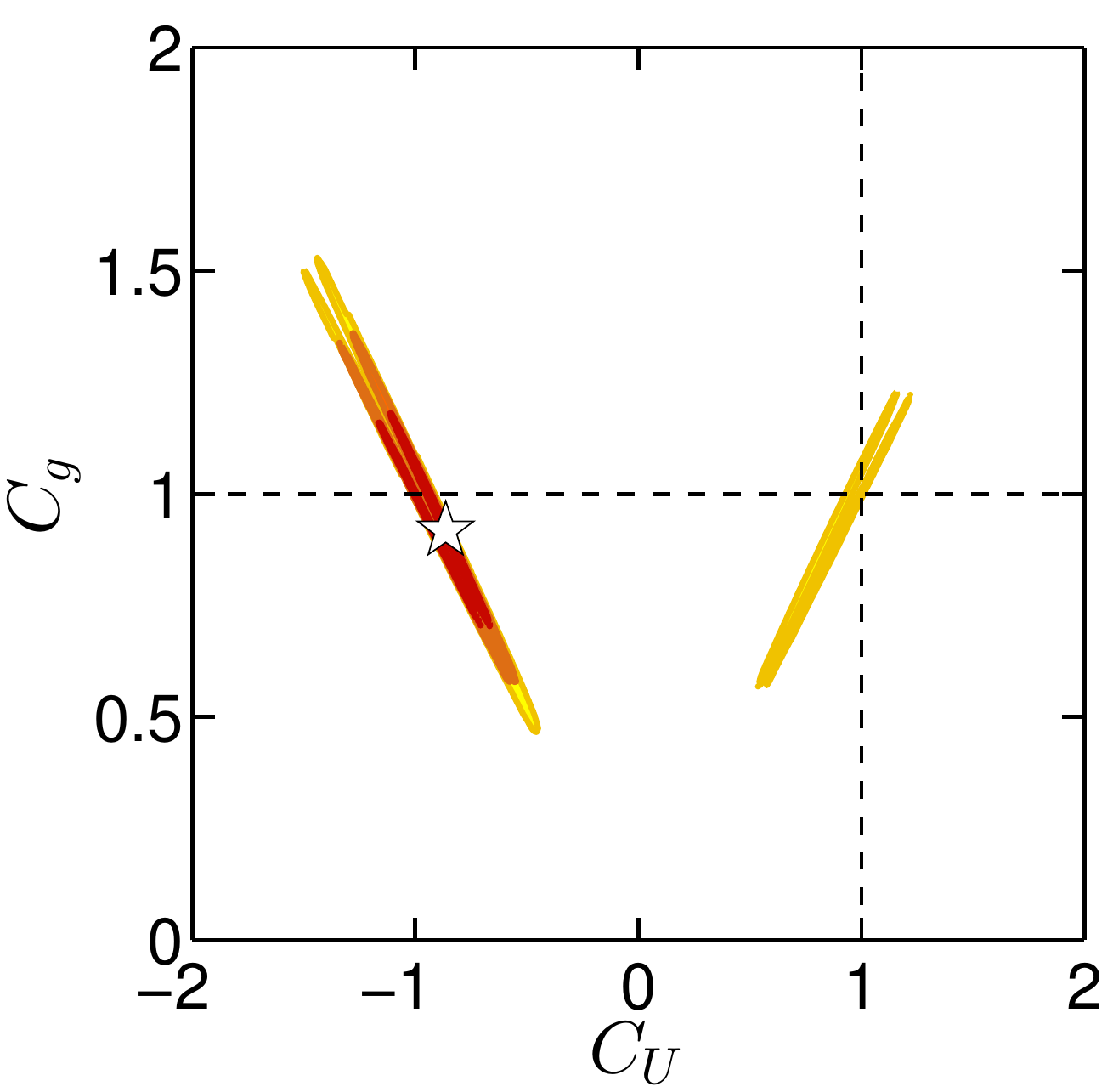}\includegraphics[scale=0.4]{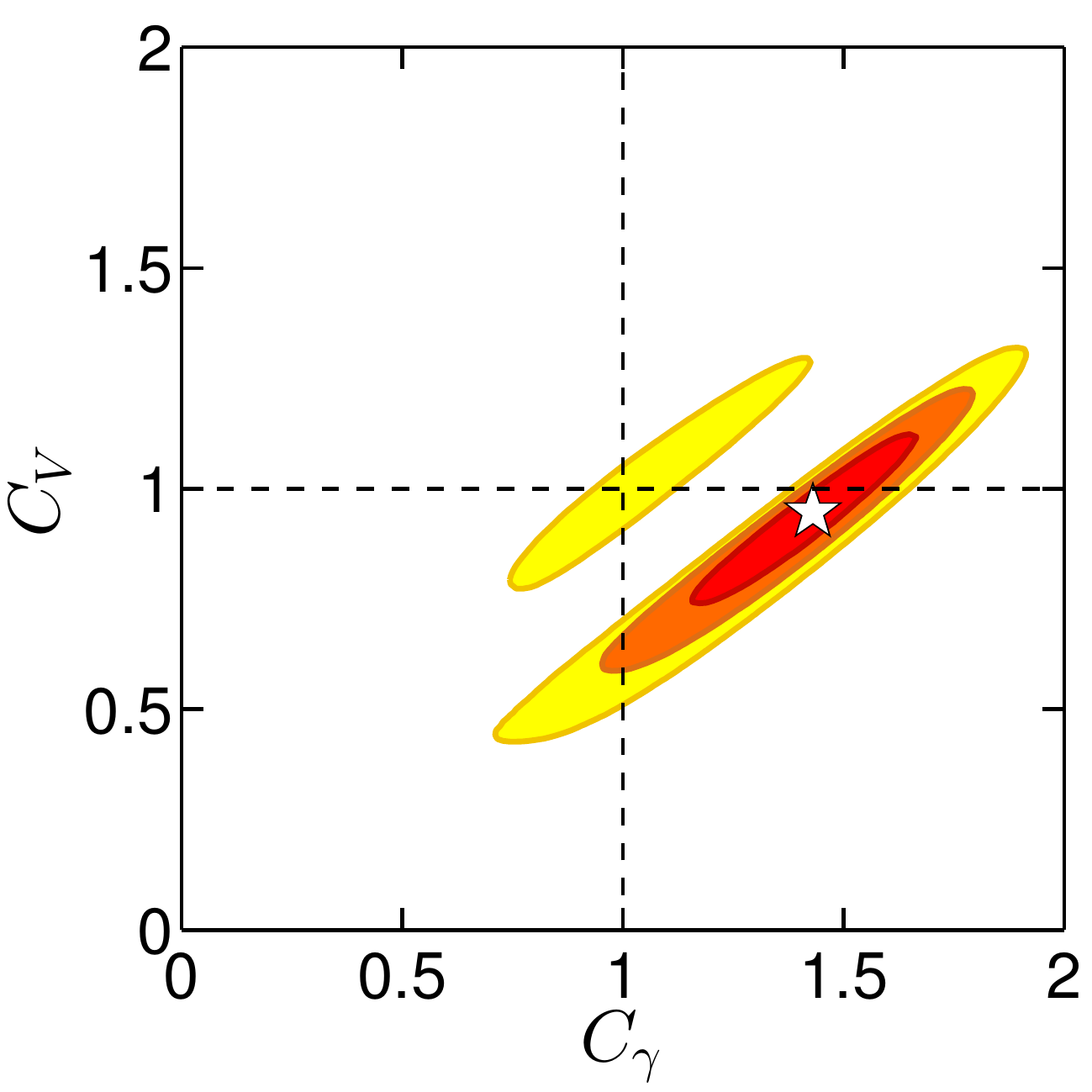}\includegraphics[scale=0.4]{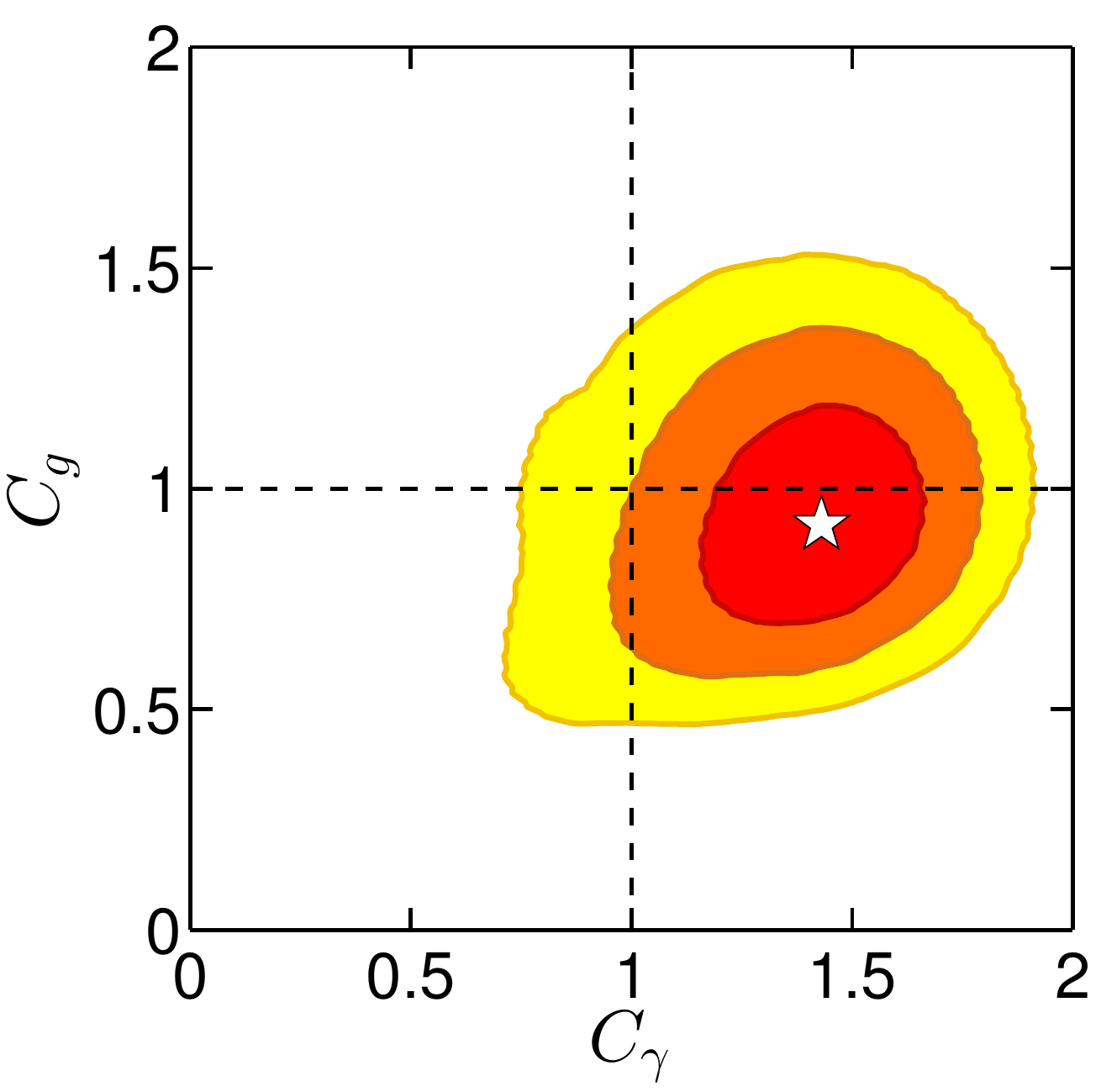}
\caption{Two-dimensional $\chisq$ distributions for the three parameter fit, Fit~{\bf II}, of $\CU$, $\CD$, $\CV$ with $\cp=\cpb$ and $\cg=\cgb$ as computed in terms of $\cu,\cd,\cv$. The red, orange and yellow ellipses show the 68\%, 95\% and 99.7\%  CL regions, respectively. The white star marks the best-fit point. Details on the minima in different sectors of the ($\cu$,\,$\cd$) plane can be found in Table~\ref{tab:fit2}. 
\label{fit2-2d} }
\end{figure}

Next, we let  $\CU$, $\CD$, $\CV$ vary,  assuming there are no new particles contributing to the effective 
Higgs couplings to gluons and photons, \ie\  we take $\dcp=\dcg=0$ implying $\CG=\anti\CG$, $\CP=\anti\CP$ 
as computed from the SM-particle loops.  
The results for the one-dimensional and two-dimensional $\chi^2$ distributions are shown in Figs.~\ref{fit2-1d} and \ref{fit2-2d}. 
The value of $\cv$ is rather well determined to be close to unity. It is intriguing that the best fit of $\cv$ is 
indeed just slightly below 1, as any model with only Higgs doublets or singlets requires $\cv\le1$. The best fit values for $\cd$ and $\cu$ are SM-like in that they have magnitudes that are close to one. However, the best fit $\cu$ value is opposite in sign to the SM Higgs case. 
The preference for $\cu<0$ is at the level of 
$2.6\sigma$ --- see the first plot in Fig.~\ref{fit2-1d}. This results from the fact that an enhanced $\gam\gam$ rate (as observed in the experimental data) is obtained by changing the sign of the top-loop contribution so that it adds, rather than subtracts, from the $W$-loop.  
In contrast, in the case of $\cd$ almost equally good minima are found with $\cd<0$ and $\cd>0$. 
Details on the minima in different sectors of the ($\cu$,\,$\cd$) plane are given in Table~\ref{tab:fit2}.
Note that, for the best fit point, the resulting $\CP$ and $\CG$ are in good agreement with the result of Fit~{\bf I} above, for which $\cp=1.43$ and $\cg=0.91$. Here, however, the enhanced $\cp$ value derives from $\cu<0$ rather than from $\dcp\neq 0$. The best fit results are again tabulated in Table~\ref{chisqmintable}.

\begin{figure}[p] 
\includegraphics[scale=0.32]{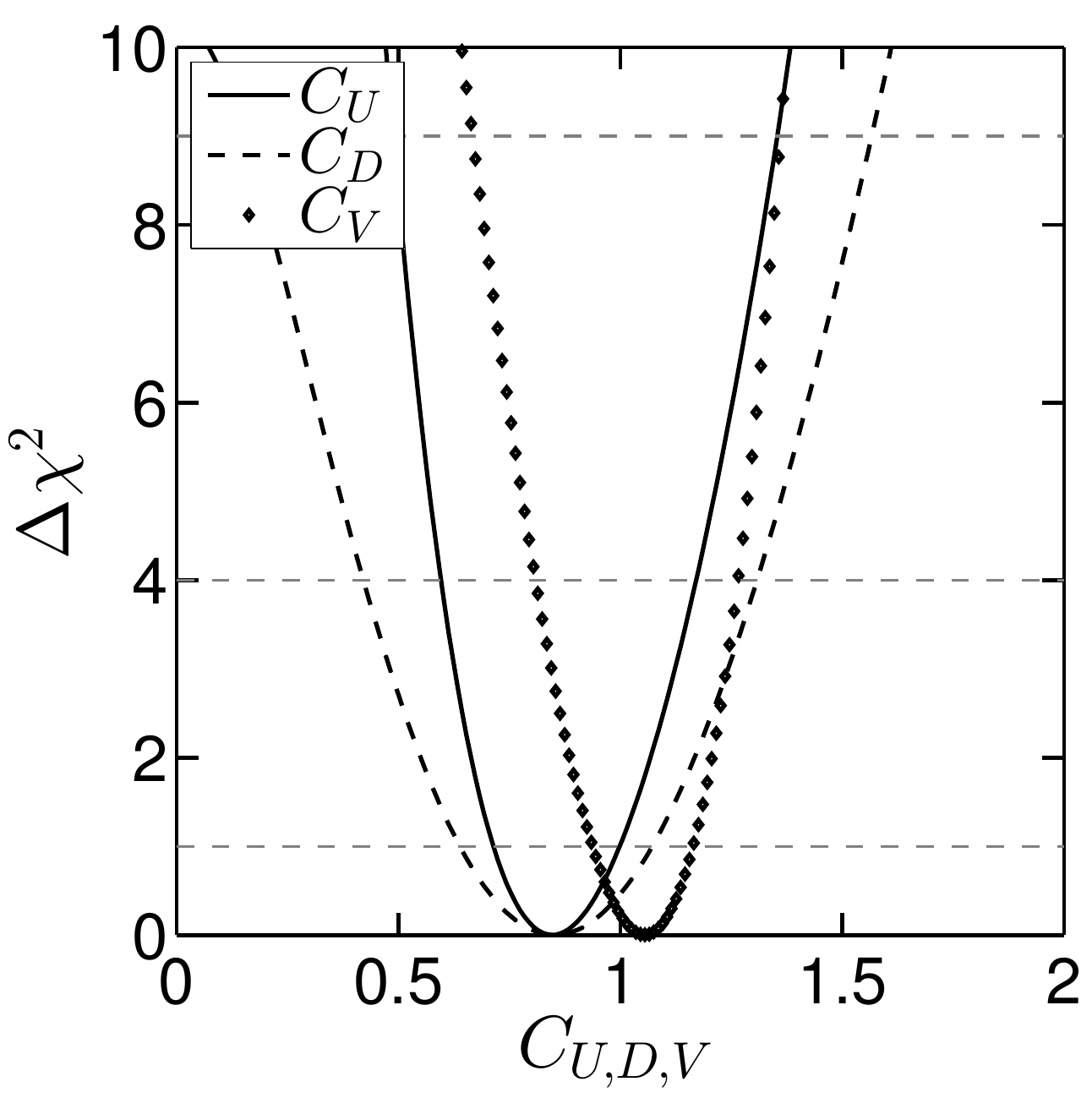}\includegraphics[scale=0.32]{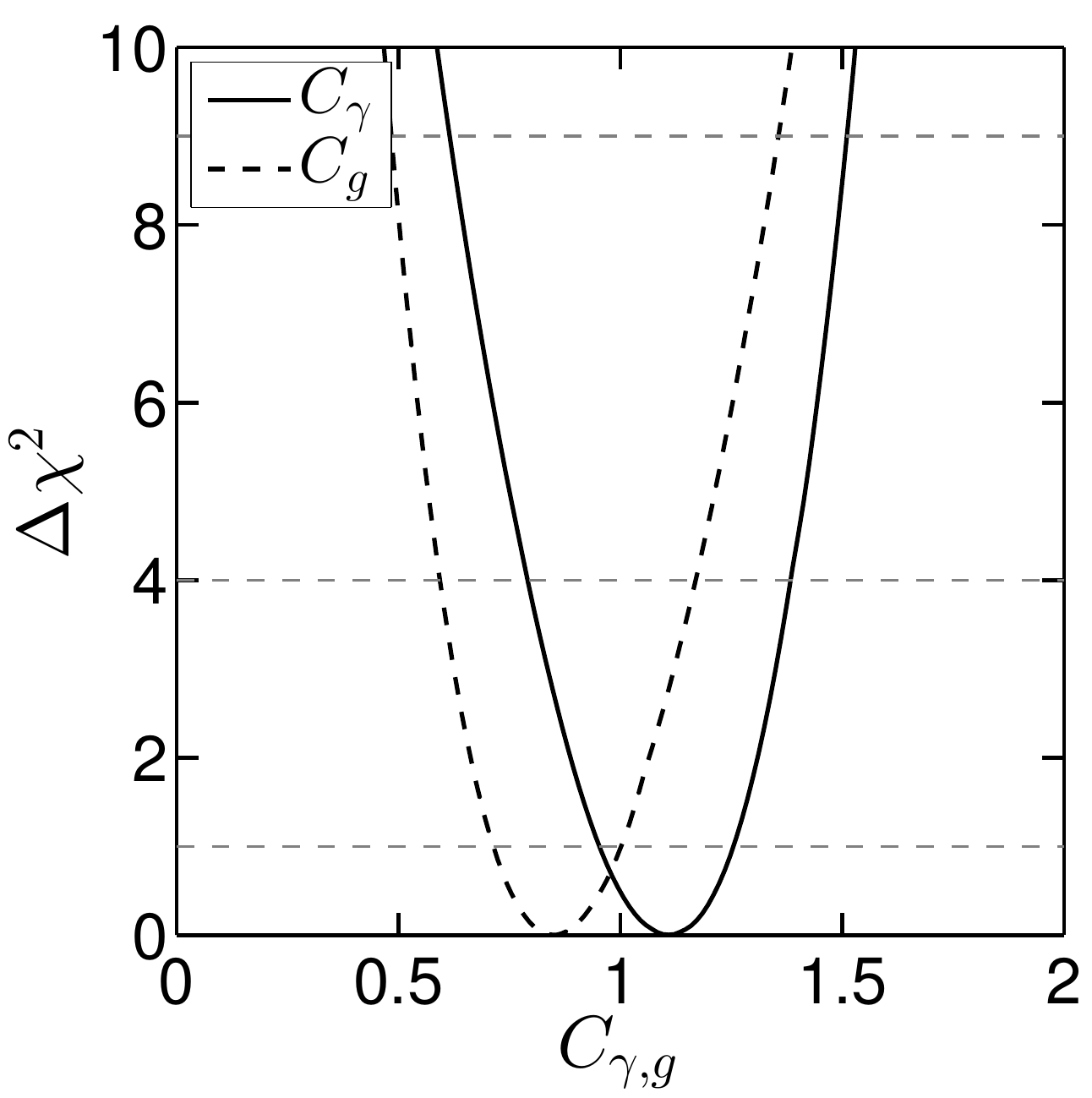}\includegraphics[scale=0.32]{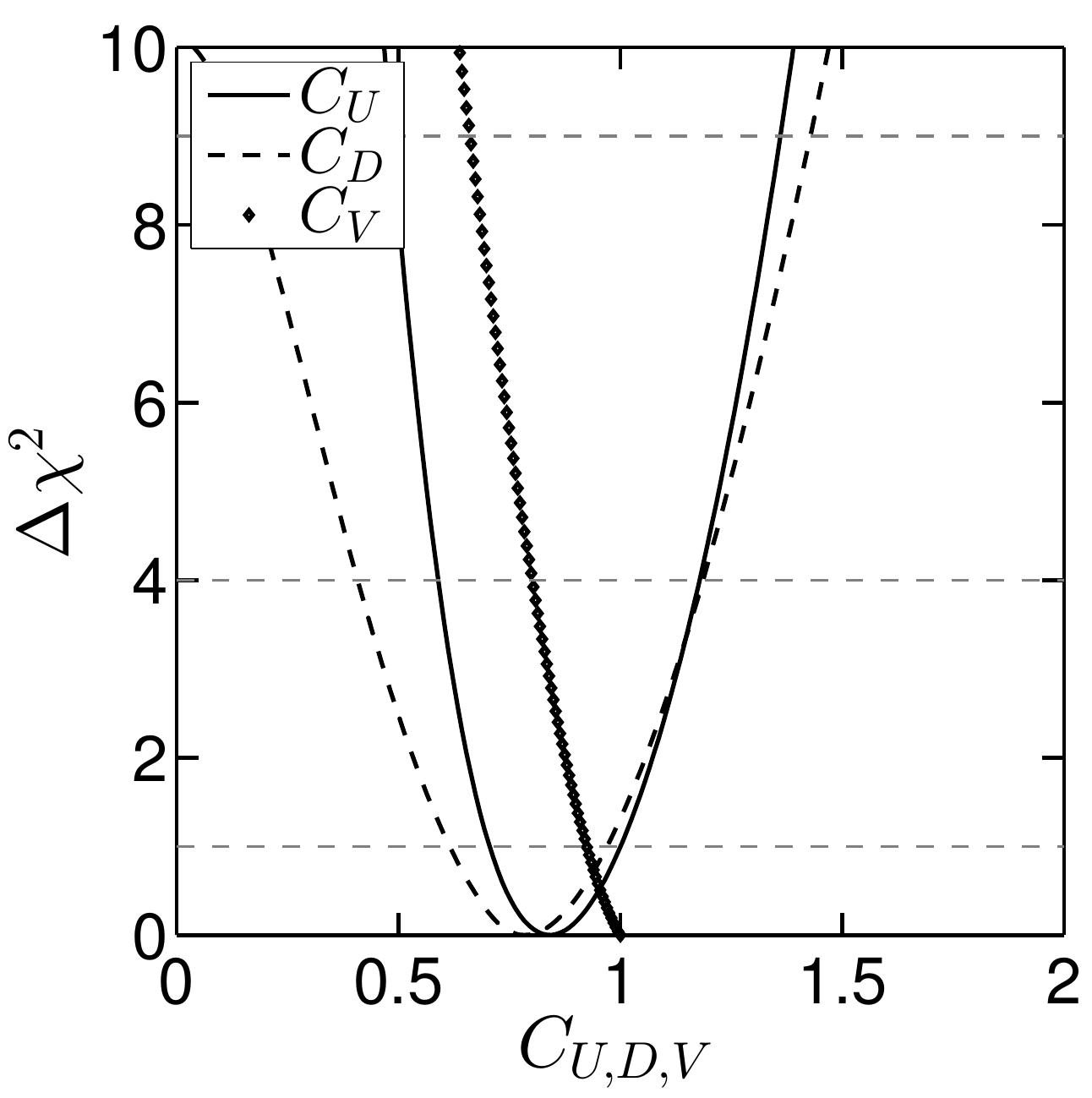}\includegraphics[scale=0.32]{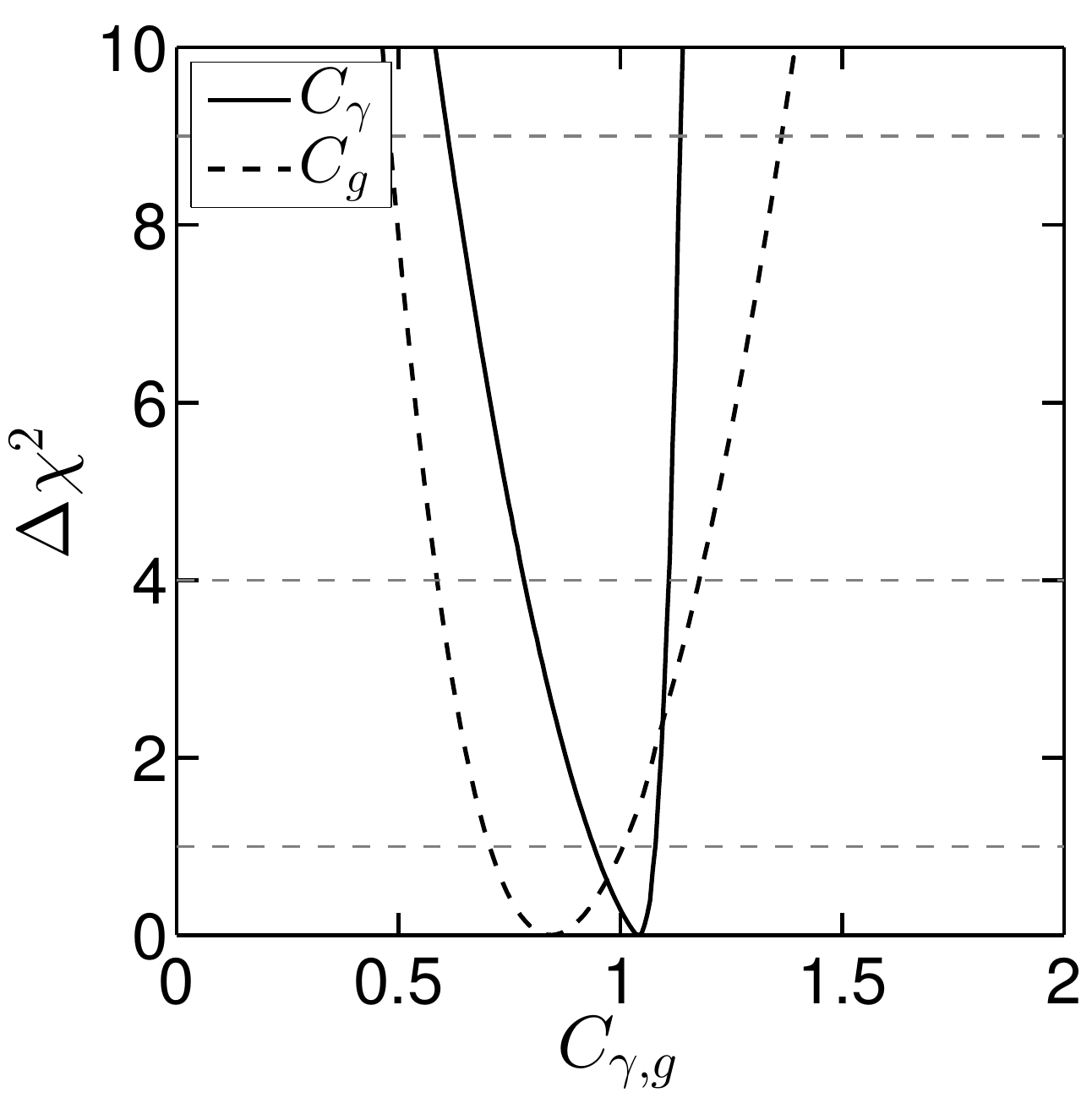}
\caption{One-dimensional $\chisq$ distributions for the three parameter fit, Fit~{\bf II},  
but imposing $\CU>0$, $\CD>0$; the left two plots allow for $\CV>1$ ($\chisq_{\rm min}=18.66$), while in the right two plots $\CV\le1$ ($\chisq_{\rm min}=18.89$). 
\label{fit2-1dpos} }
\end{figure}

\begin{figure}[p]\centering
\includegraphics[scale=0.4]{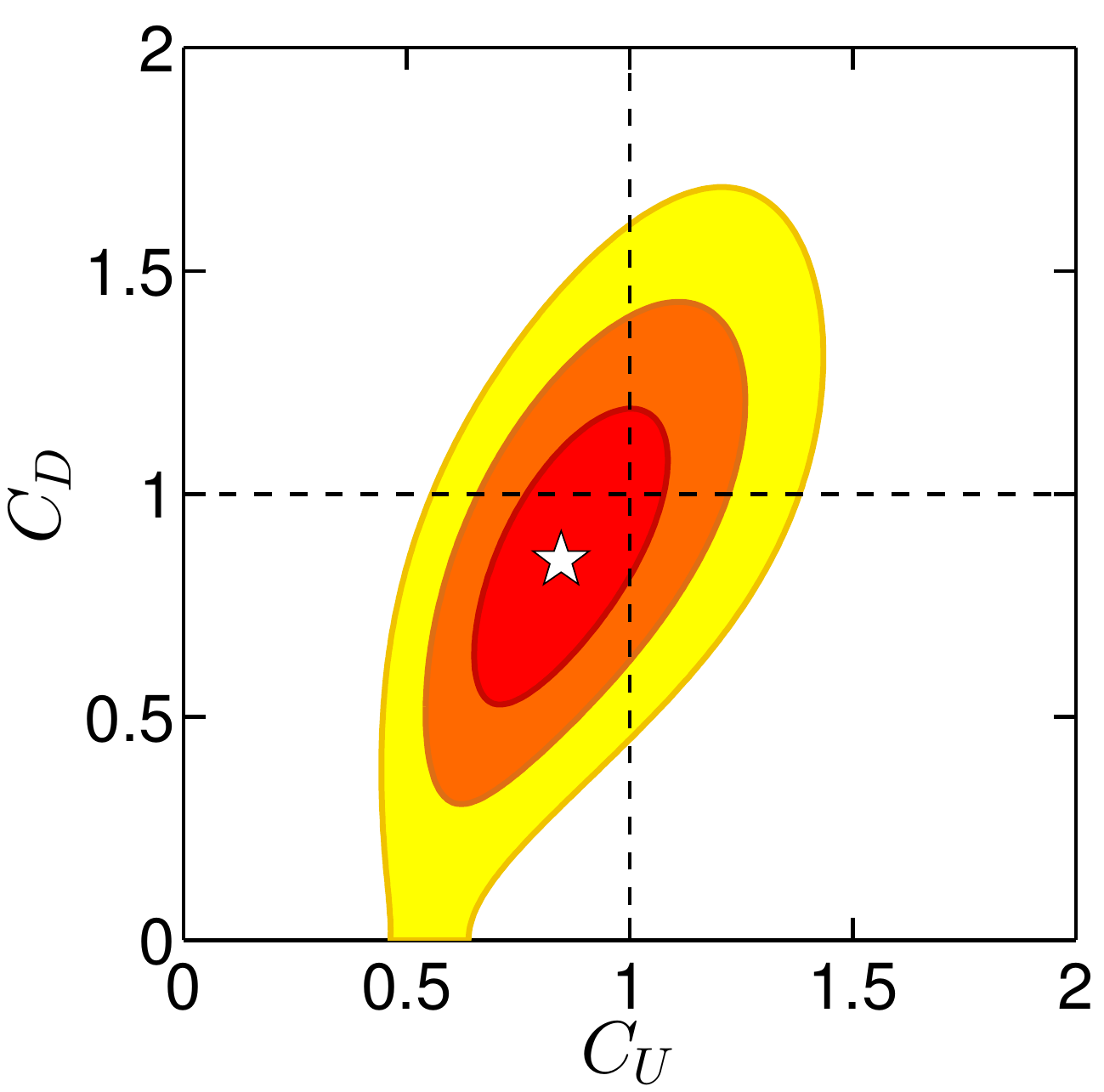}\includegraphics[scale=0.4]{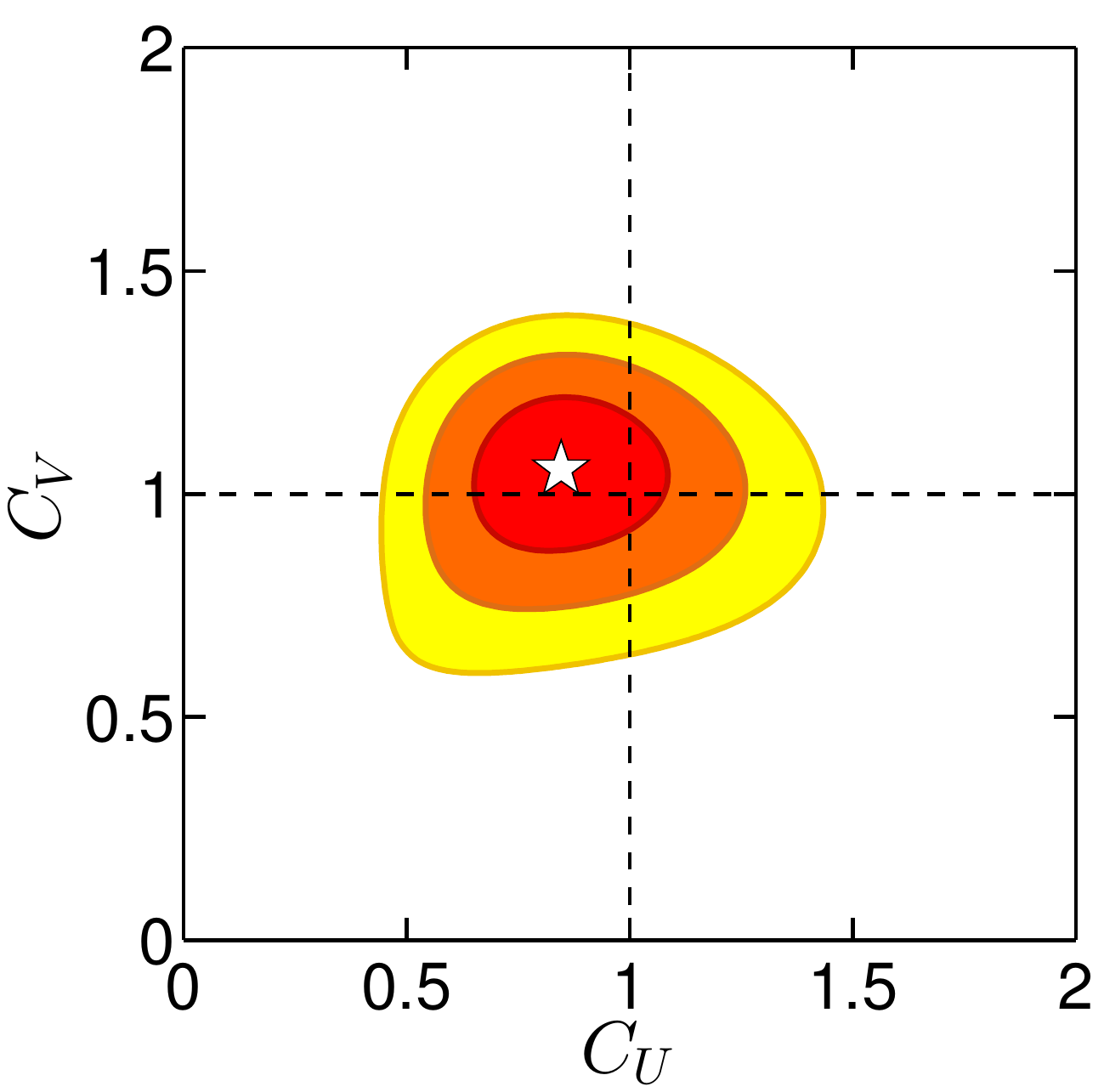}\includegraphics[scale=0.4]{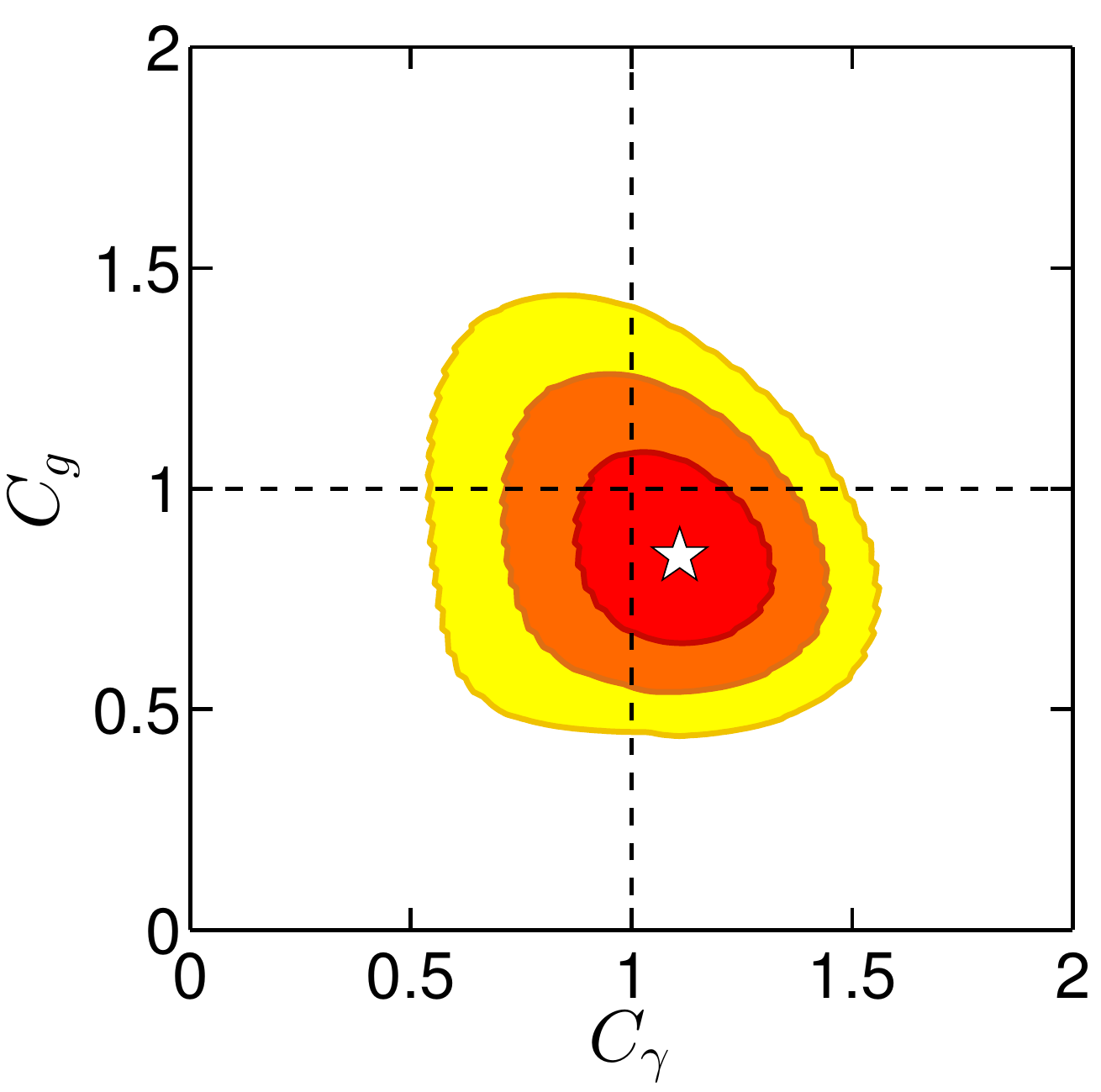}\\\includegraphics[scale=0.4]{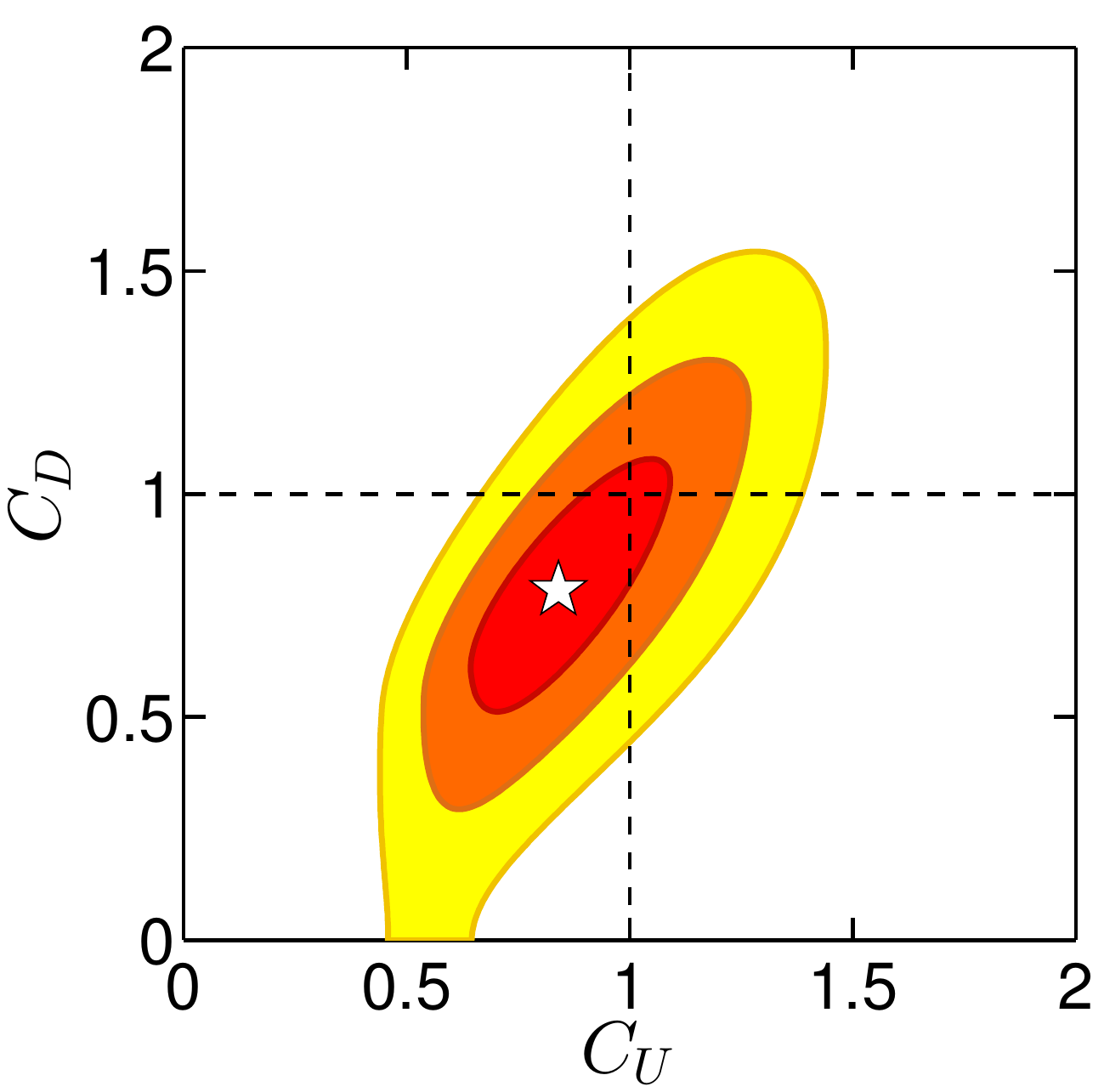}\includegraphics[scale=0.4]{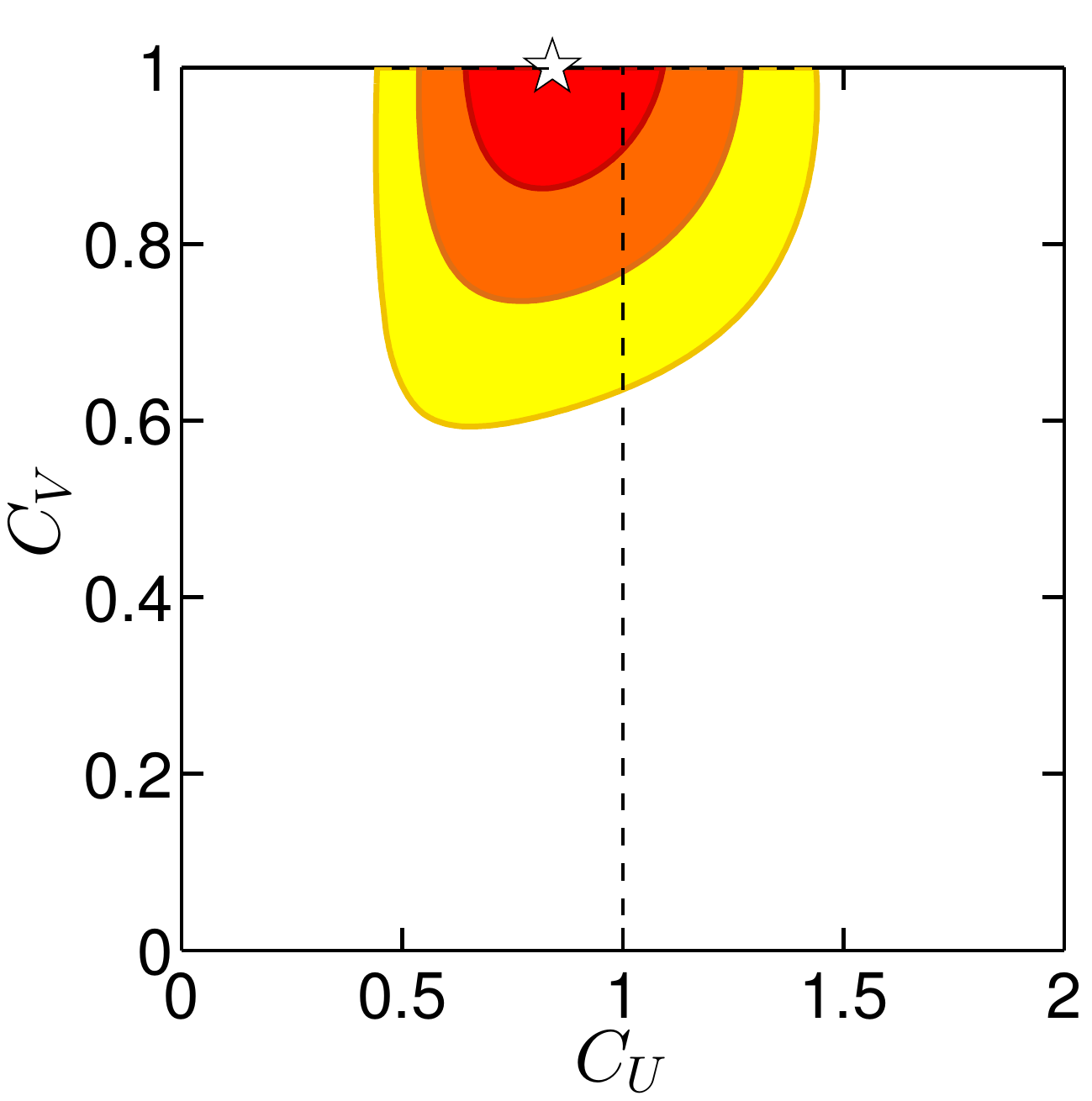}\includegraphics[scale=0.4]{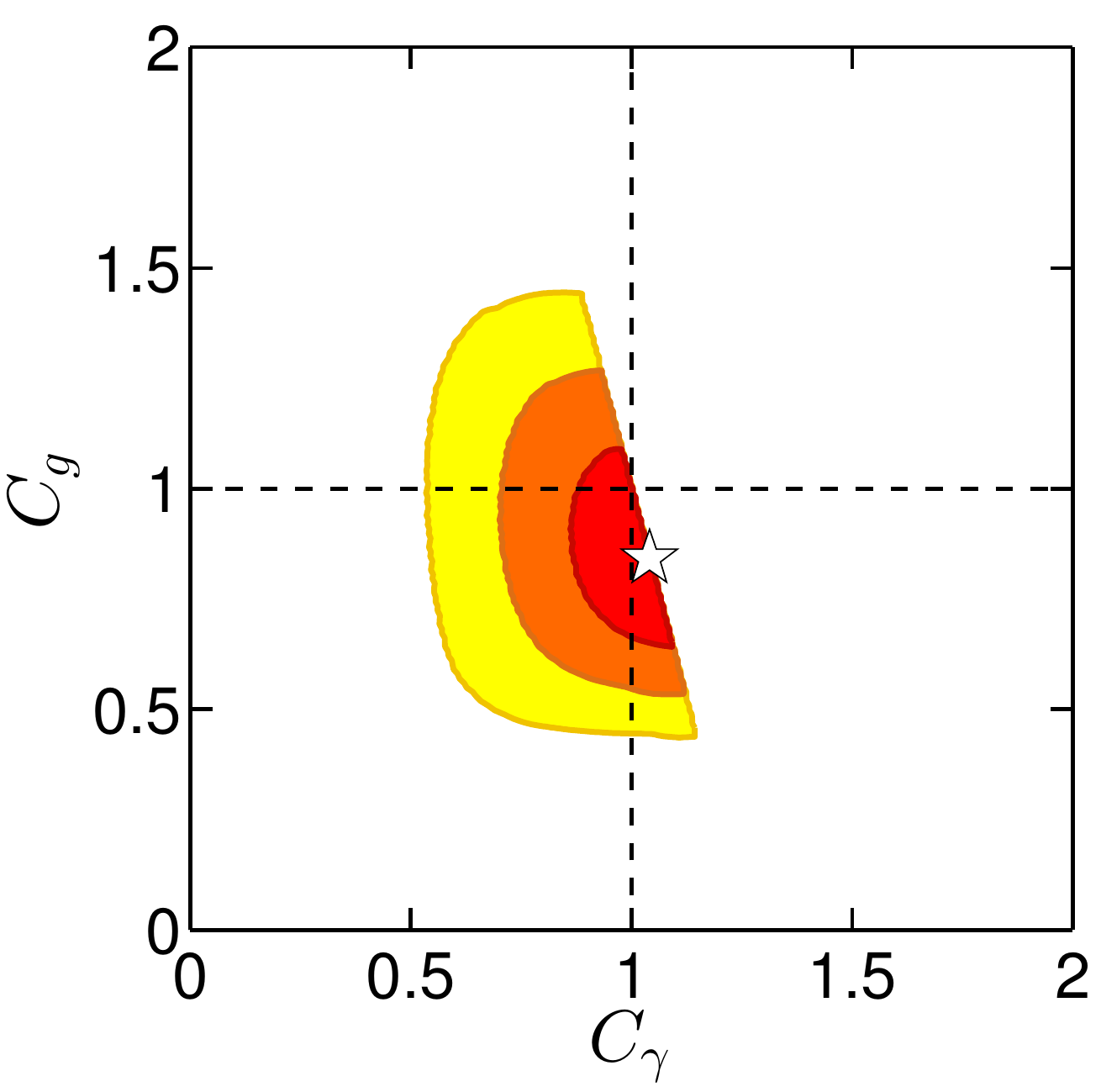}
\caption{Two-dimensional $\chisq$ distributions for the three parameter fit, Fit~{\bf II}, as in Fig.~\ref{fit2-2d} 
but with $\CU>0$, $\CD>0$, $\CV>0$. 
The upper row of plots allows for $\cv>1$, while in the lower row of plots $\cv\le1$ is imposed. 
\label{fit2-2dpos} }
\end{figure}

A negative sign of $\CU$---while maintaining a positive sign of
$m_t$---is actually not easy to achieve. (A sign change of
both $\CU$ and $m_t$ would have no impact on the top quark
induced loop amplitudes.) It would require that $m_t$ is
induced dominantly by the vev of a Higgs boson which is \emph{not} the
Higgs boson considered here. Hence, we have $\CU > 0$ in most models, implying that it
 is important to study the impact of this constraint on our fits.
The fit results when requiring $\CU,\CD > 0$ are shown in the left two plots of Fig.~\ref{fit2-1dpos} 
and the top row of Fig.~\ref{fit2-2dpos}; see also Table~\ref{tab:fit2}.  
We observe that  for this quadrant the results are consistent with SM expectations (\ie\ within $\sim1\sigma$).
Interestingly the fit is not better than the SM itself: $\chimin=18.66$ for $21-3=18$~\dof, corresponding to $p=0.41$.

Another possible model constraint is to require $\cv\leq 1$ (recall that $\cv>0$ by  convention).  
This constraint applies to any model containing only Higgs doublets and singlets. 
The 1d results for the combined requirement of $\cu,\cd>0$ and $\cv\leq 1$ are shown 
in the right two plots of Fig.~\ref{fit2-1dpos}, and  in the bottom-row plots of Fig.~\ref{fit2-2dpos}.
We observe that the best fit values for $\cu$ and $\cd$ are only slightly shifted relative those found 
without constraining $\cv$, and that accordingly the $\cp=\cpb$ and $\cg=\cgb$ at 
the best fit point are only slightly shifted.  However, the $\cv\leq 1$ constraint does 
severely change the upper bound on $\cp$, which for $\cu>0$ and $\dcp=0$ mostly depends 
on the $W$-boson loop contribution. 
The apparent sharpness of the boundary in the $\cg$ vs.\ $\cp$ plane is a result of the fact that these 
two quantities really only depend on $\cu$ for $\cv=1$. 

Finally note that it has been shown in~\cite{Biswas:2012bd,Farina:2012xp} that single top production in
association with a Higgs is greatly enhanced when $\cu$,$\cv$ have
opposite signs. Thus, the possibility of $\cu<0$ should be further
scrutinized by precision measurements of the single top production cross
section at the LHC.

\renewcommand{\arraystretch}{1.3}
\begin{table}\centering
\begin{tabular}{|c|c|c|c|c|}
\hline
Fit  & {\bf I} & {\bf II} & {\bf III}, 1st min. & {\bf III}, 2nd min. \\
\hline 
   $\CU$  &  $1$  & $-0.86_{-0.16}^{+0.14}$ & $-0.06\pm1.30$\; & $\phantom{-}0.06\pm1.30$\; \\
   $\CD$  &  $1$  & $\phantom{-}0.99_{-0.26}^{+0.28}$ & $\phantom{-}1.00_{-0.26}^{+0.28}$ &  $-1.00_{-0.28}^{+0.26}$  \\ 
   $\CV$  &  $1$  & $\phantom{-}0.95_{-0.13}^{+0.12}$ & $\phantom{-}0.93_{-0.14}^{+0.12}$ &  $\phantom{-}0.93_{-0.14}^{+0.12}$  \\
   $\Delta\CP$   &  $\phantom{-}0.43_{-0.16}^{+0.17}$ & -- & $\phantom{-}0.16_{-0.36}^{+0.38}$ & $\phantom{-}0.21_{-0.39}^{+0.37}$ \\ 
   $\Delta\CG$   &  $-0.09 \pm 0.10$ & -- & $\phantom{-}0.83_{-1.17}^{+0.24}$& $\phantom{-}0.83_{-1.17}^{+0.24}$ \\
\hline
  $\CP$ & $\phantom{-}1.43_{-0.16}^{+0.17}$ & $\phantom{-}1.43\pm 0.17$ & $\phantom{-}1.36_{-0.23}^{+0.26}$ & $\phantom{-}1.36_{-0.23}^{+0.26}$ \\ 
  $\CG$ & $\phantom{-}0.91 \pm 0.10$ & $\phantom{-}0.92_{-0.15}^{+0.17}$ & $\phantom{-}0.95_{-0.23}^{+0.26}$ &  $\phantom{-}0.95_{-0.23}^{+0.26}$\\
\hline
 $\chisq_{\rm min}$ &  $12.31$ & $11.95$  & $11.46$ & $11.46$ \\
 $\chisq_{\rm min}/\dof$ &  $0.65$ & $0.66$ & $0.72$ & $0.72$ \\
\hline
\end{tabular}
\caption{Summary of results for Fits {\bf I}--{\bf III}.  For Fit {\bf II}, the tabulated results are from the best fit, cf.\ column 1 of Table~\ref{tab:fit2}.}
\label{chisqmintable}
\end{table}

\renewcommand{\arraystretch}{1.3}
\begin{table}[t]\centering
\begin{tabular}{|c|c|c|c|}
\hline
Sector  & $\cu<0,\,\cd>0$ & $\cu,\,\cd<0$ & $\cu,\,\cd>0$ \\
\hline 
   $\CU$  & $-0.86_{-0.16}^{+0.14}$ & $-0.91_{-0.17}^{+0.15}$ & $0.85_{-0.13}^{+0.15}$ \\
   $\CD$  & $\phantom{-}0.99_{-0.26}^{+0.28}$ & $-0.98_{-0.27}^{+0.26}$ & $0.85_{-0.21}^{+0.22}$ \\ 
   $\CV$  & $\phantom{-}0.95_{-0.13}^{+0.12}$ & $\phantom{-}0.94_{-0.13}^{+0.12}$ & $1.06_{-0.12}^{+0.11}$ \\
\hline
  $\CP$ & $\phantom{-}1.43 \pm 0.17$ & $\phantom{-}1.43_{-0.17}^{+0.16}$ & $1.11_{-0.16}^{+0.15}$ \\ 
  $\CG$ & $\phantom{-}0.92_{-0.15}^{+0.17}$ & $\phantom{-}0.91_{-0.15}^{+0.17}$ & $0.85_{-0.13}^{+0.16}$ \\
\hline
 $\chisq_{\rm min}$ & 11.95 & 12.06 &  18.66 \\
 $\chisq_{\rm min}/\dof$ & 0.66 & 0.67 &  1.04 \\
\hline
\end{tabular}
\caption{Results for Fit {\bf II} in different sectors of the ($\cu$,\,$\cd$) plane.}
\label{tab:fit2}
\end{table}
\renewcommand{\arraystretch}{1.0}

\subsubsection*{\boldmath Fit III: varying $\cu$, $\cd$, $\cv$, $\dcp$ and $\dcg$}

\begin{figure}[t]
\includegraphics[scale=0.4]{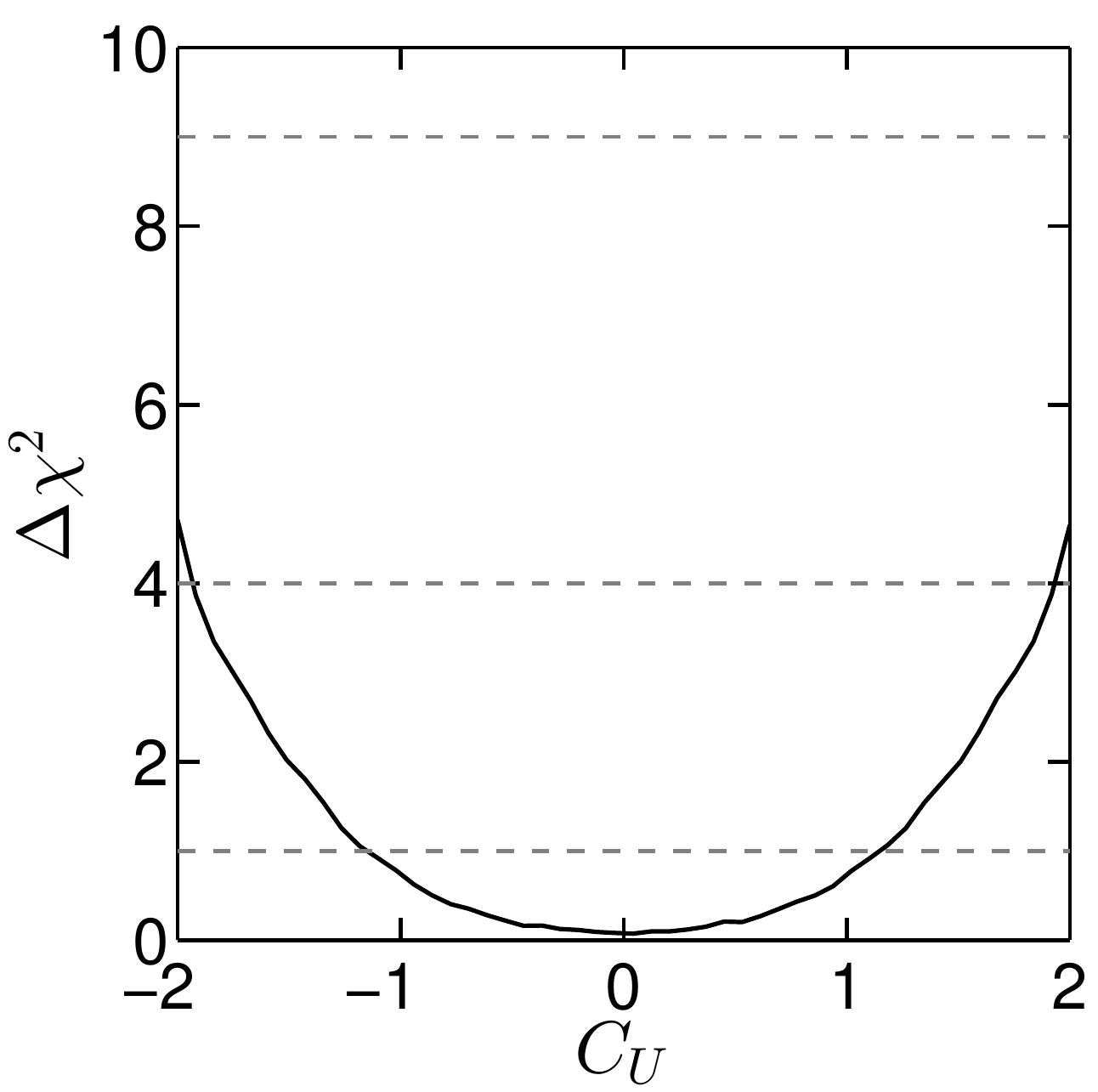}\includegraphics[scale=0.4]{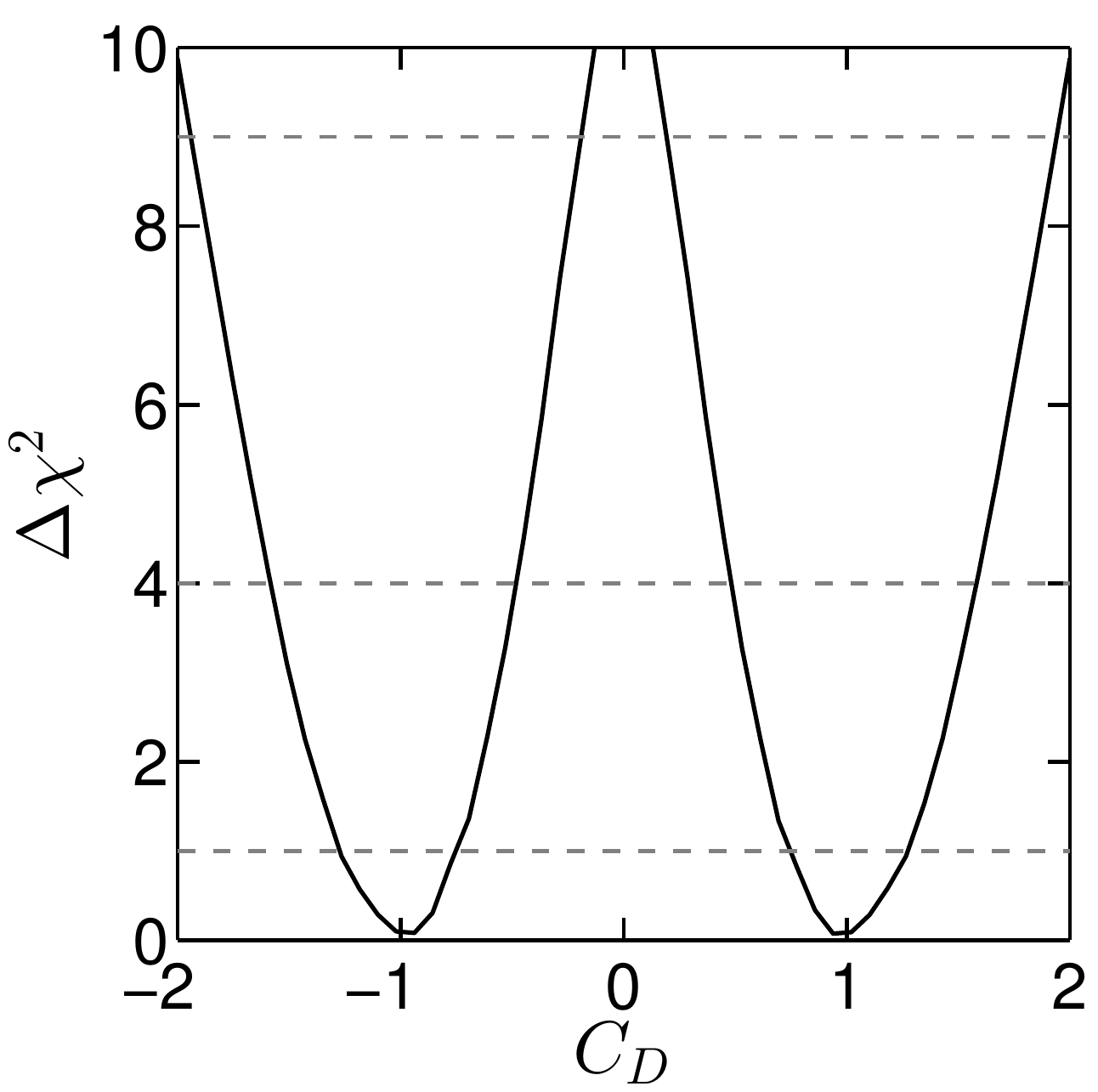}\includegraphics[scale=0.4]{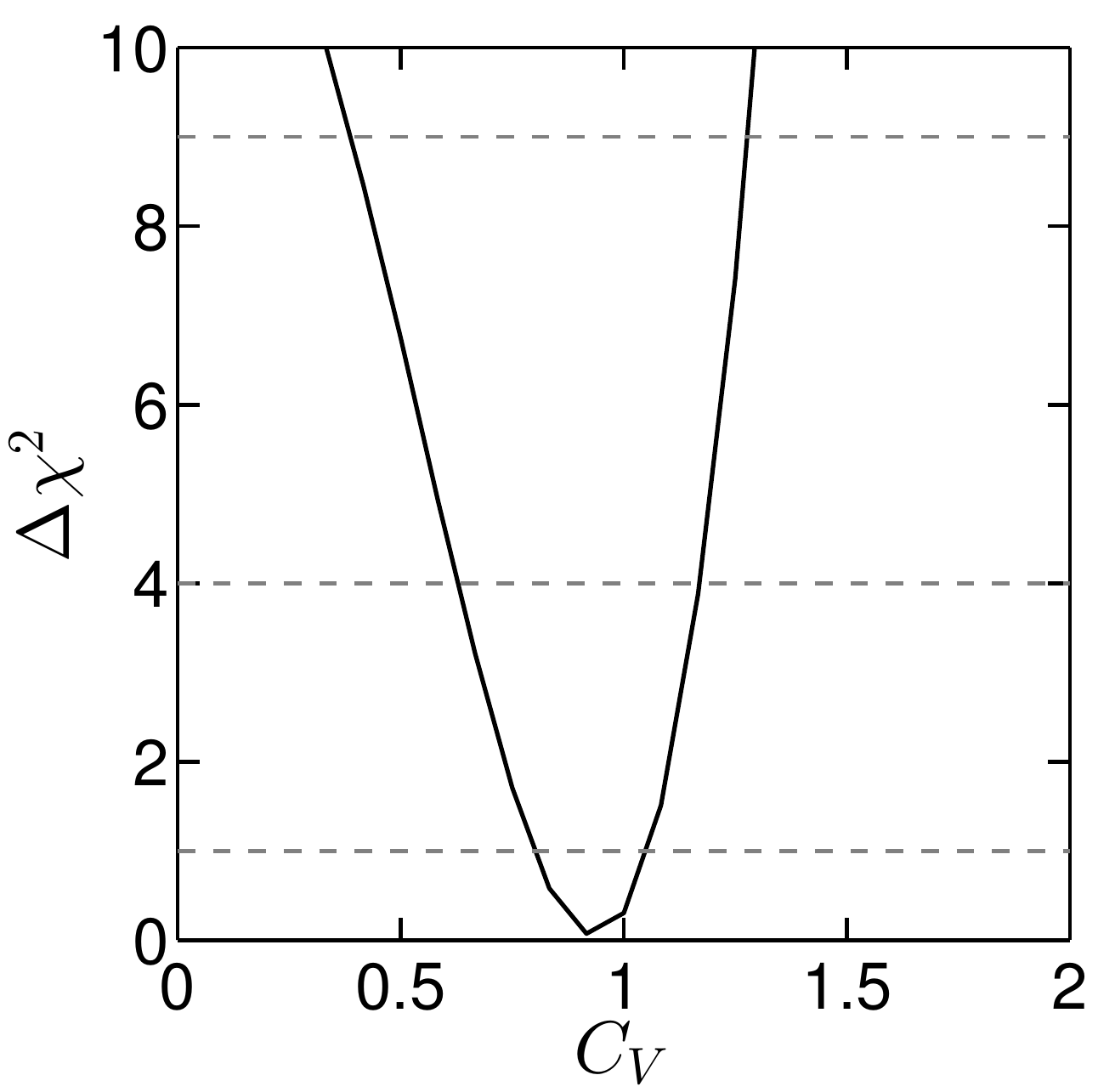}
\includegraphics[scale=0.4]{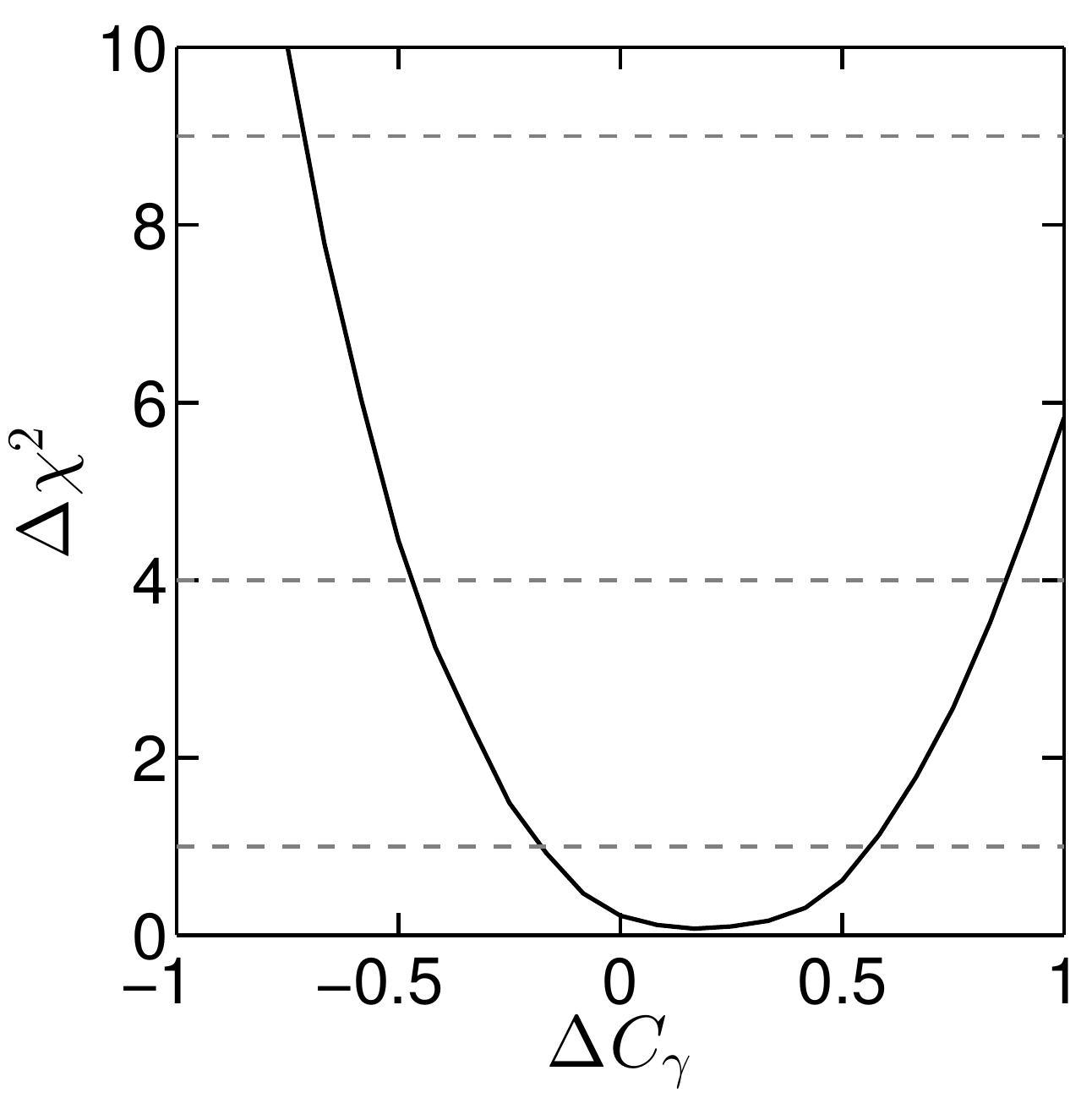}\includegraphics[scale=0.4]{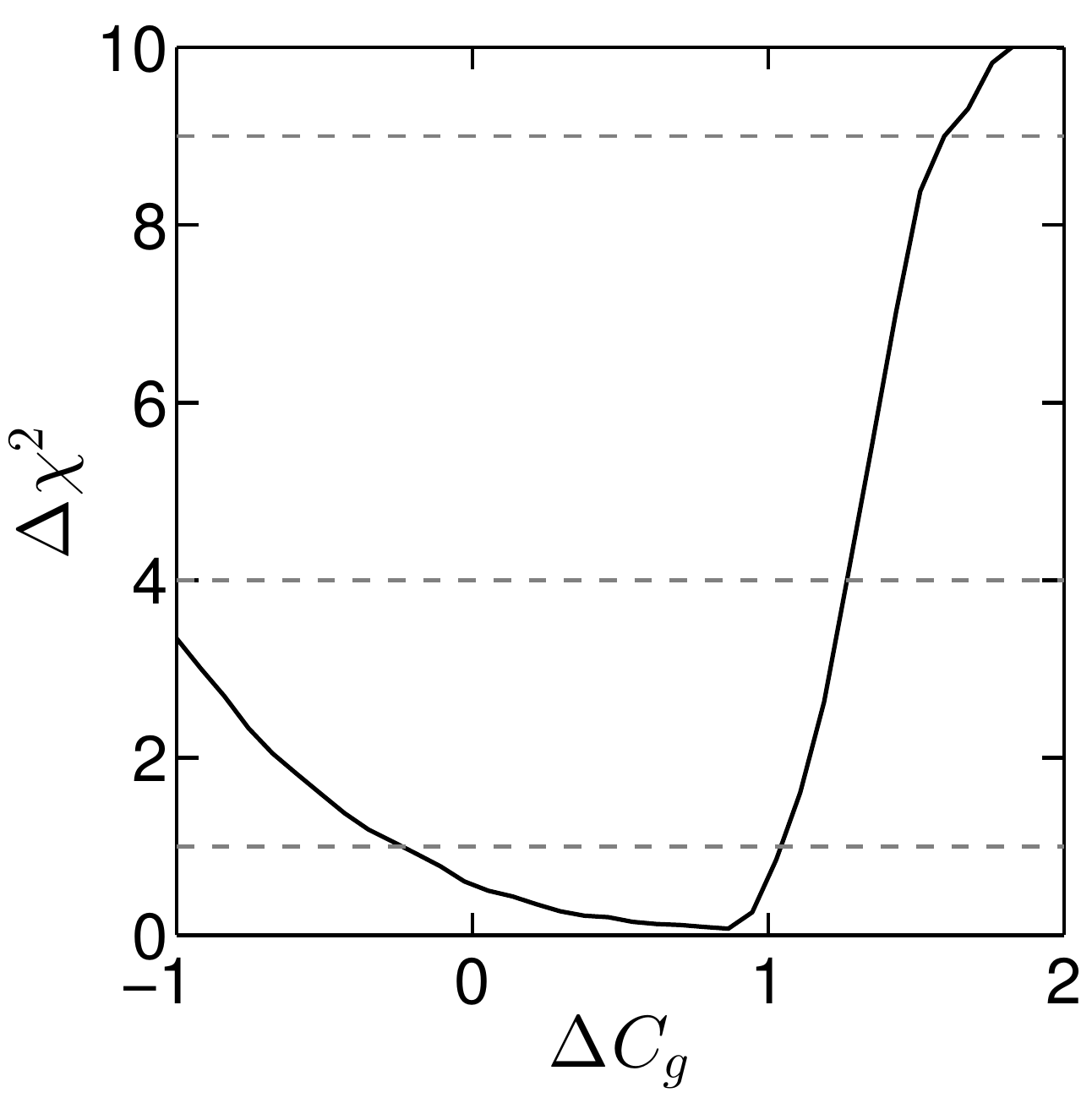}\includegraphics[scale=0.4]{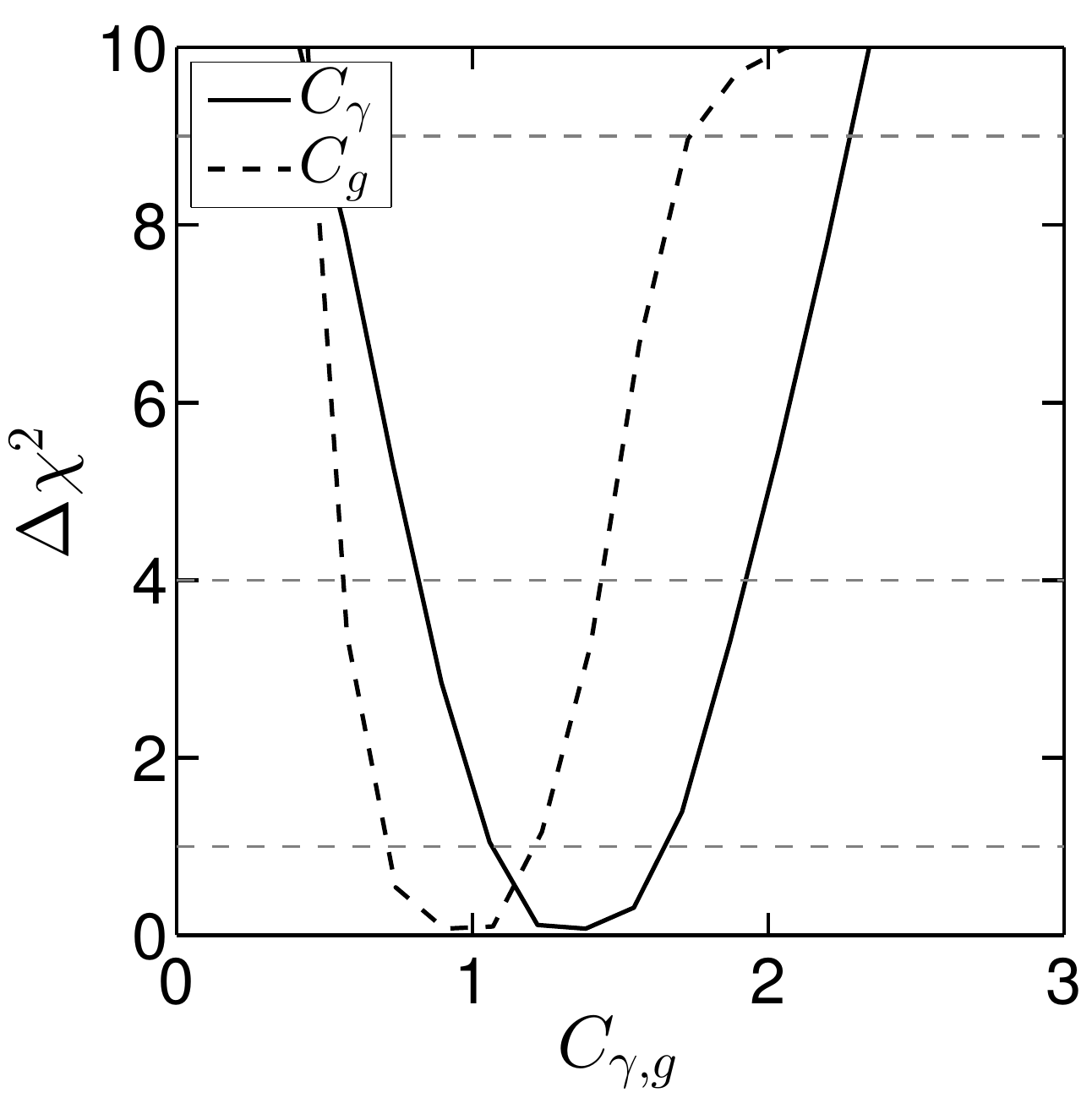}
\caption{One-dimensional $\chisq$ distributions for the five parameter fit of $\CU$, $\CD$, $\CV$, $\Delta\CP$ and $\Delta\CG$ (Fit~{\bf III}). Details regarding the best fit point are given in Table~\ref{chisqmintable}.
\label{fit3-1d} }
\end{figure}

\begin{figure}[t]\centering
\includegraphics[scale=0.4]{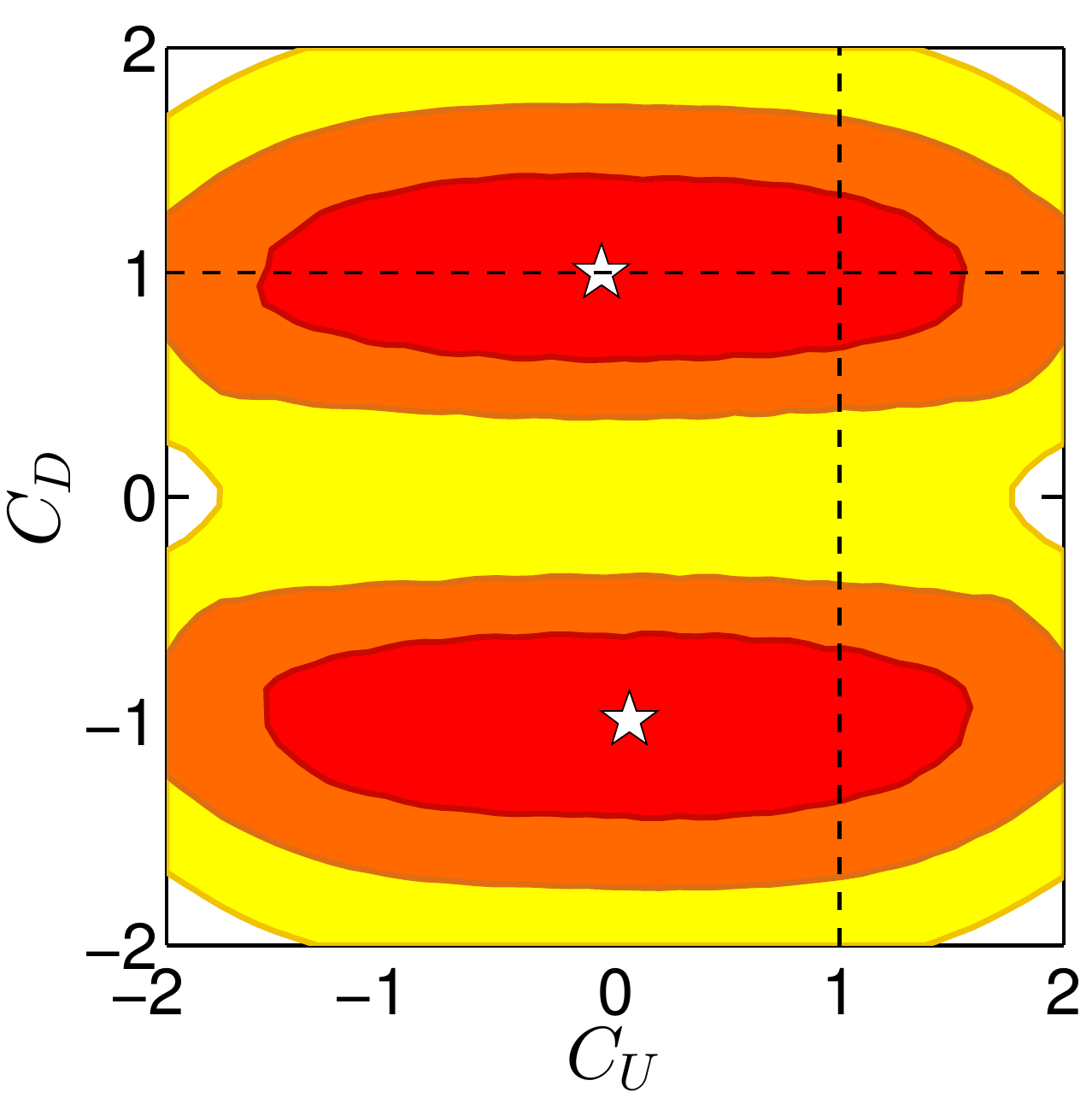}\includegraphics[scale=0.4]{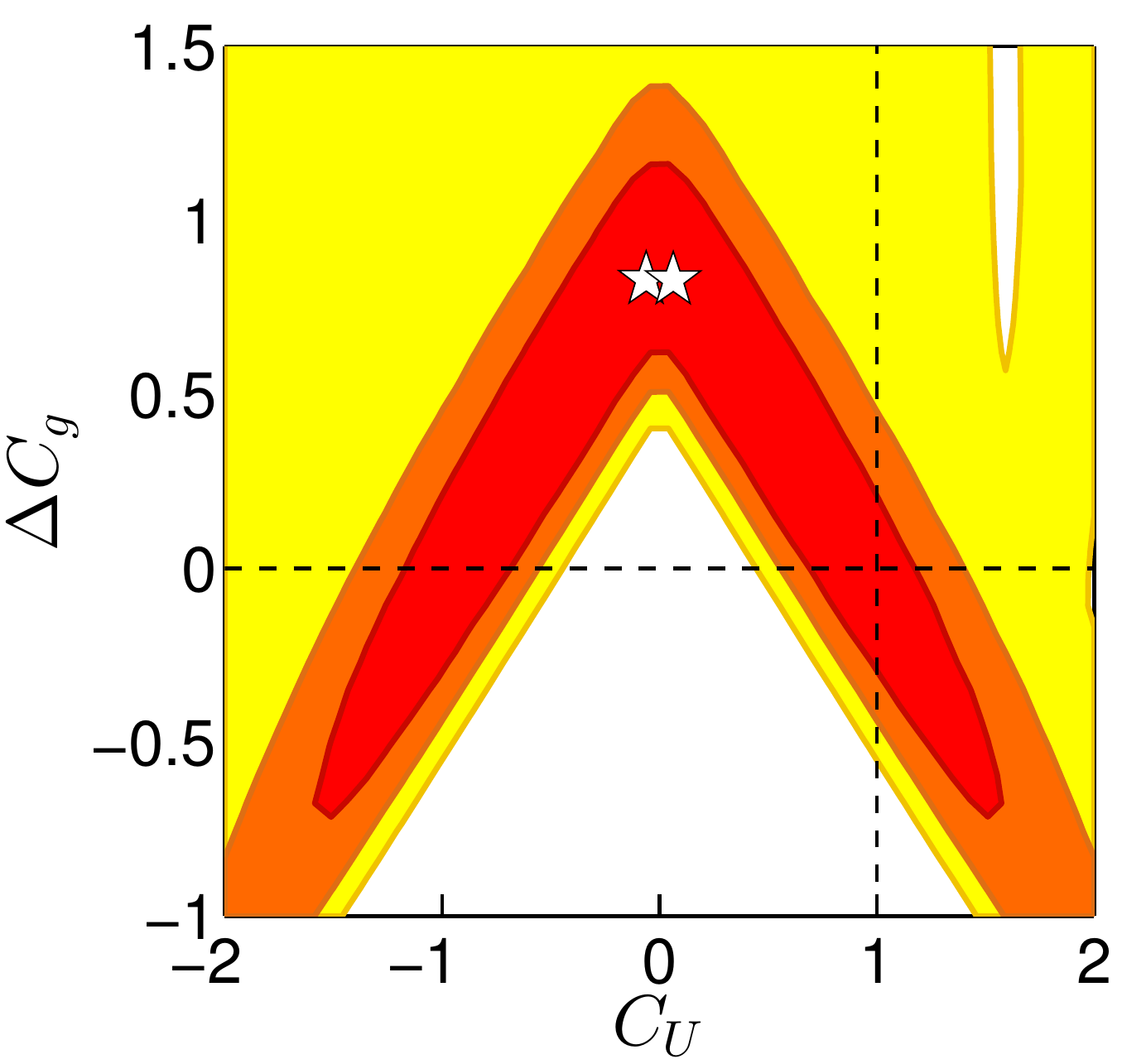}\includegraphics[scale=0.4]{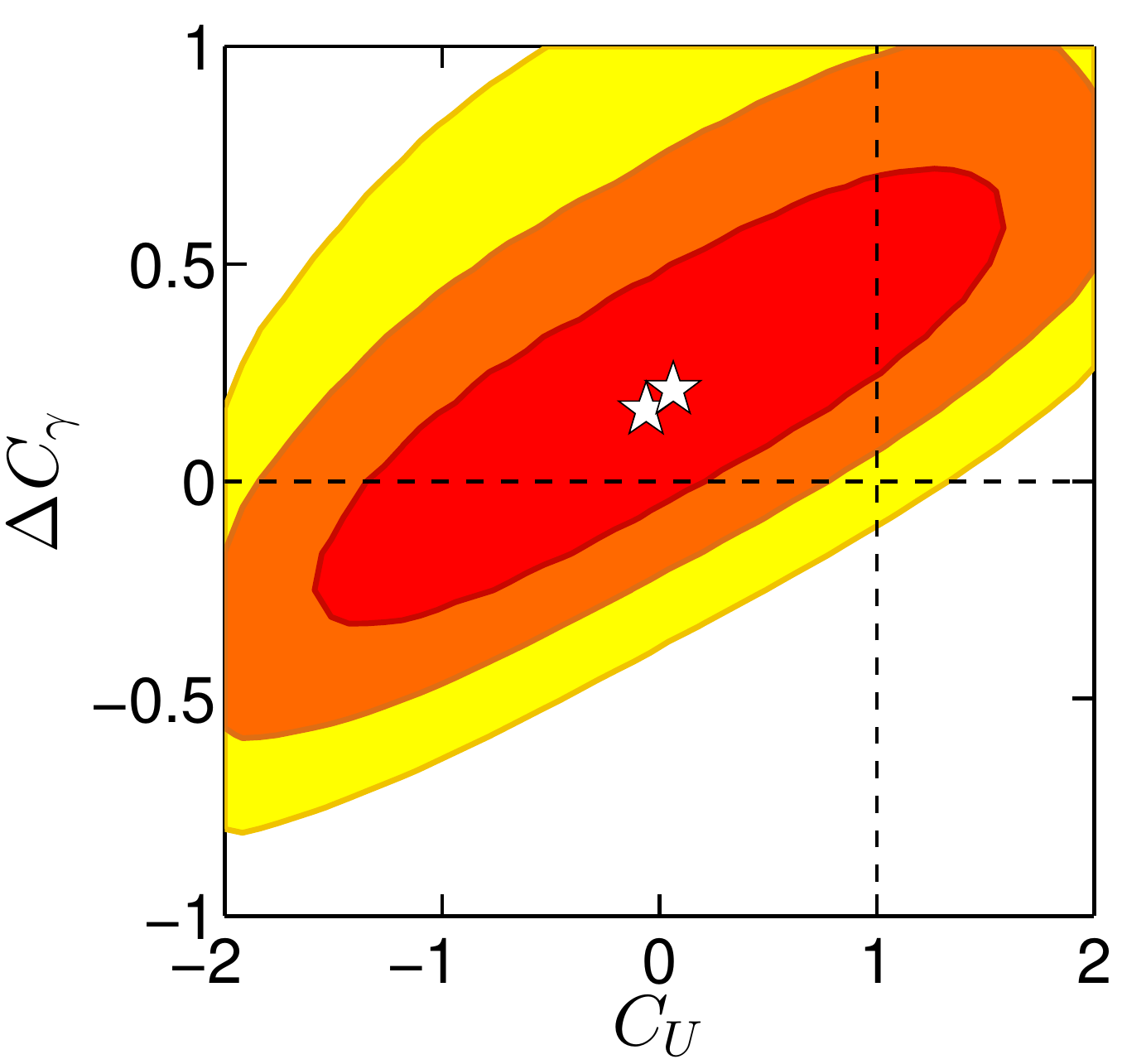}
\includegraphics[scale=0.4]{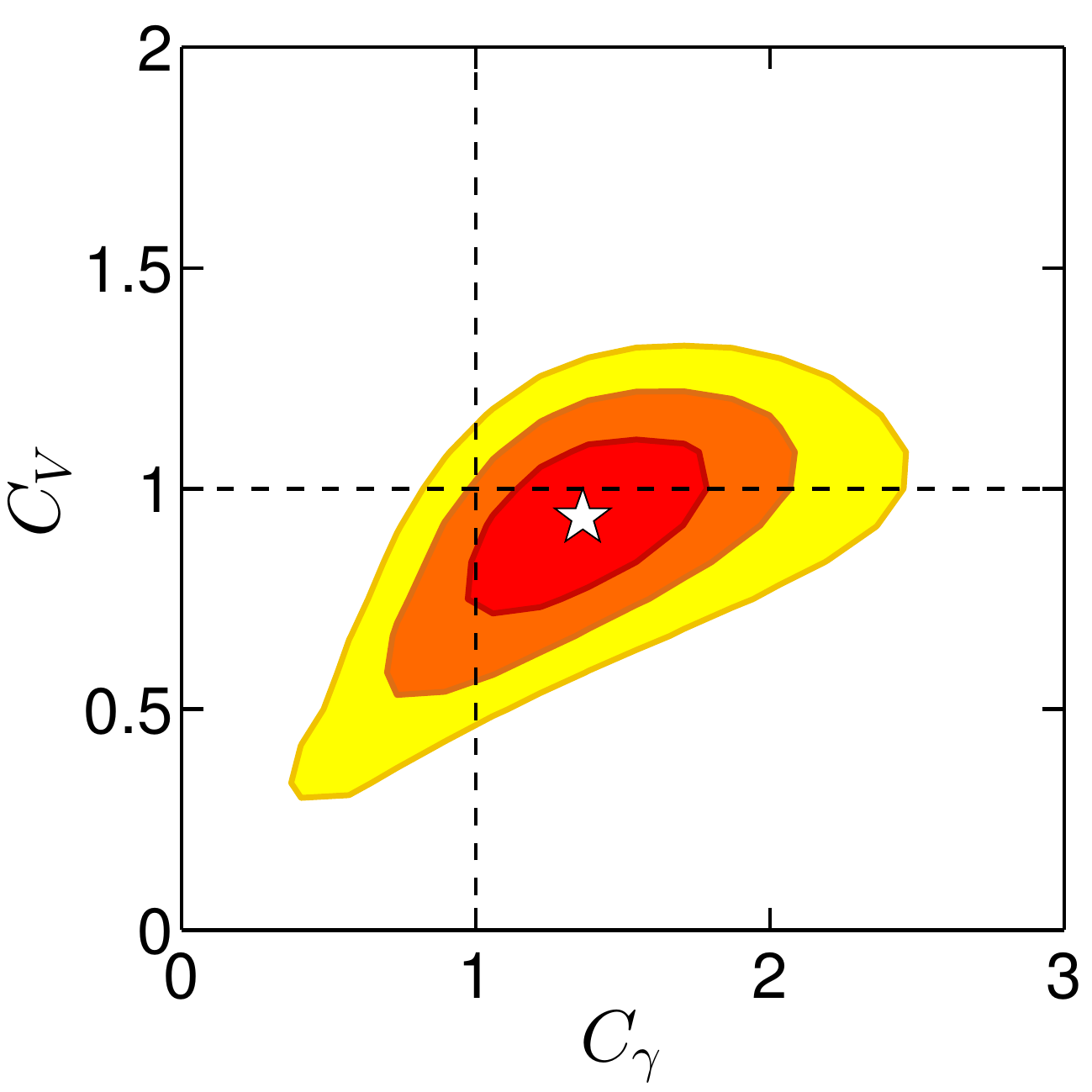}\includegraphics[scale=0.4]{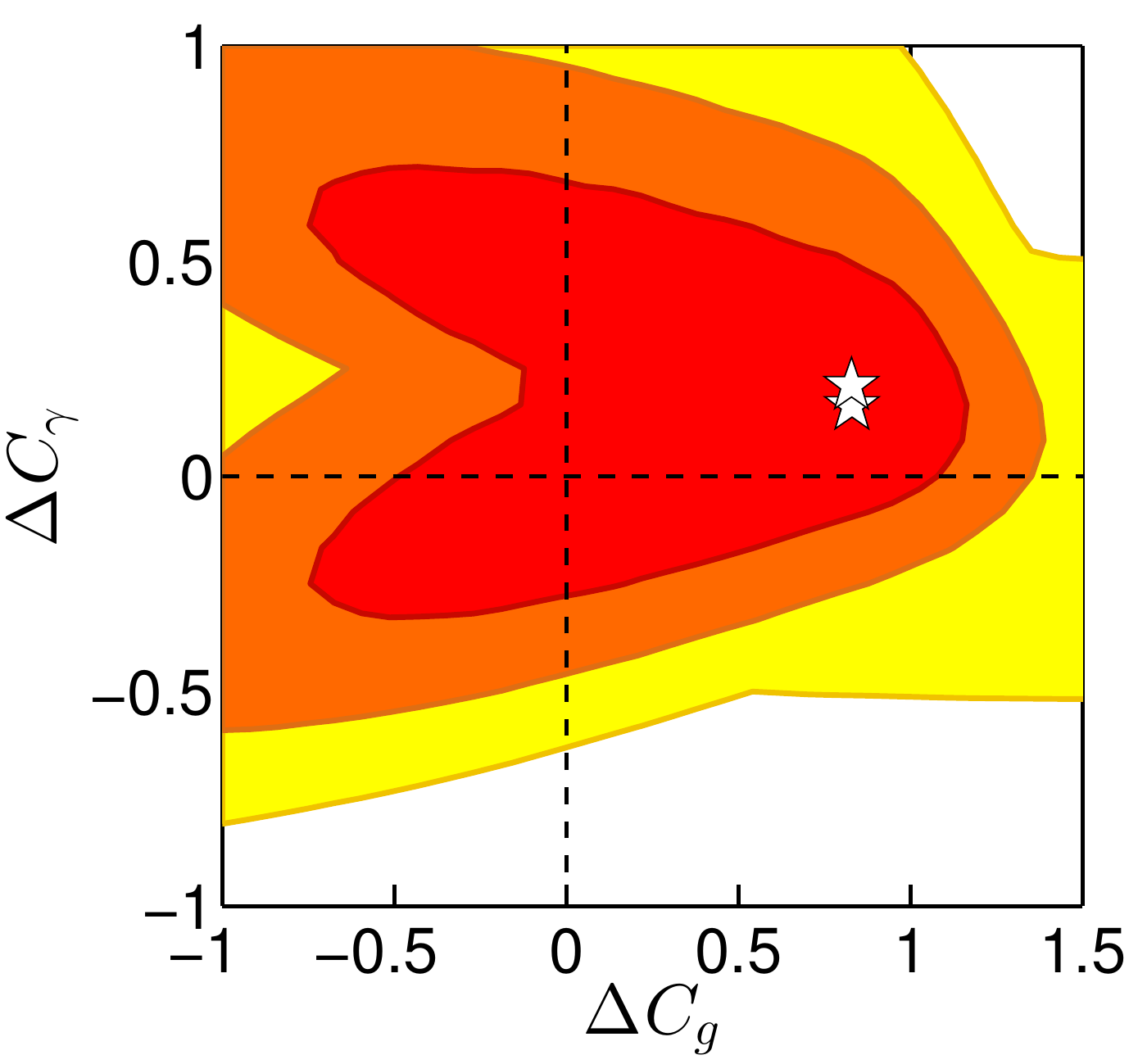}\includegraphics[scale=0.4]{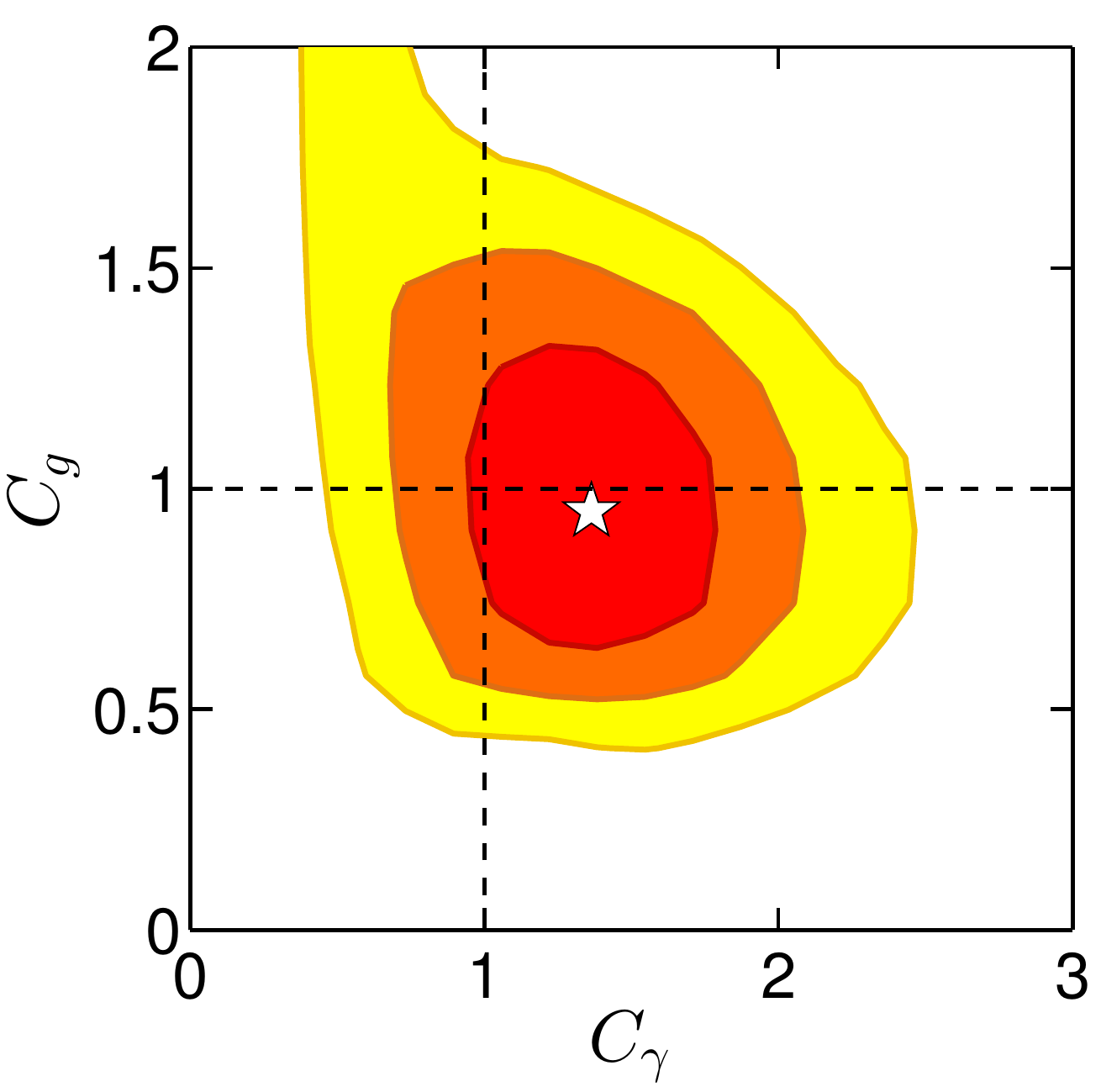}
\caption{Two-dimensional distributions for the five parameter fit of $\CU$, $\CD$, $\CV$, $\Delta\CP$ and $\Delta\CG$ (Fit~{\bf III}). Details regarding the best fit point are given in Table~\ref{chisqmintable}.
\label{fit3-2d} }
\end{figure}

Finally, in Fit~{\bf III}, we allow the $\dcg$ and $\dcp$ additions to $\cgb$ and $\cpb$, fitting 
therefore to five free parameters: $\CU$, $\CD$, $\CV$, $\dcg$, and $\dcp$. 
The associated 1d and 2d plots are given in Figs.~\ref{fit3-1d} and \ref{fit3-2d}.  
There are two main differences as compared to Fit~{\bf II}.  
On the one hand, the preference for $\cp>1$ does does not necessarily imply a negative value for $\cu$, 
since a  positive value for $\dcp$ can contribute to an increase in $\cp$ even when the top-quark loop 
interferes destructively with the $W$ loop. (This is obviously already expected from  Fit~{\bf I}.) 
On the other hand, both $\cu$ and $\dcg$ feed into the effective $\cg$, and if one of them is large the other 
one has to be small to result in a near SM-like $gg\to H$ cross section. 
This anti-correlation between $|\cu|$ and $\dcg$ can be seen in the center-top  plot in Fig.~\ref{fit3-2d}.  
The best fit is actually obtained for $\cu\approx 0$, with $\dcg\approx 1$ in order to compensate for the very 
suppressed top-loop contribution to ggF.  However, it is also apparent that the minimum at $\cu=0$ is quite 
shallow (cf.\ the top left plot in Fig.~\ref{fit3-1d}) and that a fit with $\cu\approx 1$ with small $\dcg$ is well 
within the $68\%$ contour (as should indeed be the case for consistency with Fits {\bf I} and {\bf II}).  

We also note that at the best fit, \ie\ that with $\cu\approx 0$, one finds $\cp\sim \cpb>1$ by virtue of the fact that the 
$W$ loop is not partially 
cancelled by the top loop and only a small $\dcp\sim 0.16$--$0.21$ is needed to further enhance the $\gam\gam$ final state  
and bring $\mu(\gam\gam)$ into agreement with observations; see top-right and bottom-right plots of Fig.~\ref{fit3-2d}. 
If we move to the SM value of $\cu=1$ then 
$\dcp\sim 0.45$ is needed to fit the $\gam\gam$ rate. The best fit results are tabulated in Table~\ref{chisqmintable}.

\begin{figure}[htb]\centering
\includegraphics[scale=0.4]{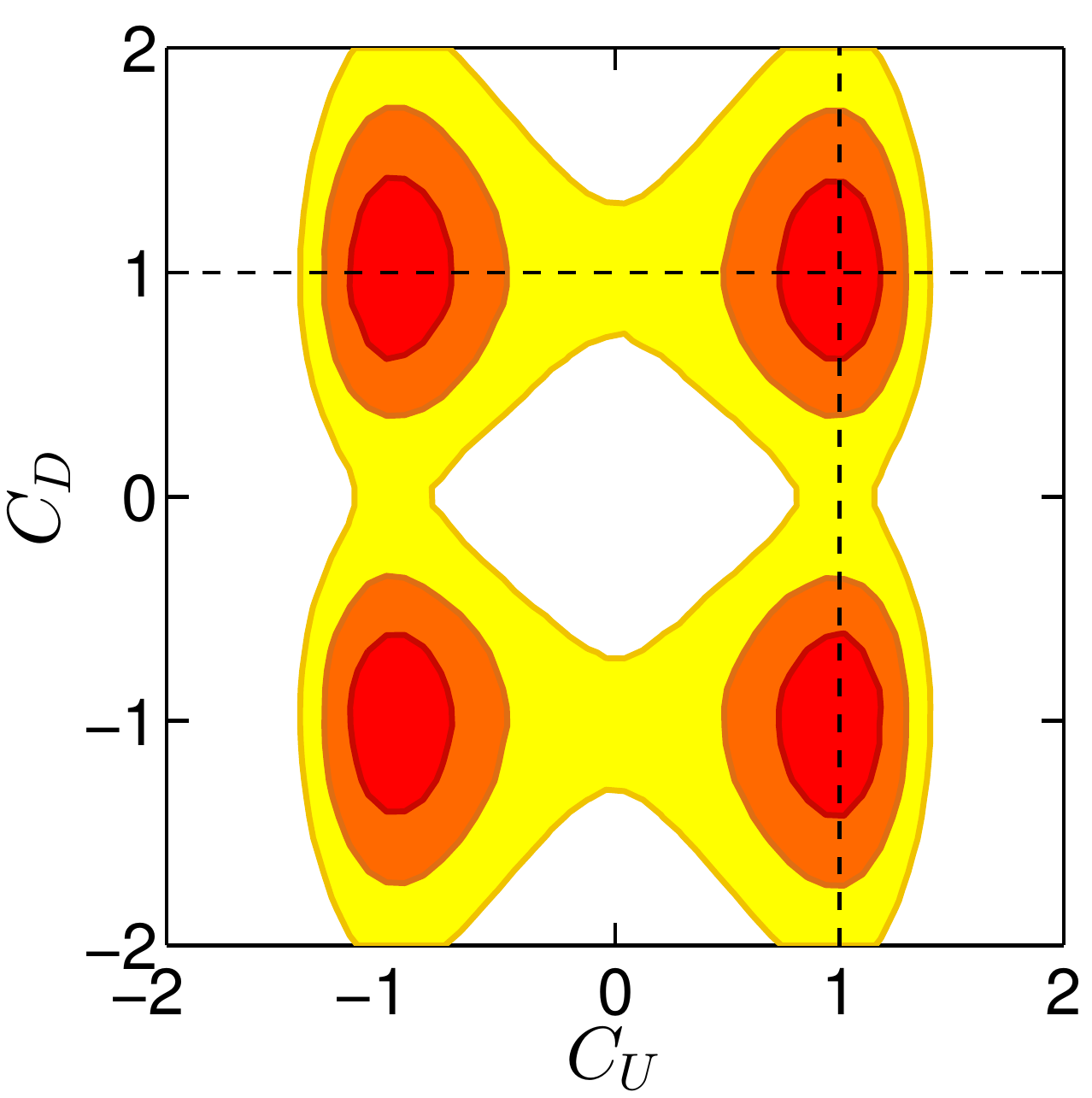}\includegraphics[scale=0.4]{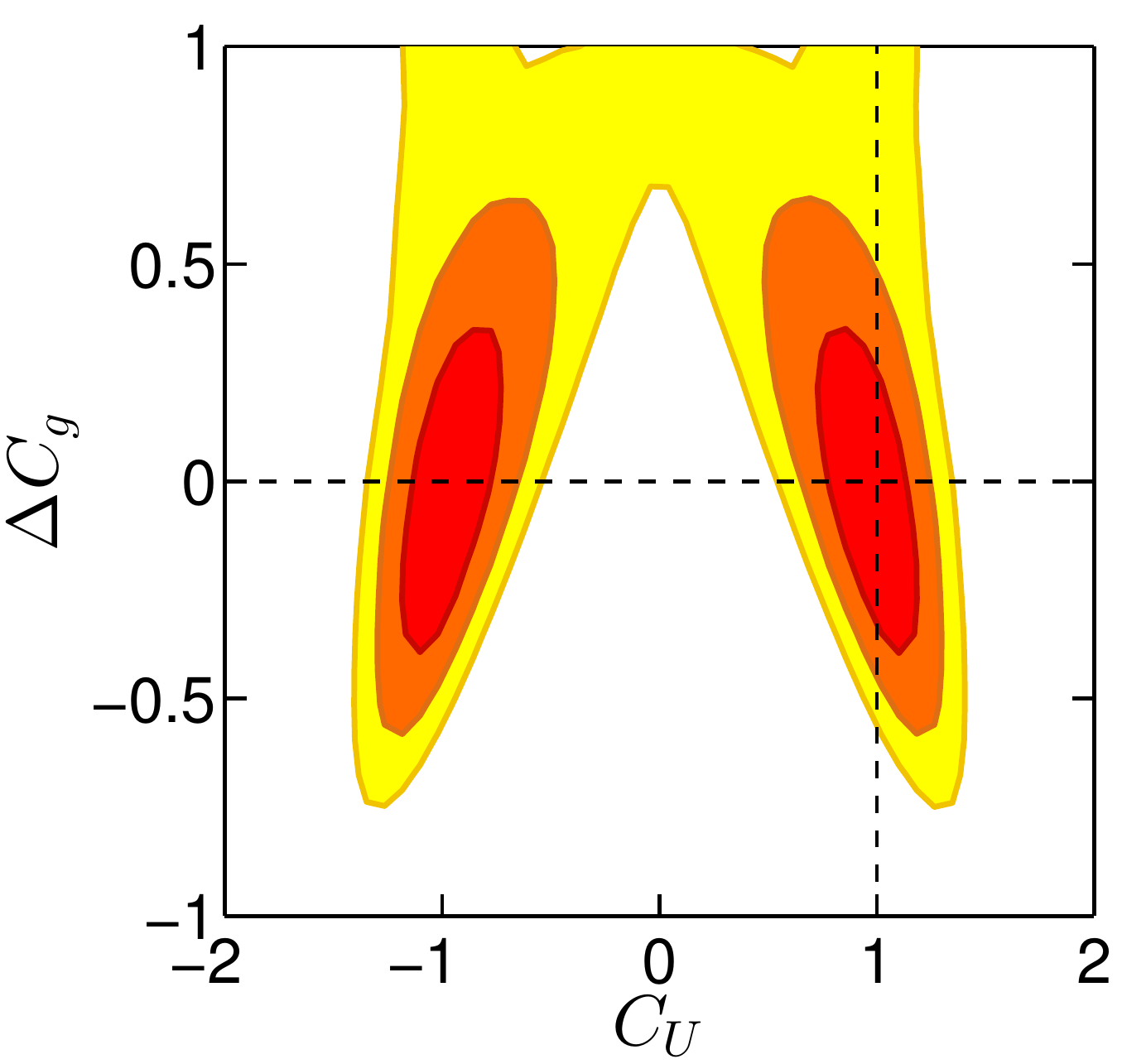}
\caption{Lifting of the degeneracy in $\cu$ and $\dcg$ in Fit~{\bf III} when $t\bar tH$ is measured to 30\% accuracy ($\mu(\tth)=1\pm 0.3$). These two plots should be compared to the top left and top middle plots of Fig.~\ref{fit3-2d}. See text for details. 
\label{fit3-2d-with-ttH} }
\end{figure}

A way to lift the degeneracy in $\cu$ and $\dcg$ would be to have an independent determination of $\cu$. 
This might be achieved by an accurate measurement of the $\tth$ channel, as illustrated in Fig.~\ref{fit3-2d-with-ttH}. 
This figure assumes that $\mu(\tth)$ will eventually be measured with 30\% accuracy --- more concretely, the figure assumes 
$\mu(\tth)=1\pm 0.3$. This is certainly a very challenging task. 
For comparison, CMS currently gives $\mu(\tth)\approx -0.8^{+2.2}_{-1.8}$~\cite{CMS-PAS-HIG-12-045}. 
Finally, as mentioned above, $\cu$ may also be constrained by the associated production of a single top and a Higgs~\cite{Biswas:2012bd,Farina:2012xp}.

\clearpage 
\subsection{Two-Higgs-Doublet Model}
 
So far our fits have been model-independent, relying only on the Lagrangian structure of \Eq{ourldef}. 
Let us now turn to the concrete examples of Two-Higgs-Doublet Models (2HDMs) of Type~I and Type~II. 
In both cases, the basic parameters describing the coupling of either the light $h$ or heavy $H$ CP-even 
Higgs boson are only two: $\alpha$ (the CP-even Higgs mixing angle) and $\tanb=v_u/v_d$, where $v_u$ 
and $v_d$ are the vacuum expectation values of the Higgs field that couples to up-type quarks and down-type 
quarks, respectively.  The Type~I and Type~II models are distinguished by the pattern of their fermionic couplings 
as given in Table~\ref{fermcoups}.  The SM limit for the $h$ ($H$) in the case of both Type~I and Type~II models 
corresponds to $\alpha=\beta-\pi/2$ ($\alpha=\beta$).  
In our discussion below, we implicitly assume that there are no contributions from non-SM particles to the loop 
diagrams for $\cp$ and $\cg$.  In particular, this means our results correspond to the case where the charged 
Higgs boson, whose loop might contribute to $\cp$, is heavy.  

\begin{table}[t]
\begin{center}
\begin{tabular}{|c|c|c|c|c|c|}
\hline
\ & Type I and II  & \multicolumn{2}{c|}  {Type I} & \multicolumn{2}{c|}{Type II} \cr
\hline
Higgs & VV & up quarks & down quarks \& & up quarks & down quarks \&  \cr
&  &  &  leptons &  &  leptons \cr
\hline
 $h$ & $\sin(\beta-\alpha)$ & $\cosa/ \sinb$ & $\cosa/ \sinb$  &  $\cosa/\sinb$ & $-{\sina/\cosb}$   \cr
\hline
 $H$ & $\cos(\beta-\alpha)$ & $\sina/ \sinb$ &  $\sina/ \sinb$ &  $\sina/ \sinb$ & $\cosa/\cosb$ \cr
\hline
 $A$ & 0 & $\cotb$ & $-\cotb$ & $\cotb$  & $\tanb$ \cr
\hline 
\end{tabular}
\end{center}
\vspace{-.15in}
\caption{Tree-level vector boson couplings $C_V^{h_i}$ ($V=W,Z$) and fermionic couplings $C^{h_i}_{F}$
normalized to their SM values for the Type I and Type II two-Higgs-doublet models. }
\label{fermcoups}
\end{table}

\begin{figure}[t]\centering
\includegraphics[scale=0.44]{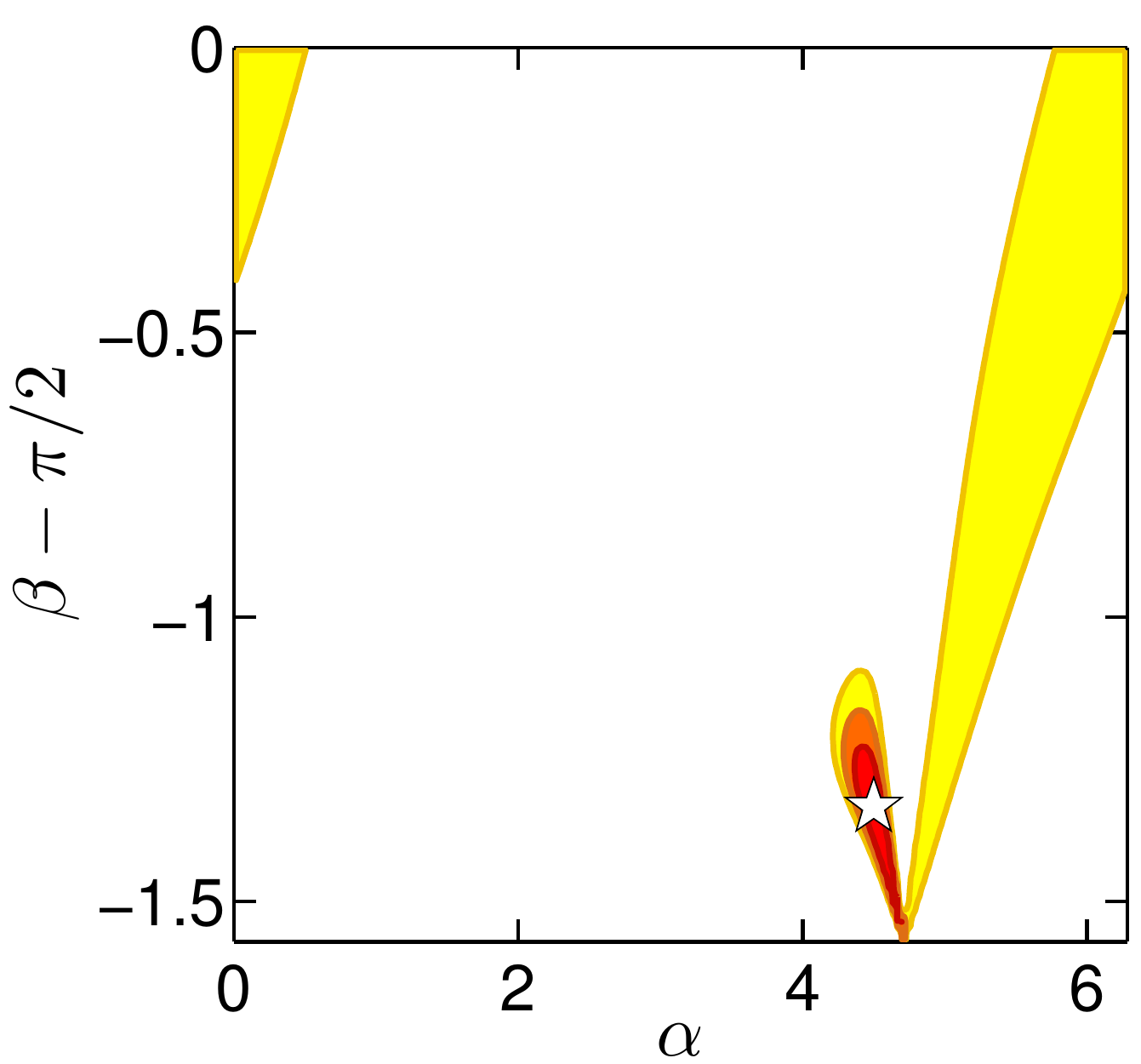}\qquad\includegraphics[scale=0.44]{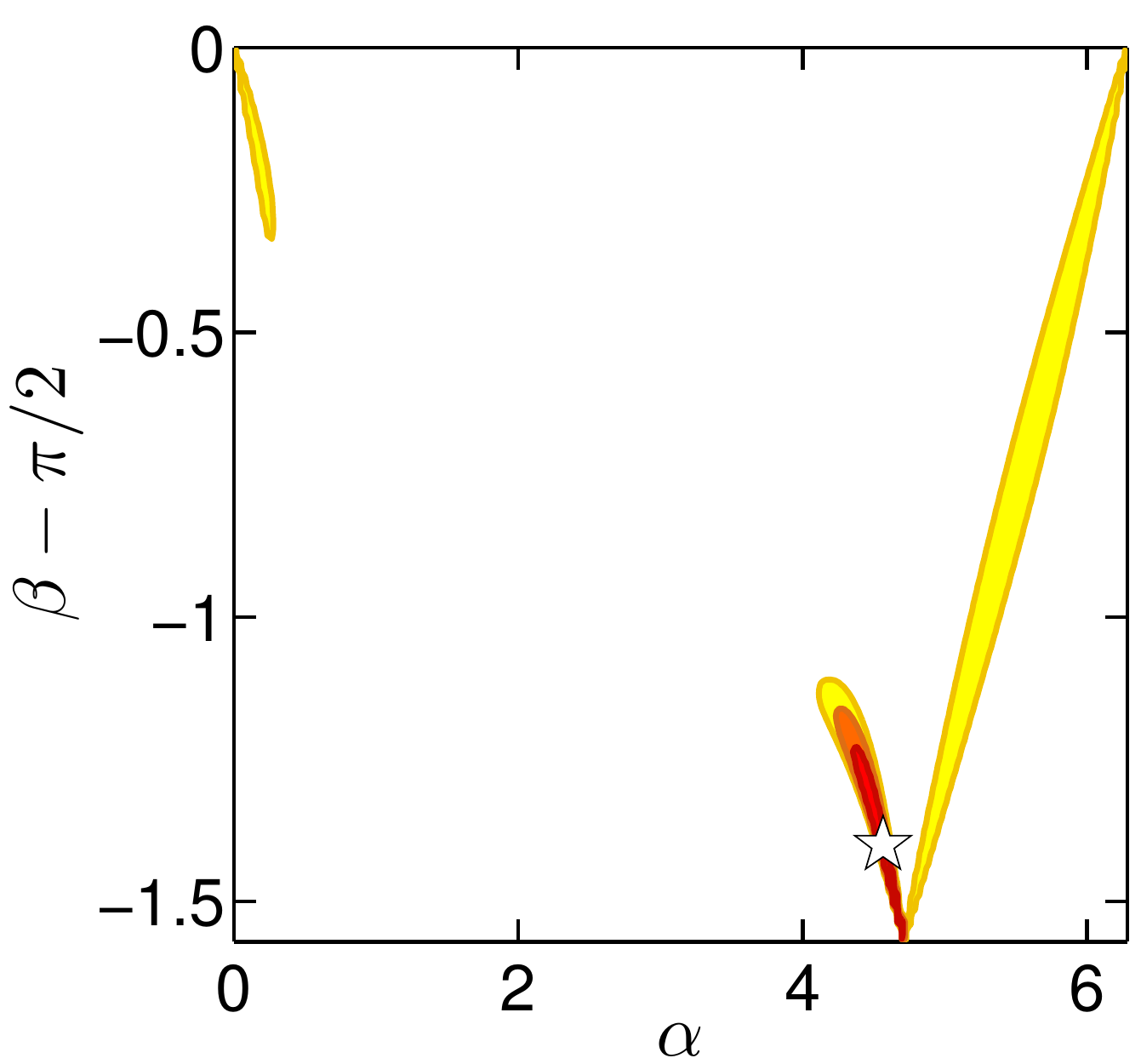}
\includegraphics[scale=0.44]{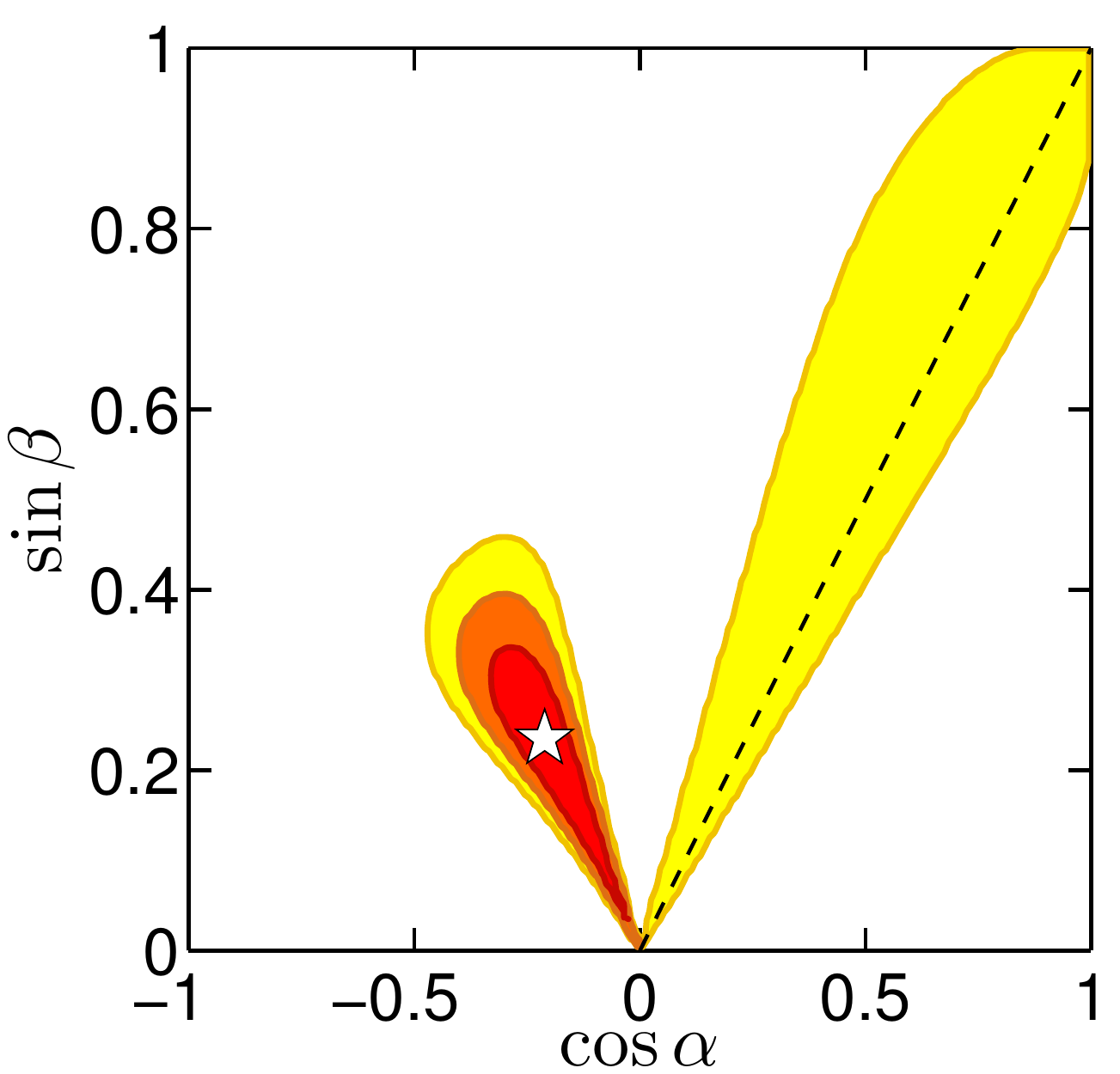}\qquad\includegraphics[scale=0.44]{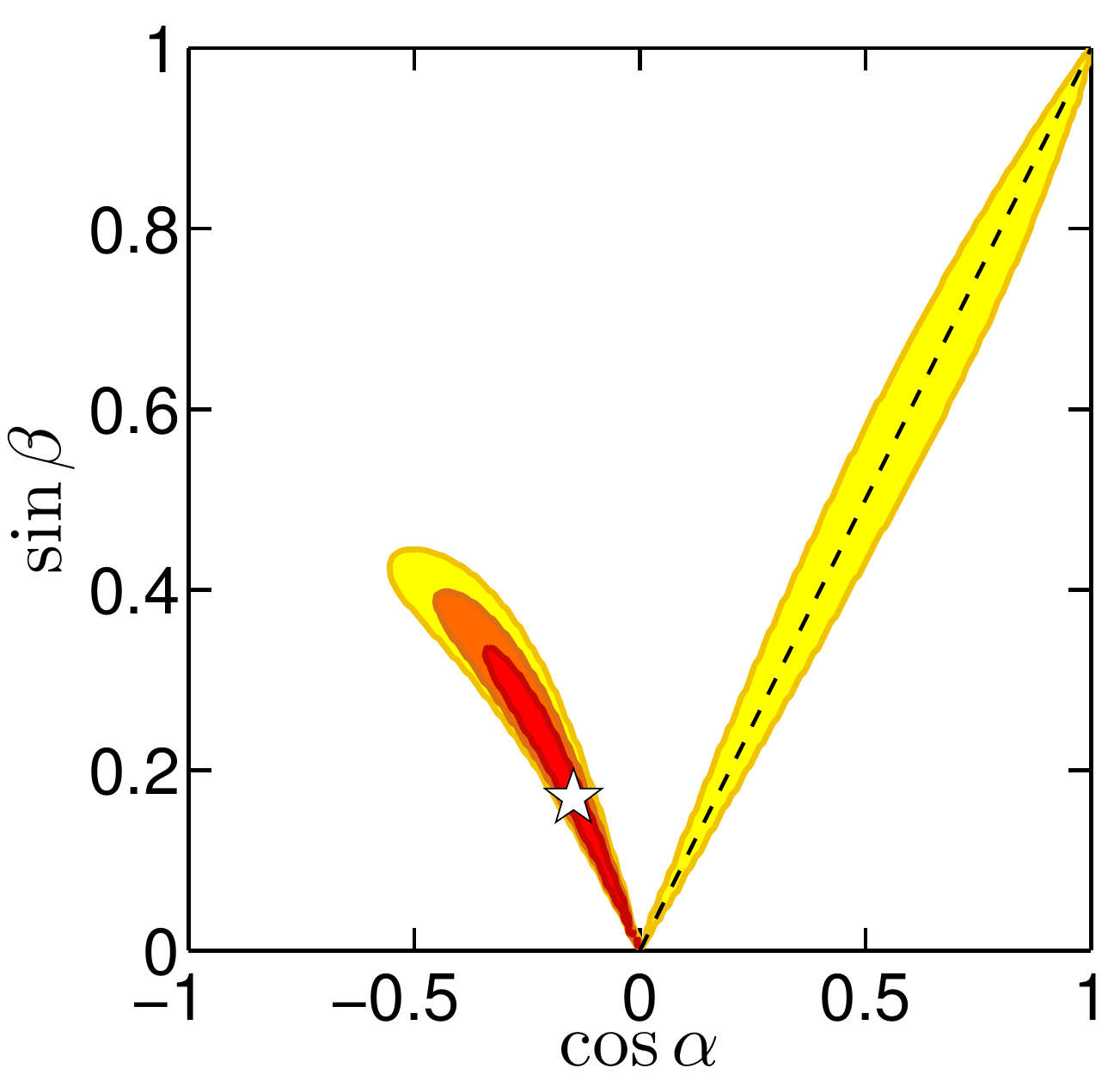}
\caption{2HDM fits for the $h$ in the Type I (left) and Type II (right) models.  
The upper row shows the fit results in the $\beta-\pi/2$ vs.\ $\alpha$ plane, while the 
lower row shows the $\sin\beta$ vs.\ $\cos\alpha$ plane. 
The dashed lines indicate the SM limit. 
The same results are obtained for the heavier $H$ with the replacements 
$\beta-\pi/2\to \beta$ and $\alpha\to \alpha+\pi/2$ ($\sin\beta\to-\cos\beta$, $\cos\alpha\to\sin\alpha$).
\label{fit-2d-2hdmh} }
\end{figure}

The results of the 2HDM fits are shown in Fig.~\ref{fit-2d-2hdmh} for the case that the state near 125~GeV
is the lighter CP-even $h$. The figure also applies for the case of the heavier $H$ being identified with the 
$\sim 125\gev$ state with the replacement rules given in the figure caption.\footnote{Since the $\sim 125\gev$ state clearly couples to $WW,ZZ$ we do not consider the case where the $A$ 
is the only state at $\sim 125\gev$.  We also do not consider the cases where the $\sim125\gev$ peak comprises 
degenerate $(h,H)$, $(h,A)$ or $(H,A)$ pairs.} 
Note that the convention $\cv>0$ implies $\sin(\beta-\alpha)>0$ for the $h$ 
and $\cos(\beta-\alpha)>0$ for the $H$.  
Moreover, the requirement $\tan\beta>0$ restricts $\beta\in [0,\pi/2]$.  
The best fit values and $1\sigma$ ranges for $\alpha$ and $\beta$, together with the corresponding 
values for $\cu$, $\cd$, $\cv$, $\cg$ and $\cg$, are listed in Table~\ref{table:THDMfits}. These numbers are  
again for the case of $h$ being the state near 125~GeV. Replacing $h$ by $H$ amounts to a shift in 
$\alpha\to \alpha+\pi/2$; thus we find $\alpha=6.07_{-0.08}^{+0.09}$ 
($\cos\alpha = 0.98 \pm 0.02$) for the 2HDM-I and $\alpha=6.14_{-0.14}^{+0.15}$ 
($\cos\alpha = 0.99_{-0.03}^{+0.01}$) for the 2HDM-II, 
while the values for $\tan\beta$, $\cu$, $\cd$, $\cv$, {\it etc.}\ do not change.

\renewcommand{\arraystretch}{1.3}
\begin{table}[h]
\centering
\begin{tabular}{|c||c|c||c|c|}
\hline
Fit  & 2HDM-I & 2HDM-II & 2HDM-I, $\tan\beta>1$  & 2HDM-II, $\tan\beta>1$  \\
 \hline 
 $\alpha$ [rad] & $\phantom{-}4.50_{-0.08}^{+0.09}$ & $\phantom{-}4.56_{-0.14}^{+0.15}$ & $5.37_{-0.13}^{+1.11}$ & $6.28_{-0.83}^{+0.17}$ \\		
 $\beta$ [rad]   & $\phantom{-}0.24_{-0.10}^{+0.07}$ & $\phantom{-}0.17_{-0.17}^{+0.12}$ & $[\pi/4,\,\pi/2]$ & $1.56_{-0.78}^{+0.01}$ \\
 \hline
 $\cos\alpha$ & $-0.21_{-0.08}^{+0.09}$ & $-0.15_{-0.13}^{+0.15}$ & $0.61_{-0.11}^{+0.39}$ & $1.00_{-0.67}$ \\		
 $\tan\beta$    & $\phantom{-}0.24_{-0.10}^{+0.08}$ & $\phantom{-}0.17_{-0.17}^{+0.13}$ & $[1,\,+\infty [$ & $[1,\,+\infty [$ \\
 \hline
 $\cu$ & $-0.90_{-0.19}^{+0.17}$ & $-0.87_{-0.13}^{+0.12}$ & $0.87_{-0.15}^{+0.17}$ & $1.02_{-0.07}^{+0.05}$ \\
 $\cd$ & $-0.90_{-0.19}^{+0.17}$ & $\phantom{-}1.00_{-0.01}$ & $0.87_{-0.15}^{+0.17}$ & $0.94_{-0.11}^{+0.13}$ \\
 $\cv$ & $\phantom{-}0.90 \pm 0.07$ & $\phantom{-}0.95_{-0.12}^{+0.05}$ & $0.99_{-0.04}^{+0.01}$ & $1.00_{-0.05}$ \\
 $\cp$ & $\phantom{-}1.37_{-0.10}^{+0.09}$ & $\phantom{-}1.44_{-0.13}^{+0.08}$ & $1.03_{-0.06}$ & $1.01_{-0.09}^{+0.01}$ \\
 $\cg$ & $\phantom{-}0.90_{-0.16}^{+0.19}$ & $\phantom{-}0.92_{-0.11}^{+0.13}$ & $0.87_{-0.15}^{+0.16}$ & $0.99_{-0.04}^{+0.08}$ \\
\hline
 $\chimin$ & 12.20 & 11.95 & 19.43 & 19.88 \\
 \hline
 \end{tabular}
\caption{Summary of fit results for the $h$ in 2HDMs of Type~I and Type~II.}
\label{table:THDMfits}
\end{table}
\renewcommand{\arraystretch}{1.0}

Note that for both the Type~I and the Type~II model, the best fits are quite far from the SM limit in parameter space.  
In particular, since we do not include any extra loop contributions to $\cp$, we end up with negative $\cu$ 
close to $-1$ as in Fit~{\bf II}.  
Demanding $\cu>0$ (\ie\ $\cos\alpha>0$ for $h$, $\sin\alpha>0$ for $H$), 
one ends up in a long `valley' along the decoupling limit 
where the Higgs couplings are SM like, see Fig.~\ref{fit-2d-2hdmh};  
this is however always more than $2\sigma$ away from the best fit.
Furthermore, solutions with very small $\tan\beta<1$ are preferred at more than $2\sigma$.  
Since such small values of $\tanb$ are rather problematic 
(in particular $\tanb<0.5$ is problematical for maintaining a perturbative magnitude for the 
top-quark Yukawa coupling) 
we also give in Table~\ref{table:THDMfits} the corresponding fit results requiring  $\tan\beta>1$. 
These results come quite close to the SM limit, and accordingly 
have a $\chimin$ of about 19--20 (recall that for the SM we find $\chi^2\simeq20.2$). 
2HDMs with $\tan\beta>1$ hence do not provide a better fit than the SM itself. 

A couple of more comments are in order. 
First, an important question that we 
leave for future work 
is whether other --- \eg\ stability, unitarity, perturbativity (SUP) and precision electroweak (PEW) --- 
constraints are obeyed at the best-fit points, or the 68\% CL regions.  
Here we just note that according to Fig.~1 of \cite{Drozd:2012vf}, the SUP and PEW constraints 
do not seem problematic for Type~II, but may play a role for Type~I models at low $\tanb$.   

Second, the best fits correspond to very small $\tanb$ (small $\beta$) values that are potentially constrained 
by limits from B-physics, in particular from $\Delta M_{Bs}$ and $Z\to b\bar b$ . 
The B-physics constraints are summarized in Figs.~15 and 18 of \cite{Branco:2011iw} 
for Type~II and Type~I, respectively.   
Figure~18 for Type~I places a lower bound on $\tanb$ as a function of the charged Higgs mass which excludes 
small $\tanb<1$ unless the charged Higgs is {\it very} heavy, something that is possible but somewhat unnatural.  
Figure~15 for Type~II places a substantial lower bound on the charged Higgs 
mass for all $\tanb$, but such a constraint does not exclude the 68\% CL region.  

\begin{figure}[h]\centering
\includegraphics[scale=0.4]{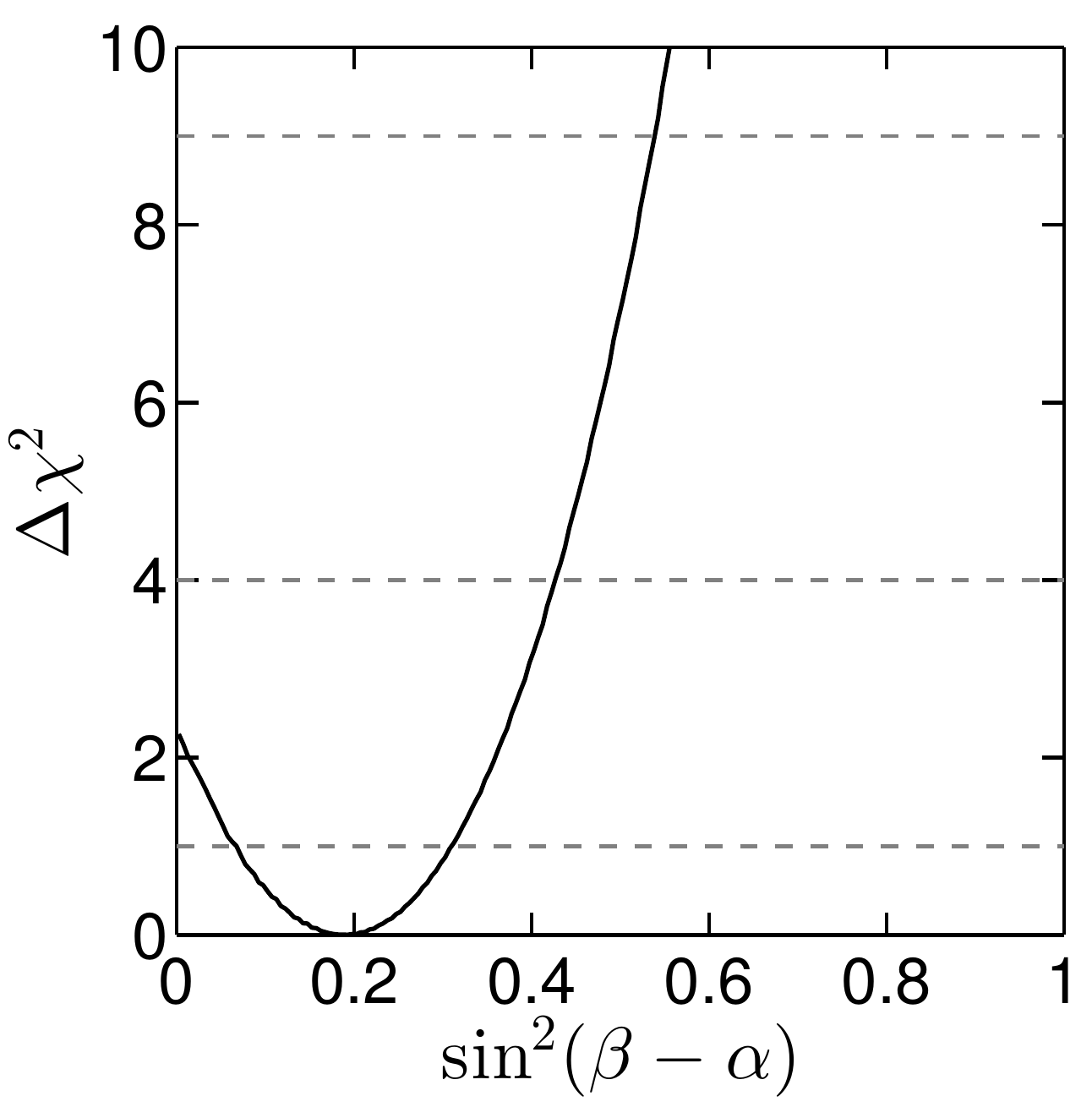}\qquad\includegraphics[scale=0.4]{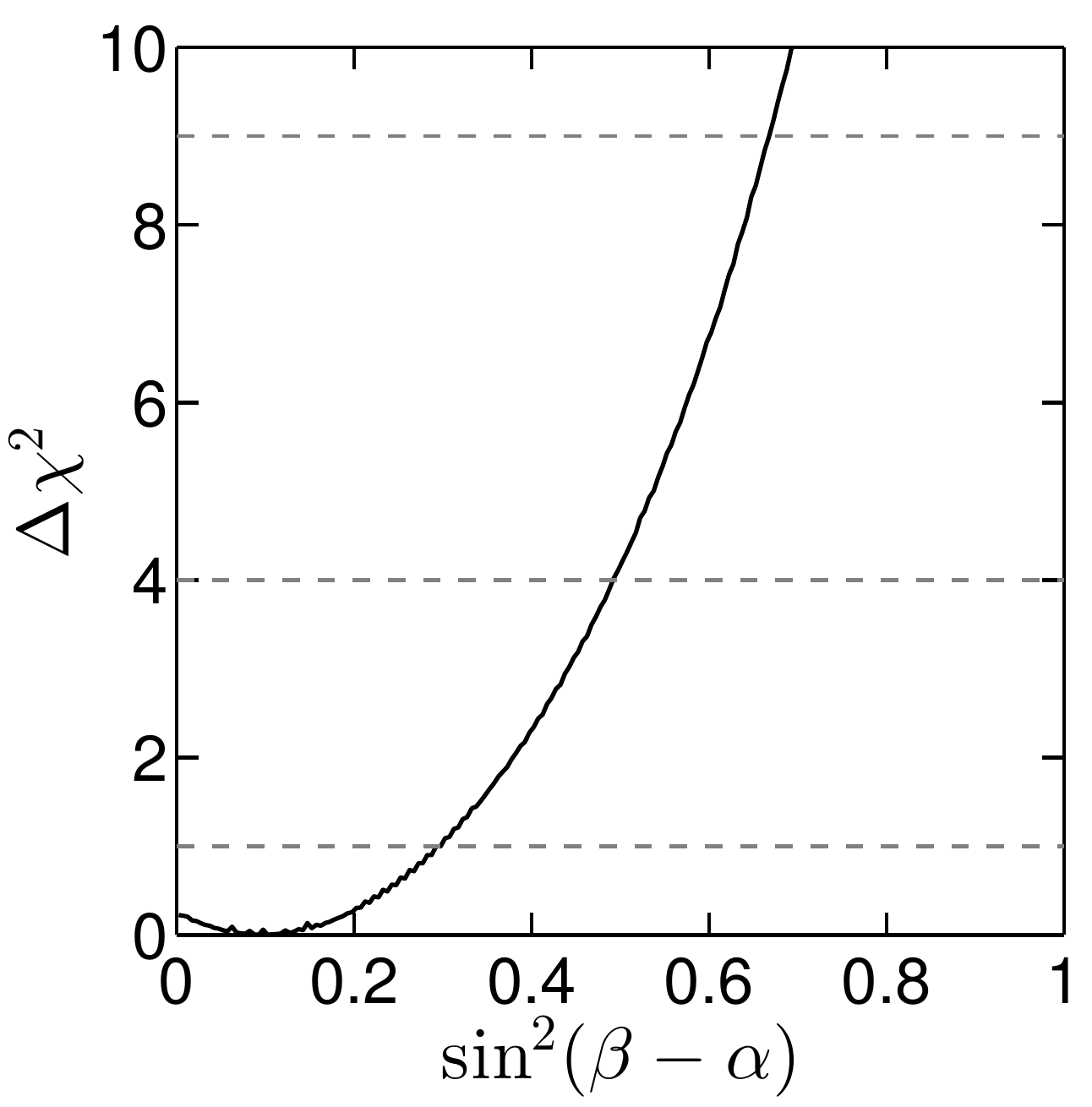}
\caption{$\Delta\chi^2$ distribution of $\sin^2(\beta-\alpha)$ in the Type I (left) and Type II (right) models 
for the case that $H$ is the observed state near 125~GeV.   
\label{sinbma} }
\end{figure}

Third, we remind the reader that in the 2HDMs, the soft $Z_2$-symmetry-breaking $m_{12}^2$ and 
the other Higgs masses ($M_h$, $M_H$ and $M_A$) are independent parameters.  
It is thus possible to have either $M_h$ or $M_H\sim 125\gev$ without violating constraints from direct 
searches for the charged Higgs whose mass is related to $m_A$. However, in the case of  $M_H\sim 125\gev$, 
one has to avoid the LEP limits for the lighter $h$, which severely constrain the $h$ coupling to $ZZ$ in case   
of $M_h<114\gev$~\cite{TeixeiraDias:2008fc}. So either $M_h\gtrsim 114 \gev$ for $M_H\approx 125\gev$, or 
$\sin^2(\beta-\alpha)$ needs to be small (\eg\ $\sin^2(\beta-\alpha)\lesssim 0.3$ for $M_h\approx 100\gev$, or 
$\sin^2(\beta-\alpha)\lesssim 0.1$ for $M_h<90\gev$). 
The $\Delta\chi^2$ distributions of $\sin^2(\beta-\alpha)$ for Type~I and Type~II with $M_H\sim 125\gev$ 
are shown in Fig.~\ref{sinbma}. Interestingly,  around the best fit the $h$ coupling to $ZZ$ is sufficiently suppressed 
to allow for $M_h$ of the order of $100\gev$ (or lower in Type~II). 

\clearpage

\section{Summary and Conclusions}

We assessed to which extend the currently available data on the Higgs-like scalar constrain the Higgs couplings. To this end we performed fits to all public data from the LHC and the Tevatron experiments. 

First, we employed a general 
parametrization of the Higgs couplings based on a SM-like Lagrangian, but allowing for extra contributions to the loop-induced couplings of the Higgs-like scalar to gluons and photons. While the SM does not provide a bad fit ($\chi^2/\dof=0.96$), 
it is more than $2\sigma$ away from our best fit solutions. The main pull comes from the enhanced $H\to\gamma\gamma$ 
rates observed by ATLAS and CMS, as well as from the Tevatron experiments. 
The best fits are thus obtained when either $\cu\sim -1$ (\ie\ opposite in sign to the SM expectation) or there is a large BSM contribution to the $\gam\gam$ coupling of the Higgs. In short, significant deviations from the SM values are preferred by the currently available data and should certainly be considered viable. Since having $\cu\sim -1$ (in the $\cv>0$ convention)  is not easy to achieve in a realistic model context, and leads to unitarity violation in $WW\rightarrow t\bar{t}$ scattering at scales that can be as low as 5~TeV~\cite{Choudhury:2012,Bhattacharyya:2012},
it would seem that new physics contributions to the effective couplings of the Higgs to gluons and photons are the preferred option.  (The possibility of a second, degenerate Higgs boson contributing to the observed signal remains another interesting option, not considered here.) 

Second, we examined how well 2HDM models of Type I and Type II fit the data. 
We found that it is possible to obtain a good fit in these models with $\sin(\beta-\alpha)$ ($\cos(\beta-\alpha)$), in the $h$ ($H$) cases, respectively, not far from $1$.  However, the best fit values for the individual $\cu$, $\cd$, $\cp$ and $\cg$ parameters lie far from their SM values. Further, the best fits give $\tan\beta<1$, which is disfavored from the theoretical point of view if we want perturbativity up to the GUT scale. 
Requiring $\tan\beta>1$ (or simply $\cu>0$) pushes the fit into the SM
`valley' and no improvement over the pure SM solution is obtained. In
particular the $\chi^2$ obtained in this region is substantially
larger than that for the best fit, and not far from the $\chi^2$ found
for the SM.

\begin{figure}[t]\centering
\includegraphics[scale=0.56]{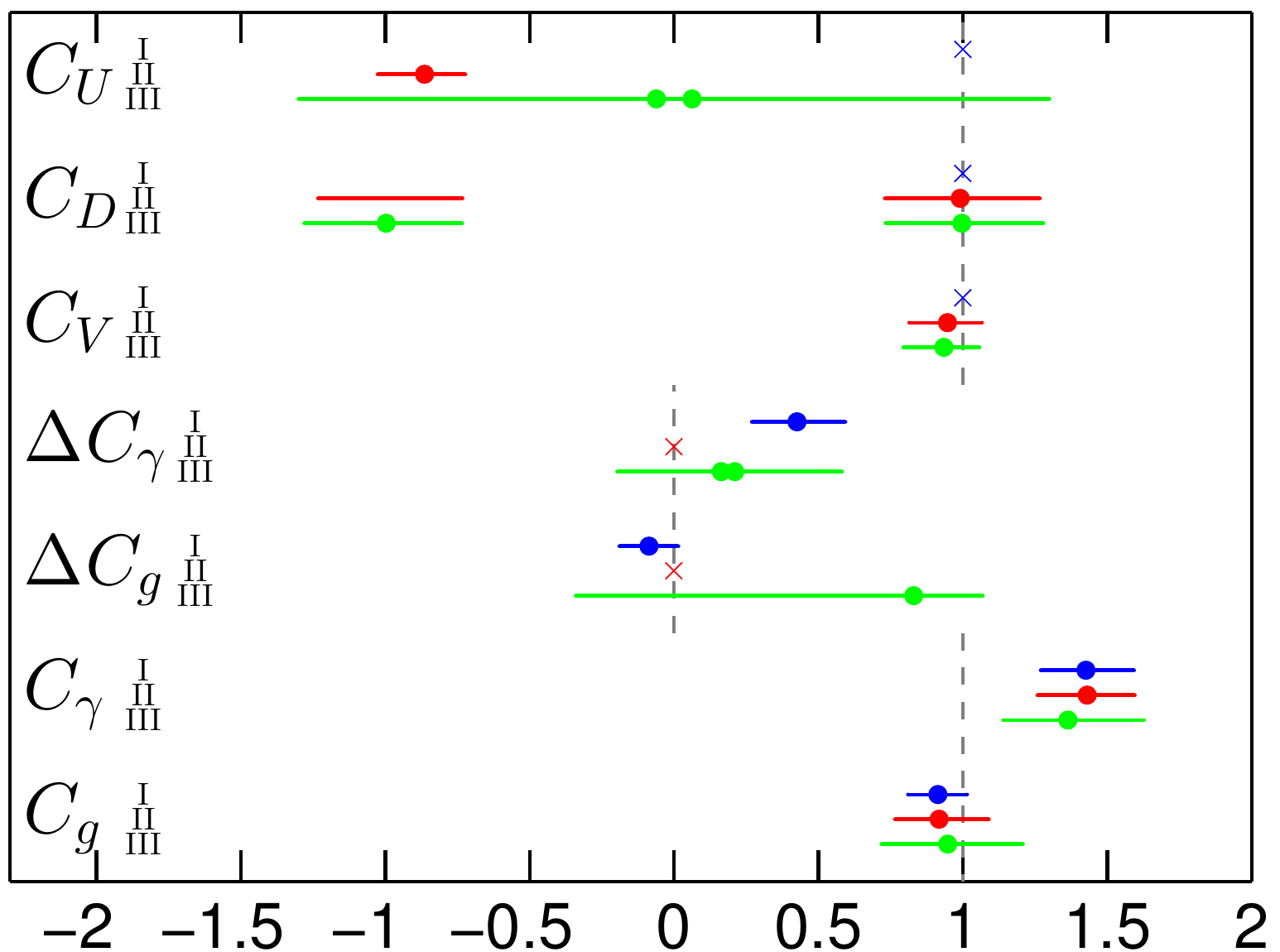}
\caption{Graphical representation of the best fit values for $\cu$, $\cd$, $\cv$, $\dcp$ and $\dcg$ of Table~\ref{chisqmintable}.   The labels refer to the fits discussed in the text.  The dashed lines indicate the SM value for the given quantity. 
The $\times$'s indicate cases where the parameter in question was fixed to its SM value. 
}
\label{bestplot}
\end{figure}

We once again refer the reader to Tables~\ref{chisqmintable}, \ref{tab:fit2} and \ref{table:THDMfits} which 
summarize the best fit values and $1\sigma$ errors for the parameters for the various cases considered. 
In Fig.~\ref{bestplot}  we show some of these results graphically.  Moreover, in order to assess the physics associated 
with our best fit points, we give in Tables~\ref{chisqminmutable} and \ref{chisqminmutableTHDM} 
the values of the derived (theory level) signal strengths $\what\mu(\ggf,\gam\gam)$, $\what\mu(\ggf,ZZ)$, $\what\mu(\ggf,b\anti b)$, $\what\mu(\vbf,\gam\gam)$, $\what\mu(\vbf,ZZ)$, and $\what\mu(\vbf,b\anti b)$  for the best fit point in the various fits we have considered.  (These are a complete set since for the models we consider $\what \mu(X,\tau\tau)=\what \mu(X,b\anti b)$, $\what \mu (X,WW)=\what\mu(X,ZZ) $ and $\what\mu(\vbf,Y)=\what\mu(\vh,Y)$.) 
We see that in the general case both $\what\mu(\ggf,\gam\gam)$ and $\what\mu(\vbf,\gam\gam)$ are enhanced by factors 1.7--2.1 (1.8--1.9 in 2HDMs), while the other signal strengths tend to be $\lesssim 1$. When demanding $\cu>0$ without allowing for extra contributions from new particles, then only very small enhancements of 
$\what\mu(\vbf,\gam\gam)$ and $\what\mu(\vbf,ZZ)$ of the order of 1.2--1.3 are found. 

Last but not least, we strongly encourage the experimental collaborations to make as complete as possible channel-by-channel information (including the important decomposition into production modes) available, in order to allow for reliable tests of non-standard Higgs scenarios. The information currently given by ATLAS and CMS for the $\gamma\gamma$ signal is an example of good practice and should become the standard for the presentation of results for all channels.  
This would immensely help interpretation efforts such as attempted in this paper. 
The ideal case would of course be if the full likelihood distributions were made available. 

\renewcommand{\arraystretch}{1.3}
\begin{table}[t!]
\centering
\begin{tabular}{|l|c|c|c|c|}
\hline
Fit  & {\bf I} & {\bf II}, $\cu<0$ & {\bf II}, $\cu>0$ & {\bf III} \\
\hline 
 $\what\mu(\ggf,\gam\gam)$ & $1.71_{-0.32}^{+0.33}$ & $1.81_{-0.41}^{+0.43}$ & $1.07 \pm 0.18$ & $1.79_{-0.34}^{+0.36}$ \\		
 $\what\mu(\ggf,ZZ)$ & $0.84_{-0.17}^{+0.18}$ & $0.79 \pm 0.15$ & $0.97 \pm 0.20$ & $0.84_{-0.18}^{+0.21}$ \\
 $\what\mu(\ggf,b\anti b)$ & $0.84_{-0.17}^{+0.18}$ & $0.87_{-0.40}^{+0.57}$ & $0.63_{-0.26}^{+0.36}$ & $0.96_{-0.43}^{+0.59}$ \\
 $\what\mu(\vbf,\gam\gam)$ & $2.05_{-0.44}^{+0.54}$ & $1.92_{-0.68}^{+0.78}$ & $1.66_{-0.63}^{+0.70}$ & $1.74_{-0.73}^{+0.84}$ \\
 $\what\mu(\vbf,ZZ)$ & $1.00 \pm 0.02$ & $0.84_{-0.36}^{+0.42}$ & $1.50_{-0.46}^{+0.50}$ & $0.82_{-0.35}^{+0.38}$ \\
 $\what\mu(\vbf,b\anti b)$ & $1.00 \pm 0.02$ & $0.92 \pm 0.30$ & $0.98 \pm 0.32$ & $0.93_{-0.29}^{+0.25}$ \\
 \hline
 \end{tabular}
\caption{Summary of $\what\mu$ results for Fits {\bf I}--{\bf III}.  
For Fit {\bf II}, the tabulated results are for the best fit with $\cu<0$, column 1 of Table~\ref{tab:fit2}, and for the 
case $\cu,\cd>0$, column 3 of Table~\ref{tab:fit2}. }
\label{chisqminmutable}
\end{table}

\begin{table}[t!]
\centering
\begin{tabular}{|l||c|c||c|c|}
\hline
Fit  & 2HDM-I & 2HDM-II & 2HDM-I, $\tan\beta>1$  & 2HDM-II, $\tan\beta>1$  \\
\hline 
 $\what\mu(\ggf,\gam\gam)$ & $1.86_{-0.38}^{+0.41}$ & $1.81_{-0.40}^{+0.41}$ & $0.97_{-0.09}^{+0.03}$ & $1.08_{-0.26}^{+0.25}$ \\
 $\what\mu(\ggf,ZZ)$ & $0.81 \pm 0.11$ & $0.79_{-0.18}^{+0.17}$ & $0.91_{-0.13}^{+0.10}$ & $1.08_{-0.21}^{+0.17}$ \\
 $\what\mu(\ggf,b\anti b)$ & $0.80_{-0.37}^{+0.54}$ & $0.88_{-0.29}^{+0.35}$ & $0.70_{-0.27}^{+0.37}$ & $0.96_{-0.14}^{+0.24}$ \\
 $\what\mu(\vbf,\gam\gam)$ & $1.87_{-0.65}^{+0.77}$ & $1.91_{-0.63}^{+0.36}$ & $1.27_{-0.34}^{+0.33}$ & $1.08_{-0.26}^{+0.16}$ \\
 $\what\mu(\vbf,ZZ)$ & $0.81_{-0.33}^{+0.41}$ & $0.84_{-0.33}^{+0.20}$ & $1.19_{-0.25}^{+0.21}$ & $1.08_{-0.22}^{+0.16}$ \\
 $\what\mu(\vbf,b\anti b)$ & $0.81 \pm 0.11$ & $0.93_{-0.20}^{+0.10}$ & $0.91_{-0.13}^{+0.10}$ & $0.95_{-0.11}^{+0.10}$ \\
 \hline
 \end{tabular}
\caption{Summary of $\what\mu$ results for the interpretation in 2HDM models.}
\label{chisqminmutableTHDM}
\end{table}
\renewcommand{\arraystretch}{1.0}

\clearpage
\section*{Acknowledgements} 

We are indebted to Guillaume Drieu La Rochelle for help in extracting 
the signal strength information 
in the ATLAS $\gamma\gamma$ and $\tau\tau$ and the CMS $\gamma\gamma$ channels, and to 
Albert de Roeck for detailed discussions on the CMS results. 
Moreover, we gratefully acknowledge discussions with 
Ada Farilla, Jonathan Hays, Marumi Kado, Tom LeCompte, Koji Nakamura and Mayda Velasco.

This work was supported in part by US DOE grant DE-FG03-91ER40674 and by IN2P3 under contract PICS FR--USA No.~5872. 
UE acknowledges partial support from the French ANR~LFV-CPV-LHC, ANR~STR-COSMO and the European Union FP7 ITN INVISIBLES (Marie Curie Actions,~PITN-GA-2011-289442).
GB, UE, JFG and SK acknowledge the hospitality and the inspiring working atmosphere  
of the Aspen Center for Physics which is supported by the National Science Foundation Grant No.\ PHY-1066293.
GB, JFG and SK also thank the Galileo Galilei Institute for Theoretical Physics for  hospitality and the INFN for partial support.
JFG acknowledges the generous hospitality while completing this work of the Kavli Institute for Theoretical Physics which is supported
by the National Science Foundation under Grant No.  NSF PHY11-25915.


\end{document}